\documentclass[preprint,superscriptaddress,amsmath,amssymb,aps]{revtex4-1}
\usepackage{graphicx}
\usepackage{dcolumn}
\usepackage{booktabs}
\usepackage{bm}
\usepackage{subfig}
\usepackage{amsmath}
\newtheorem{remark}{Remark}
\usepackage{ulem}
\usepackage{soul}
\usepackage{color,xcolor}
\sethlcolor{yellow}
\setstcolor{red}

\usepackage{cancel}

\usepackage{color}

\graphicspath{{Figs/}}

\begin{document}

\preprint{APS/123-QED}

\title{\textcolor{blue}{Discrete unified gas kinetic scheme for the conservative Allen-Cahn equation}}
\author{Chunhua Zhang}
\email{zhangch6@sustech.edu.cn}
\affiliation{
Guangdong Provincial Key Laboratory of Turbulence Research and Applications, Department of Mechanics and Aerospace Engineering, Southern University of Science
 and Technology, Shenzhen 518055, Guangdong, China}
\author{Hong Liang}
\affiliation{
Department of Physics, Hangzhou Dianzi University, Hangzhou 310018, China}
\author{Zhaoli Guo}
\email{zlguo@hust.edu.cn}
\affiliation{
State Key Laboratory of Coal Combustion, Huazhong University of Science and Technology, Wuhan 430074, China}
\author{Lian-Ping Wang}
\email{wanglp@sustech.edu.cn}
\affiliation{
Guangdong Provincial Key Laboratory of Turbulence Research and Applications, Department of Mechanics and Aerospace Engineering, Southern University of Science
 and Technology, Shenzhen 518055, Guangdong, China}
\affiliation{
 Center for Complex Flows and
Soft Matter Research, Southern University of Science
 and Technology, Shenzhen 518055, Guangdong, China}
 \affiliation{
 Guangdong-Hong Kong-Macao Joint Laboratory for Data-Driven Fluid Mechanics and Engineering Applications,
, Southern University of Science
 and Technology, Shenzhen 518055, Guangdong, China}
\date{\today}

\begin{abstract}
In this paper, the discrete unified gas kinetic scheme (DUGKS) with an improved microflux across the cell interface for the conservative Allen-Cahn equation (CACE) is proposed.
In the context of DUGKS,  the recovered kinetic equation from the flux evaluation with linear reconstruction in the previous DUGKS is analyzed. It is found that the calculated microflux across the cell interface is only the solution to the target kinetic equation with first order accuracy, which can result in an inaccurate CACE since the force term is involved or the first moment of the collision model has no conservation property.
To correctly recover the kinetic equation up to the second order accuracy, the value of the distribution function that will propagate along the characteristic line with ending point at the cell interface is appropriated by the parabolic reconstruction   instead of the linear reconstruction.
To validate the accuracy of the present DUGKS for the CACE,  several benchmark problems, including the diagonal translation of a circular interface, the rotation of a Zaleska disk and the deformation of a circular interface, have been simulated.
 Numerical results  show that the present DUGKS scheme is able to capture the interface with improved accuracy when compared with the previous DUGKS.
\end{abstract}
\maketitle

\section{Introduction}
Immiscible multiphase flows occur in nature and many engineering applications, such as petroleum industry, geological storage of carbon dioxide, bubble reactors in micro devices and liquid jets. The accurate representation and evolution of the interface between different phases are crucial in simulations of immiscible multiphase flows to mimic the surface tension force and wetting behavior. Up to now, many numerical methods for tracking interface have been developed, which can be broadly classified into two categories: sharp-interface~\cite{unverdi1992front,sussman1994level} and diffuse interface approaches~\cite{anderson1998diffuse}. In sharp interface approach, the interface between phases is represented as a surface of zero thickness incorporating the interfacial boundary condition for capillary force. It is also assumed that physical quantities, such as density and viscosity, are discontinuous across the interface generally. This treatment is prone to produce numerical instabilities when dealing with rapid topological changes and multiphase flow with large density or viscosity ratios. In contrast,  the interface in diffuse-interface approach is represented as a transition region of small but finite width. All physical quantities of the two fluids undergo rapid but smooth variation in the interfacial region. This feature provides great advantages in treating with complex topological variation of the interface~\cite{acar2009simulation}.

The phase field method, as one of the diffuse interface approaches,
has been widely used to identify and track interface between immiscible fluids and
 has also gained a great success in  multiphase flows due to its high accuracy and efficiency~\cite{gurtin1996two,jacqmin1999calculation,ding2007diffuse,kim2012phase}.
In this method, a scalar function called the order parameter is used to identify different phases. The interface can be implicitly defined as the zero-contour of the order parameter. The evolution of the order parameter is governed by the phase field equations, which  may be classified into Cahn-Hilliard type equation(CHE)~\cite{cahn1958free,zhang2019high,kim2016basic,zhang2019fractional} and Allen-Cahn type equation (ACE)~\cite{allen1979microscopic,chiu2011conservative,jeong2017conservative,ma2017numerical}. The most popular phase field equation in the simulations of multiphase flow may be the CHE that satisfies the global mass conservation naturally. However,  CHE is a fourth-order partial differential equation, which is cumbersome in numerical discretization.
In contrast,  ACE is a second-order partial differential equation that is easy to solve numerically. However, the well-known classical ACE is lack of the conservation of mass. To this end,
Rubinstein an Sternberg~\cite{rubinstein1992nonlocal} introduced a nonlocal Allen-Cahn equation with a time dependent Lagrange multipiler to  enforce the mass conservation property. However, it is found that this type ACE fails to keep small features, for example, a small droplet or bubble can dissolve into its surround region. To resolve this insufficiency, Brassel and Bretin et al.~\cite{bretin2009modified} proposed a space-time dependent Lagrange multiplier and the resulting nonlocal Allen-Cahn equation  can not only maintain the mass conservation but also present good performance of tracking geometric features of interfaces. It is still difficult for both nonlocal ACE to maintain the local mass conservation due to the Lagrange multipiler acting on the global computational domain.
 On the other hand, Chiu and Lin et al.~\cite{chiu2011conservative} proposed a local conservative Allen Cahn equation based on the work of Sun and Beckermann~\cite{sun2007sharp}. In their derivation, a anti-diffusive term is added into the general interface advection equation without considering curvature-driven interface motion in multiphase flows. Meanwhile,  the hyperbolic tangent interfacial profile of the order parameter derived from the equilibrium system equipped with  double-well free energy function  is carefully used to reformulate the phase field equation in a conservative form.

From a numerical point of view, many techniques can be used to numerically solve the interface tracking equation. In recent years, the lattice Boltzmann equation (LBE) method has gained a great attention. Compared to the traditional computational fluid dynamics,  it has demonstrated some computational advantages, such as high parallelization efficiency, simple boundary treatment and easy programming.  In the framework of LBE, the target equation is approximated by solving a simplified megascopic kinetic equation.
Therefore, many efforts have been made to construct an efficient LBE model that is able to accurately recover the phase field equation though multiscale expansions. For the CHE, Zheng et al.~\cite{zheng2005lattice} first proposed a modified LBE  that can recover the CHE through the \textcolor{red}{Chapman-Enskog} analysis. Based on the similar idea, Zu and He~\cite{zu2013phase} presented an improved LBE model for solving the CHE in terms of accuracy and stability. Liang et al.~\cite{liang2014phase} used the standard LBE and a modified force term involving  time derivative to exactly recover the CHE. For the conservative ACE, Geier et al.\cite{geier2015conservative} first proposed an LBE model with central-based collision operator, which showed better accuracy than CHE in terms of tracking the interface. However, it is found that there exist some additional terms in the recovered macroscopic equation in the Gerer's model. To remove these terms, Ren et al.~\cite{ren2016improved} introduced a time-derivative term in the source term. Similarly, Wang et al.~\cite{wang2016comparative} used the time-derivative term to remove the additional terms and offered an alternative algorithm to calculate the gradient of the order parameter using the nonequilibrium part of the distribution function.  Recently, Begmohammadi et al.~\cite{begmohammadi2020study} conducted a comparative investigation on the differences among the above LBE models and showed the additional terms can be negligible in the limit of low Mach numbers. More recently, Zu et al.~\cite{zu2020phase} proposed a modified LBE version in which the difference of equilibrium distribution function is added  to correctly recover the local conservative ACE.

Although the above LBE models for the phase field equation are able to produce improved accuracy and stability of interface tracking in different degrees, all of these methods still suffer from substantial drawbacks coming from the LBE algorithm itself, such as the uniform grid and fixed Courant-Friedrichs-Lewy (CFL) condition. To address these deficiencies, both
the finite difference  and the finite volume versions of the discrete Boltzmann equation have been developed. Among these methods, the discrete unified gas kinetic scheme (DUGKS) that is originally developed for simulating multiscale flows based on kinetic models has received particular attention in recent years~\cite{guo2013discrete,guo2021progress}. Due to its finite-volume formulation and decoupling of time step and mesh size, DUGKS is a competitive tool even for continuum flows. Zhang et al.~\cite{zhang2018discrete} first extended the DUGKS to two-phase flows in which the CHE is used to track the interface. Recently, Yang et al.\cite{yang2019phase}  extended the DUGKS to model two-phase flows with large density ratios  using the local conservative ACE for tracking the interface. However, their results shown that the DUGKS  fails to accurately identify the complicated interface patterns due to the large numerical dissipation of DUGKS compared to the LBE. In the present work, we aim to develop an improved DUGKS method for the conservative ACE to capture the interface more accurately.

The outline of the paper is as follows. In Sec. II, the conservative phase field equation   and the kinetic equation with the well-defined equilibrium distribution function and force term that can recover the phase field equation  are presented. Then  the DUGKS scheme is briefly introduced and  the recovered equation during the flux evaluation is analyzed. According to the analysis, the DUGKS scheme with corrected flux evaluation is proposed. Numerical simulations and discussions are made in Sec III. Finally, a brief summary is given in Sec. IV .

\section{Discrete unified gas kinetic scheme model for the CACE}
\subsection{Conservative Allen-Cahn equation}
In the phase field theory, the dynamics of the order parameter is governed by minimizing the following Ginzburg-Landau energy function $\Psi$~\cite{badalassi2003computation}
\begin{equation}\label{eq:free_energy}
\Psi(\phi,\nabla\phi)=\int_V\left( \epsilon(\phi)+\frac{\kappa}{2} |\nabla\phi|^2 \right) dV,
\end{equation}
where $\phi$ is the order parameter, $\epsilon=\beta(\phi^2-1)^2$ is a double-well potential function representing for the bulk free energy, the second term represents the interfacial energy. $\beta$ and $\kappa$ are parameters related to the surface tension $\sigma$ and interface thickness $W$, i.e., $\beta=3\sigma/(4W)$ and $\kappa=3W\sigma/8$.
The classical advection Allen-Cahn equation can be written as~\cite{allen1979microscopic}
\begin{equation}\label{eq:classical_ACE}
\frac{\partial \phi}{\partial t} +\bm u\cdot \nabla\phi=-M_{\phi}\mu_{\phi},
\end{equation}
where  $\bm u$ is the velocity, $M_{\phi}$ is the mobility coefficient,
$\mu_{\phi}$ is the chemical potential defined as the variational derivative of the energy functional with respect to $\phi$~\cite{dadvand2021advected},
\begin{equation}\label{eq:chemical}
\begin{aligned}
\mu_{\phi}=& \frac{\partial \epsilon}{\partial \phi} -\kappa\bm n\bm n:\nabla\phi\nabla\phi
-\kappa|\nabla\phi|\nabla\cdot\bm n,
\end{aligned}
\end{equation}
where $\bm n=\nabla\phi/|\nabla |\phi|$ is the normal unit vector. By considering a flat interface at the equilibrium state (i.e., $\nabla\cdot\bm n=0$), the chemical potential is equal to zero, which  gives rise to a hyperbolic tangent profile along the normal direction of the interface,
\begin{equation}\label{eq:tanh_phi}
\phi=\tanh\left(\frac{2s}{W}\right),
\end{equation}
where $s$ is the coordinate normal to the interface. By using Eq.~(\ref{eq:tanh_phi}), one can have
\begin{equation}\label{eq:absphi}
|\nabla\phi|=\frac{\partial \phi}{\partial s}=\frac{2(1-\phi^2)}{W}=\Theta.
\end{equation}
Since the classical ACE does not guarantee mass conservation, several conservative AC equation has been proposed. Here the CACE proposed by Chiu and Lin~\cite{chiu2011conservative} is used. For the CACE, the evolution of the order parameter could be interpreted as the minimization of the following free energy functional in $L^2$ space,
\begin{equation}\label{eq:new_free_energy}
  \Psi(\phi,\nabla\phi)=\int_V \frac{\lambda}{2}\left(|\nabla\phi|- \frac{2(1-\phi^2)}{W}\right)^2dV,
\end{equation}
where $\lambda$ is a positive parameter. The variational derivative of the above  free energy functional with respect to $\phi$ leads to
\begin{equation}\label{eq:new_chemical}
\mu_L=\frac{\delta \Psi_L}{\delta \phi}
=\lambda\left(\Theta-|\nabla \phi|\right)\frac{\partial\Theta}{\partial\phi}
-\nabla\cdot\lambda(\nabla\phi-\Theta\bm n),
\end{equation}
Then, substituting Eq.~(\ref{eq:new_chemical}) into Eq.~(\ref{eq:classical_ACE}) yields
\begin{equation}\label{eq:CAC}
\frac{\partial \phi}{\partial t}+\bm u\cdot \nabla\phi=
\nabla\cdot M_\phi(\nabla\phi-\Theta\bm n)-M_\phi\left(\Theta-|\nabla\phi|\right)\frac{\partial\Theta}{\partial \phi},
\end{equation}
where $M_\phi$ is redefined as $M_\phi\lambda$.
It is noted that Eq.~(\ref{eq:CAC}) does not precisely conserve the mass. With  Eq.~(\ref{eq:absphi}) in mind, the terms on the right hand of Eq.~(\ref{eq:CAC}) can actually enable the interface to be a hyperbolic tangent profile. Assume that the deviation of the order parameter from the equilibrium profile is small so that the second term on the right-hand side of Eq.(\ref{eq:CAC}) can be neglected. Then the above equation can be rewritten as
\begin{equation}\label{eq:targetCAC}
\frac{\partial\phi}{\partial t}+\nabla\cdot(\phi\bm u)=\nabla\cdot
M_{\phi}(\nabla\phi-\bm n\Theta),
\end{equation}
where the incompressible condition is enforced, which is consistent with the CACE in Ref.~\cite{chiu2011conservative}. It is worth noting that the resulting CACE Eq.(\ref{eq:targetCAC}) is different from the original ACE. The coefficients $\beta$ and $\kappa$ do not appear in CACE. In fact, $\Theta$ could be considered as a kernel function that enforces the order parameter across the interface to be a tanh profile.

With the reference velocity $U_c$, reference length $L_c$ for $\nabla\phi$ and reference length $L_0$ for $\Theta\bm n$, the dimensionless phase field equation can be expressed as
\begin{equation}\label{eq}
\frac{\partial \phi}{\partial t}+\nabla\cdot(\phi \bm u)=\nabla\cdot \frac{1}{\text{Pe}}\left(\nabla\phi-\text{Cn}\Theta \bm n \right),
\end{equation}
where $\text{Pe}=\frac{U_c L_c}{M_\phi}$ is the Peclet number and $\text{Cn}=L_c/L_0$ is the Cahn number.
Here we employed two reference lengths, one is introduced by considering the gradient of the order parameter and the other is used by considering the kernel function $\Theta$ acting on the whole domain. Typically, $L_c$ is set as the interface width $W$ while $L_0$ is chosen as the size of computational domain.

\subsection{Discrete velocity kinetic equation for CACE}
The discrete kinetic equation with the Bhatnagar-Gross-Krook(BGK) collision model can be written as
\begin{equation}\label{eq:kineticequaiton}
\frac{\partial f_\alpha}{\partial t} +\bm \xi_\alpha\cdot\nabla f_\alpha=
-\frac{f_\alpha-f_\alpha^{eq}}{\tau_f}+F_\alpha,
\end{equation}
where $f_{\alpha}=f(\bm x,\bm\xi_\alpha,t)$ is the particle  distribution function with discrete velocity $\bm \xi_\alpha$ at position $\bm x$ and time $t$. The subscript $\alpha$ denotes the discrete velocity along $\alpha$ direction,
$f_\alpha^{eq}$ is the equilibrium distribution function, $F_\alpha$ is the source term.
In the current study, only two-dimensional problems are considered as an illustration.
The  well-known two-dimensional-nine-velocities(D2Q9) lattice model is considered, in which
 the discrete velocities are defined as~\cite{qian1992lattice}
\begin{equation}
\bm \xi_\alpha =c\left(\begin{array}{ccccccccc}
0,1,0,-1,0,1,-1,-1,1,\\
0,0,1,0,-1,1,1,-1,-1.
\end{array} \right),
\end{equation}
where \textcolor{magenta}{$c=\sqrt{3RT}$ with $R$ being the gas constant and
$T$ being the temperature}. The equilibrium distribution function is given by
\begin{equation}\label{eq:equilibrium-feq}
f_\alpha^{eq}=\omega_\alpha \phi\left[1+\frac{\bm \xi_\alpha\cdot \bm u}{c_s^2} +\frac{(\bm \xi_\alpha\cdot\bm u)^2}{2c_s^4}-\frac{|\bm u|^2}{2c_s^2} \right],
\end{equation}
where $c_s=c/\sqrt{3}$ is the sound speed. The source term $F_\alpha$ is defined as
\begin{equation}\label{eq:force-gks}
F_\alpha=
\omega_i\Theta\bm \xi_\alpha\cdot \bm n,
\end{equation}
The order parameter is updated by
\begin{equation}\label{eq:macro-gks}
\phi(\bm x,t)=\sum_\alpha f_\alpha.
\end{equation}

The recovered equation at the continuum level from the above model is
\begin{equation}\label{eq:recovered_equation}
\partial_t \phi+\nabla\cdot (\phi \bm u)=\nabla\cdot M_\phi(\nabla\phi -\Theta\bm n)
 + \frac{M_\phi}{c_s^2}\nabla\cdot\left[
 \partial_t (\phi \bm u)+\nabla\cdot(\phi \bm u\bm u) \right],
\end{equation}
where $M_{\phi}=c_s^2 \tau_f$ is the mobility. Details of the derivatives are provided in Appendix~\ref{ap:dukgs-ace}.
Compared with (\ref{eq:targetCAC}),  the additional terms are $\frac{M_\phi}{c_s^2}\nabla\cdot\left[
 \partial_t (\phi \bm u)+\nabla\cdot(\phi \bm u\bm u) \right]$,
which is order of $\text{Ma}^2/\text{Pe}$ with $\text{Ma}=U_c/c_s$ being the Mach number based on the dimensional analysis.
Because the equilibrium distribution function is valid for small Mach number (i.e., $\text{Ma}\le0.3$), these additional terms can be negligible, which will be proved in the next section.

In Ref.~\cite{yang2019phase}, the distribution function is defined as
\begin{equation}\label{eq:feq-second-DBE}
f_\alpha^{eq}=\omega_\alpha \phi\left(1+\frac{\bm \xi_\alpha\cdot \bm u}{c_s^2}\right),
\end{equation}
and the force term is given by
\begin{equation}\label{eq:Fi-second-DBE}
F_\alpha=
\omega_i\Theta\bm \xi_\alpha\cdot \bm n+\omega_i\frac{\bm\xi_\alpha\cdot\partial_t(\phi\bm u)}{c_s^2}.
\end{equation}
The order parameter is still updated by Eq.(\ref{eq:macro-gks}).
It can be proved that  Eqs.(\ref{eq:kineticequaiton}), (\ref{eq:feq-second-DBE}) and (\ref{eq:Fi-second-DBE})  are able to recover  the correct CACE with the second-order accuracy in $\Delta t$. However, the leading error terms from both kinetic methods are different.
To understand the effects of the additional terms, both kinetic models are calculated by the proposed DUGKS scheme that will be given later.

\subsection{Discrete unified gas-kinetic scheme}~\label{sec:DUGKS}
\begin{figure}
  \centering
\subfloat[]{%
  \includegraphics[width=0.4\textwidth]{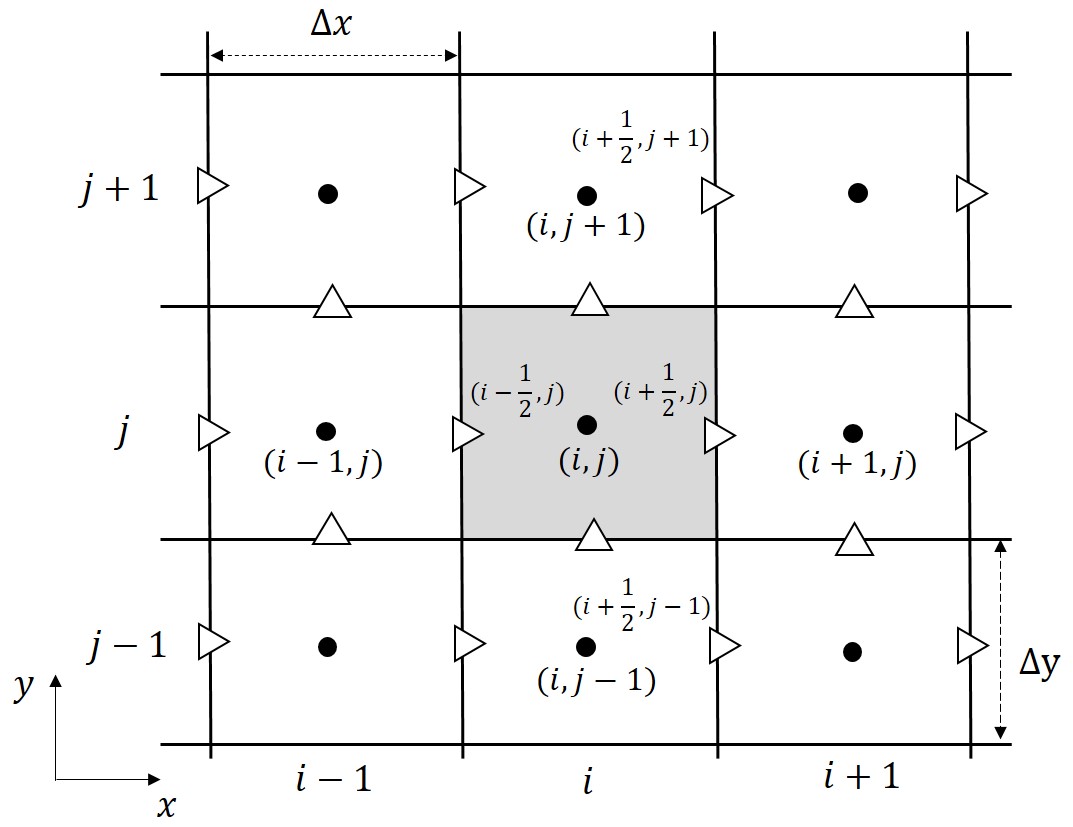}~
  }~
\subfloat[]{%
  \includegraphics[width=0.4\textwidth]{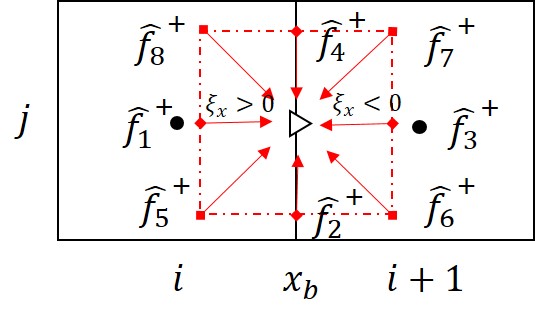}
  }
  \caption{(a) A typical control volume cell and the notation used for a Cartesian 2D grid and (b) distribution functions at the cell face.}\label{mesh}
\end{figure}

Without loss of generality, we divide the spatial domain into discrete rectangular control volumes with equal grid spacing $\Delta x$ and $\Delta y$ in  x and y directions respectively, as shown Fig.~\ref{mesh}(a).
First, integrating Eq.~(\ref{eq:kineticequaiton}) over a control volume and applying the divergence theorem to the second term yield
\begin{equation}\label{eq:1-integrating}
\frac{\partial \bar{f}_\alpha}{\partial t}
+\frac{1}{|V_{ij}|}\oint (\bm \xi_\alpha f_\alpha)\cdot \bm ndA=\bar{\Omega}_\alpha+\bar{F_\alpha},
\end{equation}
where $\bar{\Omega}_\alpha=-(\bar{f}_\alpha-\bar{f}_\alpha^{eq})/\tau_f$,
$V_{ij}$ is the control volume centered at the node $(i\Delta x,j\Delta y)$ that is coordinate index in the computational domain, $A$ is the cross-sectional area of the control volume face, $\bm n$ is a unit vector normal to the surface and pointing outward, $\bar{f}_\alpha$, $\bar{f}_\alpha^{eq}$ and $\bar{F}_\alpha$ are  cell-averaged values of the corresponding distribution functions, which are defined as
\begin{equation}\label{eq:spatial_average}
\begin{aligned}
\bar{f}_\alpha =& \frac{1}{|V_{ij}|}\int_{V_{ij}} f_\alpha d\bm x, \\
\bar{f}_\alpha^{eq}=&\frac{1}{|V_{ij}|}\int_{V_{ij}} f_\alpha^{eq}d \bm x, \\
\bar{F}_\alpha=& \frac{1}{|V_{ij}|} \int_{V_{ij}} F_\alpha d\bm x.
\end{aligned}
\end{equation}
Then, integrating Eq.~(\ref{eq:1-integrating}) from time level $n\Delta t$ to the next time level $(n+1)\Delta t$ and
using the trapezoidal rule  for the collision term and external force term and applying the midpoint rule for the integration of the convection term, one can obtain
\begin{equation}\label{eq:2-integrating}
\bar{f}_\alpha^{n+1}-\bar{f}_\alpha^{n}+\frac{\Delta t}{|V_{ij}|}\oint(\bm \xi_\alpha\cdot\bm n){f}_\alpha^{n+1/2} d\partial A=\frac{\Delta t}{2}(\bar{\Omega}_\alpha^{n+1}
+\bar{\Omega}_\alpha^{n})
+\frac{\Delta t}{2}( \bar{F}_\alpha^n+\bar{F}_\alpha^{n+1}).
\end{equation}

To remove the implicity, the following auxiliary distribution functions are introduced,
\begin{subequations}\label{auxftilde}
\begin{align}\label{eq:first_auxiliary1}
\widetilde{f}_\alpha= & \bar{f}_\alpha-\frac{\Delta t}{2}\bar{\Omega}_\alpha-\frac{\Delta t}{2} \bar{F}_\alpha
=\frac{2\tau_f+\Delta t}{2\tau_f}\bar{f}_\alpha-\frac{\Delta t}{2\tau_f}\bar{f}_\alpha^{eq}-\frac{\Delta t}{2}\bar{F}_\alpha,
 \\
\widetilde{f}_\alpha^+=& \bar{f}_\alpha+\frac{\Delta t}{2}\bar{\Omega}_\alpha  + \frac{\Delta t}{2} \bar{F}_\alpha
=\frac{2\tau_f-\Delta t}{2\tau_f}\bar{f}_\alpha+\frac{\Delta t}{2\tau_f}\bar{f}_\alpha^{eq}+\frac{\Delta t}{2}\bar{F}_\alpha,
\label{eq:first_auxiliary2}
\end{align}
\end{subequations}
Substituting Eqs.(\ref{eq:first_auxiliary1}),(\ref{eq:first_auxiliary2}) into Eq.(\ref{eq:2-integrating}), one can obtain
\begin{equation}\label{eq:evolution-dugks}
\widetilde{f}_\alpha=\widetilde{f}_\alpha^{+}-\frac{\Delta t}{|V_{ij}|}\oint(\bm \xi_\alpha \cdot \bm n)f_\alpha^{n+1/2}dA,
\end{equation}
The second term on the right-hand side of Eq.(\ref{eq:evolution-dugks}) can be approximated by different integration formulas.
Generally, the midpoint rule is used for  the cell face line integrals. Then, Eq.(\ref{eq:evolution-dugks}) can be rewritten as

\begin{equation}\label{eq:evolution-dugks-midpoint}
\begin{aligned}
\widetilde{f}_\alpha=&\widetilde{f}_\alpha^{+}-\frac{\Delta t}{|V_{ij}|}\times \\
&\left[
\left( (\bar{f}_\alpha)_{i+\frac{1}{2},j}-(\bar{f}_\alpha)_{i-\frac{1}{2},j}  \right)\xi_{x,\alpha} \Delta y
+
\left((\bar{f}_\alpha)_{i,j+\frac{1}{2}}-(\bar{f}_\alpha)_{i,j-\frac{1}{2}}\right)\xi_{y,\alpha} \Delta x \right],
\end{aligned}
\end{equation}
where
\begin{equation}\label{eq:cell-face-integral}
\begin{aligned}
(\bar{f}_{\alpha})_{i+\frac{1}{2},j}= \frac{1}{\Delta y}\int_{y_{j-\frac{1}{2}}}^{y_{j+\frac{1}{2}}}
f_{\alpha}(x_{i+\frac{1}{2}},y)  dy,
\qquad
(\bar{f}_{\alpha})_{i,j+\frac{1}{2}}= \frac{1}{\Delta x}\int_{x_{i-\frac{1}{2}}}^{x_{i+\frac{1}{2}}}
f_\alpha(x,y_{j+\frac{1}{2}})  dx,
\end{aligned}
\end{equation}
Based on Eq.(\ref{eq:first_auxiliary1}), the conserved parameter is calculated by
\begin{equation}\label{eq}
\phi=\sum_\alpha \widetilde{f}_\alpha.
\end{equation}
As a result, we only need to track the distribution function $\widetilde{f}_\alpha$ instead of $\bar{f}_\alpha$.

In Eq.~(\ref{eq:evolution-dugks-midpoint}), the  unknown parameters are the interface-averaged values at the cell face. Fortunately, the pointwise values at the cell face can be obtained by using the discrete kinetic equation itself. Once the pointwise values at the cell face are determined, the interface-averaged values can be reconstructed by some means.  As shown in Fig.~\ref{mesh}(b), integrating Eq.(\ref{eq:kineticequaiton}) within a half time step $s=\Delta t/2$ along the characteristic line and assuming the end point located at the cell interface $\bm x_b$ (e.g, $x_i+\Delta x/2$ or $y_j+\Delta y/2$) lead to
\begin{equation}\label{eq:flux_evolution}
f_\alpha^{n+s}(\bm x_b)-f_\alpha^{n}(\bm x_b-\bm \xi_\alpha s)
=\frac{s}{2}[\Omega_\alpha^{n+s}(\bm x_b)+\Omega_\alpha^n(\bm x_b-\bm \xi_\alpha s)]+\frac{s}{2}[F_\alpha^{n+s} (\bm x_b)+F_\alpha^{n}(\bm x_b-\bm \xi_\alpha s)].
\end{equation}
Similarly, to eliminate the implicity, another two auxiliary distribution functions $\widehat{f}_i$
 and $\widehat{f}_i^{+}$ are introduced,
\begin{subequations}\label{auxfhat}
 \begin{align}\label{second0-auxiliary}
\widehat{f}_\alpha=f_i-\frac{s}{2}\Omega_\alpha-\frac{s}{2}F_\alpha=
\frac{2\tau_f+s}{2\tau_f} f_\alpha-\frac{s}{2\tau_f}f_\alpha^{eq}-\frac{s}{2}F_\alpha,
\\
\widehat{f}_\alpha^+=f_\alpha+\frac{s}{2}\Omega_\alpha+\frac{s}{2}F_\alpha=
\frac{2\tau_f-s}{2\tau_f} f_\alpha+\frac{s}{2\tau_f}f_\alpha^{eq}+\frac{s}{2}F_\alpha,
\label{second1-auxiliary}
 \end{align}
 \end{subequations}
 Then, Eq.(\ref{eq:flux_evolution}) can be rewritten as
 \begin{equation}\label{eq:flux_evolution2}
 \begin{aligned}
 \widehat{f}_\alpha^{n+s}(\bm x_b)= \widehat{f}_\alpha^{n,+}(\bm x_b-\bm \xi_\alpha s).
 \end{aligned}
 \end{equation}
From Eq.(\ref{eq:flux_evolution2}), the point value of $\widehat{f}_\alpha^{n,+}(\bm x_b-\bm \xi_\alpha s)$ must be reconstructed in advance.
In the published DUGKS methods, $\widehat{f}_\alpha^{n,+}(\bm x_b-\bm \xi_\alpha s)$ is approximated by the piecewise linear reconstruction, namely,
\begin{equation}\label{eq:flux_evolution3}
  \widehat{f}_\alpha^{n}(\bm x_b-\bm \xi_\alpha s)
 = \widehat{f}_\alpha^{n,+}(\bm x_b)
 -s \bm \xi_\alpha\cdot \nabla \widehat{f}_\alpha^{n,+}(\bm x_b).
 \end{equation}
 The values of $\widehat{f}_\alpha^{n,+}(\bm x_b) $ and its gradient $\nabla \widehat{f}_\alpha^{n,+}(\bm x_b)$ at the cell face can be obtained by different difference schemes and will be given later.
From Eq.(\ref{second0-auxiliary}), the original distribution function can be calculated by
\begin{equation}\label{eq:reconstruction-point-average}
f_\alpha^{n+s}=\frac{2\tau_f}{2\tau_f+s} \widehat{f}_\alpha+\frac{s}{2\tau_f+s}f_\alpha^{eq}+\frac{\tau_f s}{2\tau_f+s}F_\alpha.
\end{equation}

Due to the second-order accuracy of the present DUGKS,
there is no differentiation between the pointwise quantities and the face-averaged quantities (see the Appendix~\ref{ap:dugks-reconstruction}),
\begin{equation}\label{eq:reconstruction_DUGKS}
(\bar{f}_\alpha)_{ij}=(f_\alpha)_{ij},\qquad  (\bar{f}_{\alpha})_{i+\frac{1}{2},j}=(f_{\alpha})_{i+\frac{1}{2},j},
\qquad (\bar{F}_\alpha)_{ij}=(F_\alpha)_{ij}.
\end{equation}
As a result, the following formulations can be obtained by combining Eqs.(\ref{eq:reconstruction-point-average}),(\ref{auxftilde}) and (\ref{auxfhat}),
\begin{equation}\label{eq}
\widehat{f}_\alpha^+=\frac{2\tau_f-s}{2\tau_f+\Delta t}\widetilde{f}_\alpha+\frac{3s}{2\tau_f+\Delta t}\bar{f}_\alpha^{eq}
+\frac{3\tau_f s}{2\tau_f+\Delta t} \bar{F}_\alpha,
\end{equation}
\begin{equation}
\widetilde{f}_\alpha^+=\frac{4}{3}\widehat{g}_\alpha^+-\frac{1}{3}\widetilde{f}_\alpha
=\frac{2\tau_f-\Delta t}{2\tau_f+\Delta t} \widetilde{f}_\alpha
+\frac{2\Delta t}{2\tau_f+\Delta t}\bar{f}_\alpha^{eq}
+\frac{2\tau_f\Delta t}{2\tau_f+\Delta t}\bar{F}_\alpha.
\end{equation}

In addition, the time step $\Delta t$ is determined by \textcolor{red}{Courant-Friedrichs-Lewy} (CFL) condition,
\begin{equation}\label{eq}
  \Delta t=\text{CFL}\frac{\Delta x_{min}}{|\bm \xi_\alpha|_{max}+|\bm U_c|_{max}}
\end{equation}
where $\text{CFL}$ lies in $0$ and $1$, $\Delta x_{min}$ denotes the minimal grid spacing and $\bm U_c$ is the characteristic velocity.

The values of the order parameter at time  $n\Delta t$ and $(n+1/2)\Delta t$ levels are respectively updated by
\begin{equation}\label{eq}
\phi^n=\sum_\alpha \widetilde{f}_\alpha^n, \quad\phi^{n+s}=\sum_\alpha\widehat{f}_\alpha^{n+s}.
\end{equation}

\begin{remark}
In the calculations, the cell-averaged values of distribution function $\bar{f}^{eq}_\alpha$ given by Eq.(\ref{eq:feq-second-DBE}) is calculated by
\begin{equation}\label{eq}
\bar{f}^{eq}_\alpha(\bm u,\phi)=\omega_\alpha\left(\bar{\phi}+\frac{\bm \xi_\alpha}{c_s^2}\cdot\overline{\phi \bm u}\right),
\end{equation}
\textcolor{magenta}{It can be found that the average of the product of the order parameter and the velocity $\bar{\phi\bm u}$ is required. The product rules for fourth accuracy are given by
\begin{equation}\label{eq}
\overline{\phi u_d}=\bar{\phi}\bar{u}_d
+
\frac{(u_d)_y\phi_y+(u_d)_x\phi_x}{12}h^2 +\frac{u_d\phi_{xx}+u_d\phi_{yy}+ (u_d)_{xx}\phi+(u_d)_{yy}\phi }{24}
h^2+O(h^4),
\end{equation}
where $d=x, y$. Due to the DUGKS with the second order accuracy, the second term on the right hand can be neglected.} Then, the equilibrium distribution function becomes
\begin{equation}\label{eq}
\bar{f}^{eq}_\alpha(\bar{\bm u},\bar{\phi})=\omega_\alpha\left(\bar{\phi}+\frac{\bm \xi_\alpha}{c_s^2}\cdot\bar{\phi}\bar{\bm u}\right),
\end{equation}
\end{remark}
\begin{remark}
The treatment of integrating Eq.(\ref{eq:flux_evolution}) along the characteristic line is similar to that in the LBE~\cite{he1997theory}. However, LBM consists of collision and propagation processes in each time step. Especially,  the distribution functions streams from one node to its neighboring nodes perfectly. In DUGKS, a discrete time step consists of the evolution of $\widetilde{f}_\alpha$ and the  calculation of micro flux at the cell faces. During flux evaluation process, the point values at cell center need to be approximated by the cell averaged values. Then the point value of the distribution function $\hat{f}_\alpha^+(\bm x_b-\bm \xi_\alpha s)$ must be reconstructed accurately enough to prepare for the calculation of the original distribution function at the cell face.
After that,  the  point values  at cell faces should be converted into the cell face averaged values to calculate the interfacial flux.
\end{remark}
\begin{remark}
In DUGKS, there are two types of strategy to approximate the value of $\widehat{f}_\alpha^+(\bm x_b-\bm \xi_\alpha s)$. As given by Eq.(\ref{eq:flux_evolution3}), one is to assume the shape of $\widehat{f}_\alpha^+(\bm x_b-\bm \xi_\alpha)$ around the cell interface center $\bm x_b$~\cite{guo2013discrete}. The other is to assume the shape of $\widehat{f}_\alpha^+(\bm x_b-\bm \xi_\alpha)$ in the upstream cell~\cite{guo2015discrete},
\begin{equation}\label{eq}
\widehat{f}_\alpha^+(\bm x_b-\bm \xi_\alpha s,\bm \xi_\alpha )=\begin{cases}
 \widehat{f}_\alpha^+(\bm x_j,\bm \xi_\alpha)+(\bm x_b-\bm \xi_\alpha s-\bm x_j)\cdot\nabla \widehat{f}_\alpha^+(\bm x_j,\bm \xi_\alpha), & \text{$\bm\xi_\alpha >0$} \\
  \widehat{f}_\alpha^+(\bm x_{j+1},\bm \xi_\alpha)+(\bm x_b-\bm \xi_\alpha s-\bm x_{j+1})\cdot\nabla \widehat{f}_\alpha^+(\bm x_{j+1},\bm \xi_\alpha),   & \text{$\bm \xi_\alpha <0$}
\end{cases}
\end{equation}
\textcolor{magenta}{For smooth flows, both  methods are identical. However,
the second method may smear the phase interface  when the distribution function(or the order parameter) has a large gradient near the cell face.}
\end{remark}
\begin{remark}
 The accuracy of DUGKS depends on the accuracy of the integration of distribution function, the accuracy of the area-averaged fluxes at the cell faces and the accuracy of dealing with the gradient operator in the discrete force term.
 Overall, the present DUGKS scheme is a  second-order finite volume method in space due to its accuracy of the numerical integration.
\end{remark}

\subsection{Analysis of the original DUGKS}
Theoretically, the recovered equations from Eqs.(\ref{eq:flux_evolution2}) and (\ref{eq:1-integrating})
should be consistent with Eq.(\ref{eq:targetCAC}). Now, we will derive the recovered equation from Eq.(\ref{eq:flux_evolution2}).
First, we consider a Taylor series expansion of the following parameters about the time level $t=n\Delta t$:
\begin{equation}\label{eq:taylor_time}
\begin{aligned}
\bar{f}_{\alpha}^{n+1}&=\bar{f}_{\alpha}^n+\Delta t \partial_t \bar{f}_{\alpha}^n+\frac{\Delta t^2}{2}\partial_t^2 \bar{f}_{\alpha}^n+O(\Delta t^3\partial^3),\\
\bar{\Omega}_{\alpha}^{n+1}&=\bar{\Omega}_{\alpha}^n+\Delta t \partial_t \bar{\Omega}_{\alpha}^n+\frac{\Delta t^2}{2}\partial_t^2 \bar{\Omega}_{\alpha}^n+O(\Delta t^3\partial^3),\\
\bar{F}_{\alpha}^{n+1}&=\bar{F}_{\alpha}^n+\Delta t \partial_t \bar{F}_{\alpha}^n+\frac{\Delta t^2}{2}\partial_t^2 \bar{F}_{\alpha}^n+O(\Delta t^3\partial^3),
\end{aligned}
\end{equation}
where $\partial$ denotes temporal derivative or spatial derivative.
Inserting Eq.~(\ref{eq:taylor_time}) into Eq.(\ref{eq:2-integrating}) leads to
\begin{equation}\label{eq:recovered-eq-evolution}
\partial_t \bar{f}_\alpha^n+\nabla\cdot(\xi \bar{f}_\alpha^n)=\bar{\Omega}_{\alpha}^n+\bar{F}_{\alpha}^n+O(\Delta t^2\partial^3),
\end{equation}
which is consistent with Eq.(\ref{eq:kineticequaiton}) up to second order accuracy that is able to recover the target phase field equation, as presented in Appendix~\ref{ap:dukgs-ace}.

Then,  we consider the following Taylor series expansion of the following point values of discrete distribution function about the cell face $\bm x_b$:
\begin{equation}\label{eq:taylor_space}
\begin{aligned}
(f_{\alpha}^n)_{\bm x_b-\bm \xi_\alpha s}&=(f_{\alpha}^n)_{\bm x_b}-\bm \xi_\alpha s\cdot \nabla (f_{\alpha}^n)_{\bm x_b}+O(s^2\partial^2),\\
(f^{eq,n}_{\alpha})_{\bm x_b-\bm \xi_\alpha s}&=(f^{eq,n}_{\alpha})_{\bm x_b}-\bm \xi_\alpha s\cdot \nabla (f^{eq,n}_{\alpha})_{\bm x_b}+O(s^2\partial^2),\\
(F_{\alpha}^n)_{\bm x_b-\bm \xi_\alpha s}&=(F_{\alpha}^n)_{\bm x_b}-\bm \xi_\alpha s\cdot \nabla (F_{\alpha}^n)_{\bm x_b}+O(s^2\partial^2). \\
\end{aligned}
\end{equation}
Inserting Eq.(\ref{eq:taylor_space}) into Eq.(\ref{eq:flux_evolution2}) leads to
\begin{equation}\label{eq}
\begin{aligned}
& \left[f+\frac{s}{2}\frac{f-f^{eq}}{\tau_f}-\frac{s}{2}F  \right]_{\bm x_b}^n
+s\partial_t\left[f+\frac{s}{2}\frac{f-f^{eq}}{\tau_f}-\frac{s}{2}F \right]_{\bm x_b}^n
\\
&
=\left[
f-\frac{s}{2}\frac{f-f^{eq}}{\tau}+\frac{s}{2}F
\right]_{\bm x_b}^n
-\bm \xi_\alpha s\cdot \nabla \left[
f-\frac{s}{2}\frac{f-f^{eq}}{\tau}+\frac{s}{2}F
\right]_{\bm x_b}^n+O(s^2\partial^2)
\end{aligned}
\end{equation}
After simplifying, one can have
\begin{equation}\label{eq:recovered_Eq_flux}
\partial_t f_\alpha+\bm \xi_\alpha \cdot \nabla f_{\alpha}=-\frac{f_{\alpha}-f_\alpha^{eq}}{\tau_f}+F_\alpha
+\Xi^{(1)}+O(s^2\partial^2).
\end{equation}
with
\begin{equation}\label{eq:err_of_flux}
\begin{aligned}
\Xi^{(1)} & =\frac{s}{2}
\partial_t\left[-\frac{f_\alpha-f_\alpha^{eq}}{\tau_f}+F_\alpha\right]
-\frac{s}{2}\bm \xi\cdot\nabla\left[\frac{f_\alpha^{eq}-f_\alpha}{\tau}+F_\alpha \right] \\
& =\frac{s}{2}\partial_t^2 f_\alpha-\frac{s}{2}\bm \xi_\alpha\bm \xi_\alpha:\nabla\nabla f_\alpha+O(s^2\partial^2) \\
& =\frac{s}{2}\partial_t^2 f_\alpha^{eq}-\frac{s}{2}\bm \xi_\alpha\bm \xi_\alpha:\nabla\nabla f_\alpha^{eq}+O(s^2\partial^2)+O(\tau_f s^2\partial^2),
\end{aligned}
\end{equation}
where  $\partial_t f_\alpha+\bm \xi_\alpha \cdot \nabla f_{\alpha}=-\frac{f_{\alpha}-f_\alpha^{eq}}{\tau_f}+F_\alpha+O(s\partial)$ is used.
When the external force term is absent and the zeroth and first moments of the collision term are conserved, the term $\Xi$ in Eq.(\ref{eq:recovered_Eq_flux}) can be neglected. Otherwise, the term $\Xi$ may have an important effect on the numerical results. In view of the present DUGKS for CACE,
it is easy to obtain that $\sum_\alpha f_\alpha^{eq}=\sum_\alpha f_\alpha=\phi$, $\sum_\alpha \bm \xi_\alpha f_\alpha^{eq}=\phi\bm u$, $\sum_\alpha \bm \xi_\alpha \bm \xi_\alpha f_\alpha^{eq}=c_s^2\phi\bm I+\phi\bm u\bm u$, and $\sum_\alpha F_\alpha=0$, $\sum_\alpha \bm \xi_\alpha F_\alpha =c_s^2\Theta\bm n$ based on  Eqs.~(\ref{eq:equilibrium-feq}),(\ref{eq:force-gks}) and (\ref{eq:macro-gks}). With these moments,
the zero-order moment of Eq.(\ref{eq:recovered_Eq_flux}) becomes
\begin{equation}\label{eq:sum_flux_equation}
\begin{aligned}
\partial_t\phi+\nabla\cdot \sum_\alpha(\bm \xi_{\alpha}f_\alpha) =-\frac{s}{2}\partial_t^2\phi-\frac{s}{2}\nabla^2c_s^2\phi-\frac{s}{2}\nabla\cdot(\phi\bm u\bm u)+O(s^2\partial^2).
\end{aligned}
\end{equation}
From Eq.(\ref{eq:recovered_Eq_flux}), one can have
\begin{subequations}\label{eq:simplify_f-feq}
\begin{align}
f_\alpha&=f_\alpha^{eq}+\tau_f F_\alpha+O(\tau_f\partial),\\
f_{\alpha}&=f_{\alpha}^{eq}+\tau_f F_{\alpha}-\tau_f\left[\partial_t f_{\alpha}^{eq}+\bm\xi_\alpha \cdot \nabla f_{\alpha}^{eq} \right]-\tau_f\left[\partial_t F_\alpha+\bm \xi_\alpha\cdot \nabla F_\alpha\right]+O(\tau_f^2\partial )+O(\tau_f s\partial),
\\
\bm \xi_\alpha f &=\bm \xi_\alpha f_{\alpha}^{eq}+\tau_f\bm \xi_\alpha F_{\alpha}-\tau_f\bm \xi_\alpha \left[\partial_t f_{\alpha}^{eq}+\bm\xi_\alpha \cdot \nabla f_{\alpha}^{eq} \right]-\tau_f\left[\partial_t \bm\xi_\alpha F_\alpha+\bm \xi_\alpha\cdot \nabla F_\alpha\right]+O(\tau_f^2\partial )+O(\tau_f s\partial),
\end{align}
\end{subequations}


Substituting Eq.(\ref{eq:simplify_f-feq}) into Eq.(\ref{eq:sum_flux_equation}) gives
\begin{equation}\label{eq}
\begin{aligned}
\partial_t \phi+\nabla\cdot (\phi \bm u)=& \nabla\cdot\left[c_s^2\tau_f (\nabla \phi-\Theta\bm n) \right] \\
&+\frac{s}{2}\partial_t^2\phi-c_s^2\frac{s}{2}\nabla^2\phi-\frac{s}{2}\nabla\nabla:\phi\bm u\bm u +c_s^2\tau_f \partial_t \nabla\cdot \Theta\bm n \\
&+O(s^2\partial^2)+O(\tau_f^2\partial^2)+O(\tau_f s\partial^2),
\end{aligned}
\end{equation}
where the additional terms $\tau_f\nabla\cdot(\partial_t(\phi\bm u)+\nabla\cdot(\phi \bm u\bm u)$ are neglected as discussed earlier.
It can be seen that the terms on the second line are unnecessary compared with the target CACE. In particular,
 when the value of the relaxation time $\tau_f$ is usually much smaller than the time step $\Delta t$ (or the Pecletnumber is large), the error term $\frac{s}{2}\nabla^2\phi$ could play an important affect on the results.

\subsection{DUGKS with improved flux evaluation}
As presented in the previous subsection, the reconstructed distribution function $  \widehat{f}_\alpha^{n,+}(\bm x_b-\bm \xi_\alpha s,\bm \xi_\alpha)$ is not exact solution of discrete Boltzmann equation with BGK collision model. To overcome this deficiency,
we still start from Eq.(\ref{eq:flux_evolution2}).
Instead of the linear reconstruction,  we use the following parabolic reconstruction
\begin{equation}\label{eq:taylor-2order-flux}
  \widehat{f}_\alpha^{n,+}(\bm x_b-\bm \xi_\alpha s,\bm \xi_\alpha)
 = \widehat{f}_\alpha^{n,+}(\bm x_b,\bm \xi_\alpha)
 -s \bm \xi_\alpha\cdot \nabla \widehat{f}_\alpha^{n,+}(\bm x_b,\bm \xi_\alpha)+\frac{s^2}{2}\bm \xi_\alpha\bm \xi_\alpha :\nabla\nabla \widehat{f}_\alpha^{n,+}(\bm x_b,\bm \xi_\alpha).
 \end{equation}
\textcolor{magenta}{ It is worth pointing out that estimates of edge values should be at least third-order accurate to arrive a piecewise parabolic reconstruction.}
Inserting Eqs.(\ref{eq:taylor_space}) and (\ref{eq:taylor_time}) into (\ref{eq:taylor-2order-flux}), we have
\begin{equation}\label{eq:recovered_Eq_flux-2order}
\partial_t f_\alpha+\bm \xi_\alpha\cdot \nabla f_{\alpha}=-\frac{f_{\alpha}-f_\alpha^{eq}}{\tau_f}+F_\alpha
+\Xi^{(2)}+O(s^2\partial^3).
\end{equation}
where
\begin{equation}\label{eq:err_of_flux-2order}
\begin{aligned}
\Xi^{(2)}&=
\frac{s}{2}\partial_t\left(
-\frac{f_\alpha -f_\alpha^{eq}}{\tau_f}+F_\alpha  \right)
+\frac{s}{2}\bm \xi_\alpha \cdot\nabla\left(
\frac{f_\alpha -f_\alpha^{eq}}{\tau_f}-F_\alpha \right)
\\
&
-\frac{s}{2}\partial_t^2\left[
f_\alpha+\frac{s}{2}\frac{f_\alpha -f_\alpha^{eq}}{\tau_f}-\frac{s}{2}F_\alpha
\right]
+\frac{s}{2}\bm \xi_\alpha\bm \xi_\alpha:\nabla\nabla\left[
f_\alpha-\frac{s}{2}\frac{f_\alpha -f_\alpha^{eq}}{\tau_f}+\frac{s}{2}F_\alpha
\right]
\end{aligned}
\end{equation}
From Eq.(\ref{eq:recovered_Eq_flux-2order}), we can have
\begin{equation}\label{eq:taylor2order-simplified-s}
\partial_t f_\alpha+\bm \xi_\alpha \cdot \nabla f_{\alpha}=-\frac{f_{\alpha}-f_\alpha^{eq}}{\tau_f}+F_\alpha
+O(s\partial).
\end{equation}
Then, the error term $\Xi^{(2)}$ can be simplified into
\begin{equation}\label{eq}
\begin{aligned}
\Xi^{(2)}&=-\frac{s}{2}\partial_t\left[
\partial_t f_\alpha+\frac{f_\alpha-f_\alpha^{eq}}{\tau_f}-F_\alpha
\right]
+\frac{s}{2}\bm \xi_\alpha \cdot\nabla \left[
\bm \xi_\alpha \cdot\nabla f_\alpha+\frac{f_\alpha-f_\alpha^{eq}}{\tau_f}- F_\alpha,
\right]+O(s^2\partial^3)\\
&=\frac{s\tau_f}{2}\partial_t^2F_\alpha-\frac{s\tau_f}{2}\bm\xi_\alpha\bm \xi_\alpha:\nabla\nabla F_\alpha+O(s^2\partial^3).
\end{aligned}
\end{equation}
Using Eqs.(\ref{eq:equilibrium-feq}) and (\ref{eq:force-gks}) and taking the zeroth moments of Eq.(\ref{eq:recovered_Eq_flux-2order}), we have
\begin{equation}\label{eq}
\partial_t\phi+\nabla\cdot \sum_\alpha(\bm \xi_\alpha f_\alpha)=0,
\end{equation}
By using (\ref{eq:simplify_f-feq}), the resulting continuum equation becomes
\begin{equation}\label{eq}
\partial_t\phi +\nabla\cdot(\phi \bm u)=\nabla\cdot M\left(
\nabla\phi-\Theta\bm n
\right)+O(s^2\partial^3),
\end{equation}
where the addition term $\tau_f\nabla\cdot(\partial_t(\phi\bm u)+\nabla\cdot(\phi \bm u\bm u))$ is neglected again.
This implies that the correct CACE can be recovered from Eq.(\ref{eq:taylor-2order-flux}).
In Eq.(\ref{eq:taylor-2order-flux}), both the first and second derivatives of $\widehat{f}_\alpha^{n,+}(\bm x_b)$ at the cell face are required. We take  these derivatives in the x-direction as an example, which can be calculated as follows
 \begin{equation}\label{eq}
 \begin{aligned}
 \frac{\partial (\widehat{f}_\alpha^{+})_{i+1/2,j}  }{\partial x} &=\frac{
 (\widehat{f}_\alpha^{+})_{i-1,j}-15(\widehat{f}_\alpha^{+})_{i,j}+15(\widehat{f}_\alpha^{+})_{i+1,j}
-(\widehat{f}_\alpha^{+})_{i+2,j}
 }{12\Delta x}, \\
  \frac{\partial (\widehat{f}_\alpha^{+})_{i+1/2,j}  }{\partial y} &=
 \frac{ 8 (\widehat{f}_\alpha^{+})_{i+1/2,j+1}- 8(\widehat{f}_\alpha^{+})_{i+1/2,j-1}-(\widehat{f}_\alpha^{+})_{i+1/2,j+2}+ (\widehat{f}_\alpha^{+})_{i+1/2,j-2}
 }{12\Delta y},
 \end{aligned}
 \end{equation}
\begin{equation}\label{eq}
\begin{aligned}
\frac{\partial^2 (\widehat{f}_\alpha^{+})_{i+1/2,j}}{\partial x^2}&=\frac{
(\widehat{f}_\alpha^{+})_{i+2,j}-(\widehat{f}_\alpha^{+})_{i,j}+(\widehat{f}_\alpha^{+})_{i-1,j}-(\widehat{f}_\alpha^{+})_{i+1,j}
}{2(\Delta x)^2}
\\
\frac{\partial^2 (\widehat{f}_\alpha^{+})_{i+1/2,j}}{\partial y^2}&=\frac{
(\widehat{f}_\alpha^{+})_{i+\frac{1}{2},j+1}-2(\widehat{f}_\alpha^{+})_{i+\frac{1}{2},j}+(\widehat{f}_\alpha^{+})_{i+\frac{1}{2},j-1}
}{(\Delta y)^2}
\\
\frac{\partial^2 (\widehat{f}_\alpha^{+})_{i+1/2,j}}{\partial x\partial y}&=\frac{\partial^2 (\widehat{f}_\alpha^{+})_{i+1/2,j}}{\partial y\partial x}=
\frac{
(\widehat{f}_\alpha^{+})_{i+1,j+1}-(\widehat{f}_\alpha^{+})_{i+1,j-1}-(\widehat{f}_\alpha^{+})_{i,j+1}+(\widehat{f}_\alpha^{+})_{i,j-1}
}{2\Delta x \Delta y}.
\\
\end{aligned}
\end{equation}

\section{Numerical Results and discussion}
In this section,
the two kinetic equations are solved numerically by the present DUGKS scheme. For brevity, the current DUGKS for Eqs.(\ref{eq:kineticequaiton}),(\ref{eq:equilibrium-feq}) and (\ref{eq:force-gks}) is referred as DUGKS-I and the present DUKGS for Eqs.(\ref{eq:kineticequaiton}) (\ref{eq:feq-second-DBE}) and (\ref{eq:Fi-second-DBE}) is referred as DGUKS-II. The first derivative in the force term is calculated by the second-order isotropic central difference formulas~\cite{lee2005stable}.
Several benchmark problems, including diagonal motion of a circular interface,
Zalesak's rotating disk and single vortex deformation of a circular interface are simulated to assess the performance of the proposed DUGKS for capturing the interface.
In the simulations,  $\text{Pe}=60$ and $L_c=W$. The interface width is fixed at $W=4$. The uniform grid is used and $\Delta x=\Delta y=1$, $RT=1/3$ unless otherwise specified.  The time step is determined by CFL condition.  For convenience, CFL is redefined as $\chi|\bm \xi_\alpha+\bm U_c|_{max}$ such that the time step is given by $\Delta t=\chi\Delta x$. As a result, the time step is determined by $\chi$.
The results obtained by the proposed DUGKS models  will be compared with the theoretical results and those obtained by  the DUGKS model (labeled as DUGKS-AC) in Ref.\cite{yang2019phase} and the LBE model (labeled as LBE-AC) in Ref.~\cite{geier2015conservative} in the following discussion.
To quantitatively measure the accuracy of all models, the $L_2$-norm relative errors of the order parameter are calculated by
 \begin{equation}\label{eq:L2-errors}
 ||\delta \phi||_2=\sqrt{\frac{\sum_{\bm x}|\phi(\bm x,nT)-\phi(\bm x,0)|^2}{\sum_{\bm x}|\phi(\bm x,0)|^2}},
 \end{equation}
 where $\phi(\bm x,0)$ is the initial values of the order parameter and $\phi(\bm x,nT)$ is the numerical result at period $nT$ with $n$ being positive integer.

\subsection{Diagonal translation of a circular interface }
In this subsection, we consider the motion of a circular interface due to a constant velocity field $\bm u=(u,v)=(U_0,U_0)$. Initially, a circular interface with radius $R=0.25L_0$ is placed in the middle of a domain of size $L_0\times L_0$. Periodic boundary conditions are applied to all boundaries. Under such flow, the circular interface will move back to its initial position after  $T=L_0/(U_0 \Delta t)$ time. The parameters are set as $L_0=100, U_0=0.02$ and $\chi=0.5$.

\begin{figure}
  \centering
\subfloat[DUGKS-AC]{%
\includegraphics[width=0.25\textwidth,trim=20 2 40 10,clip]{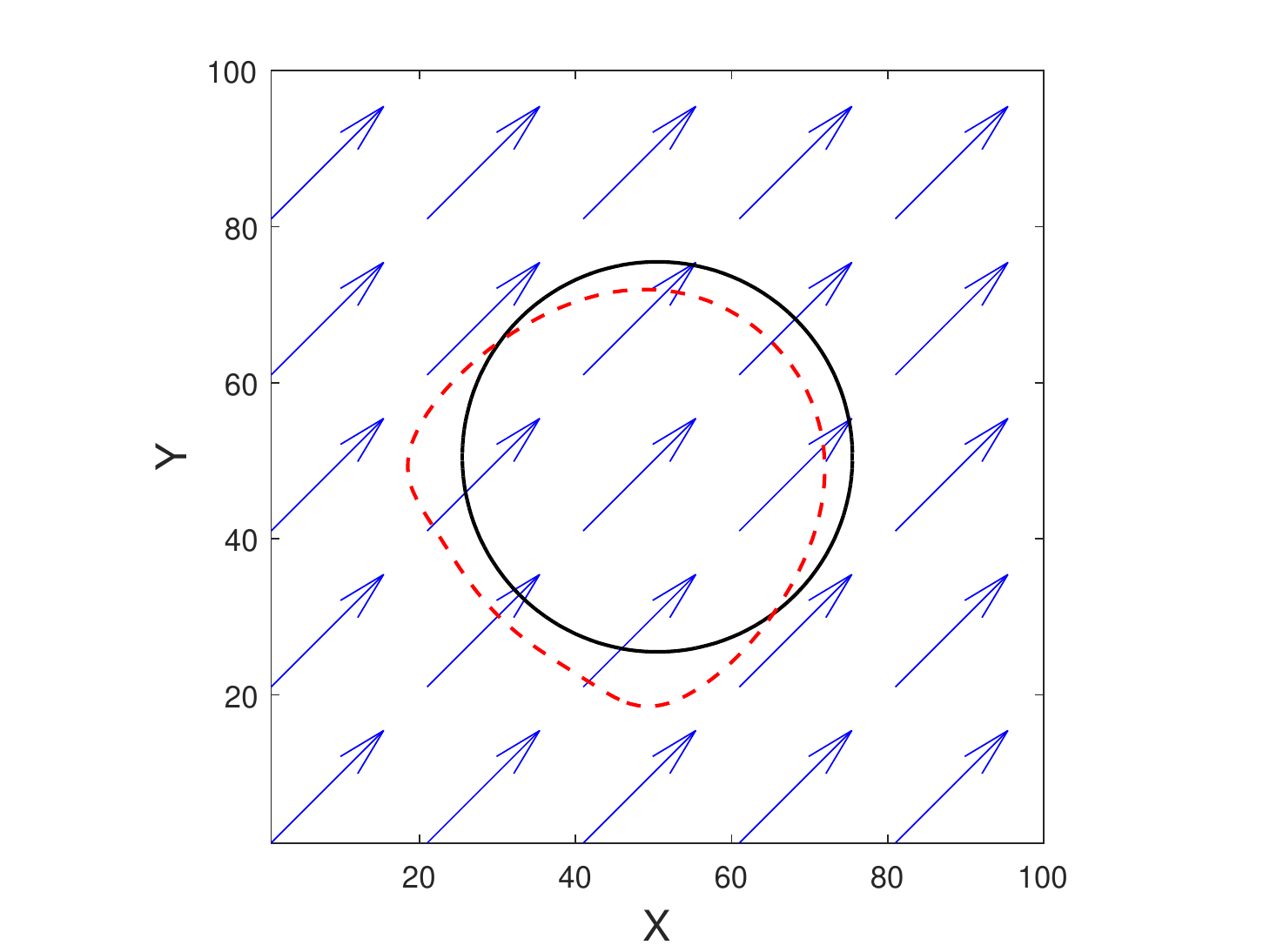}~
\includegraphics[width=0.25\textwidth,trim=20 2 40 10,clip]{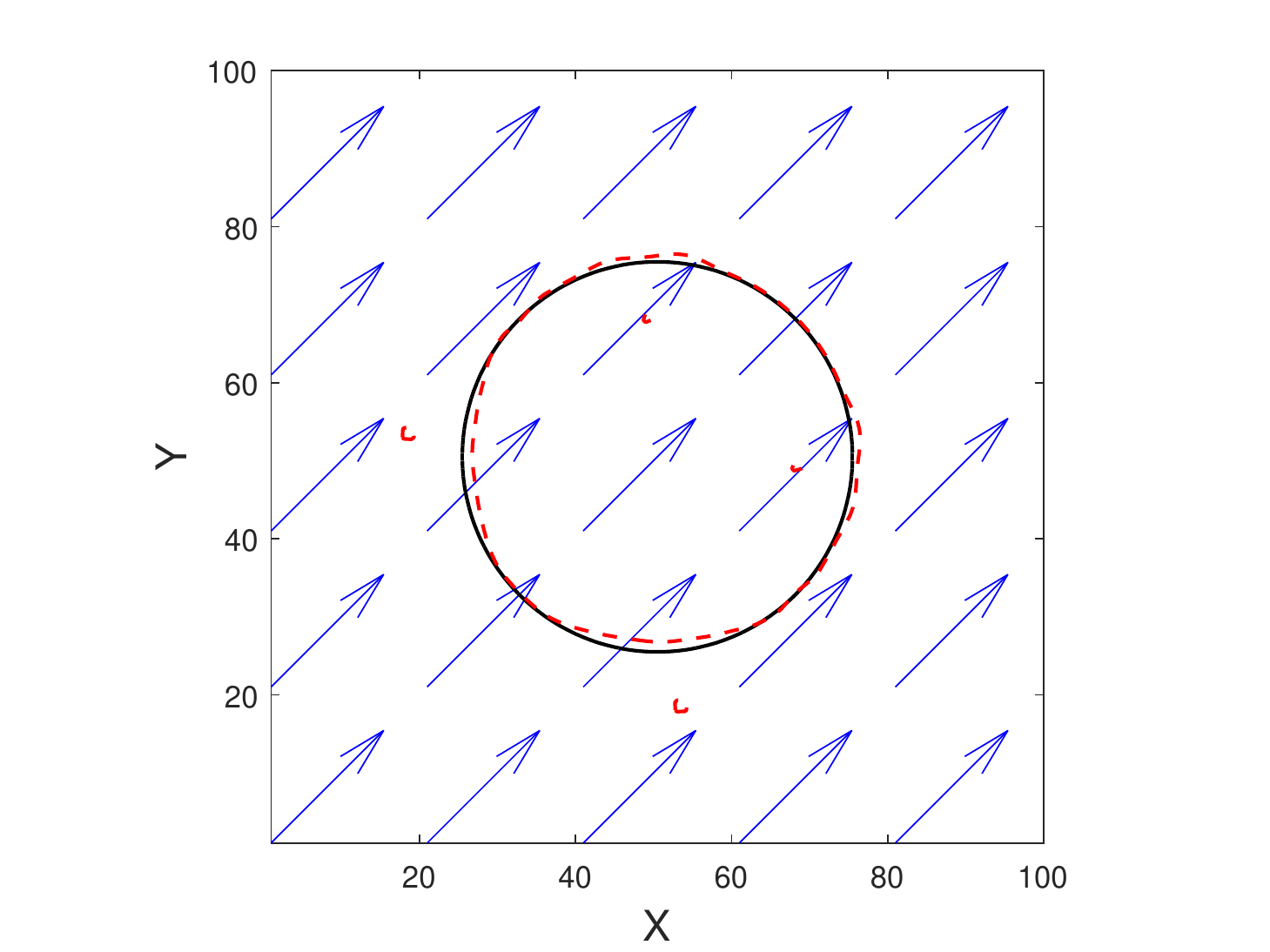}~
\includegraphics[width=0.25\textwidth,trim=20 2 40 10,clip]{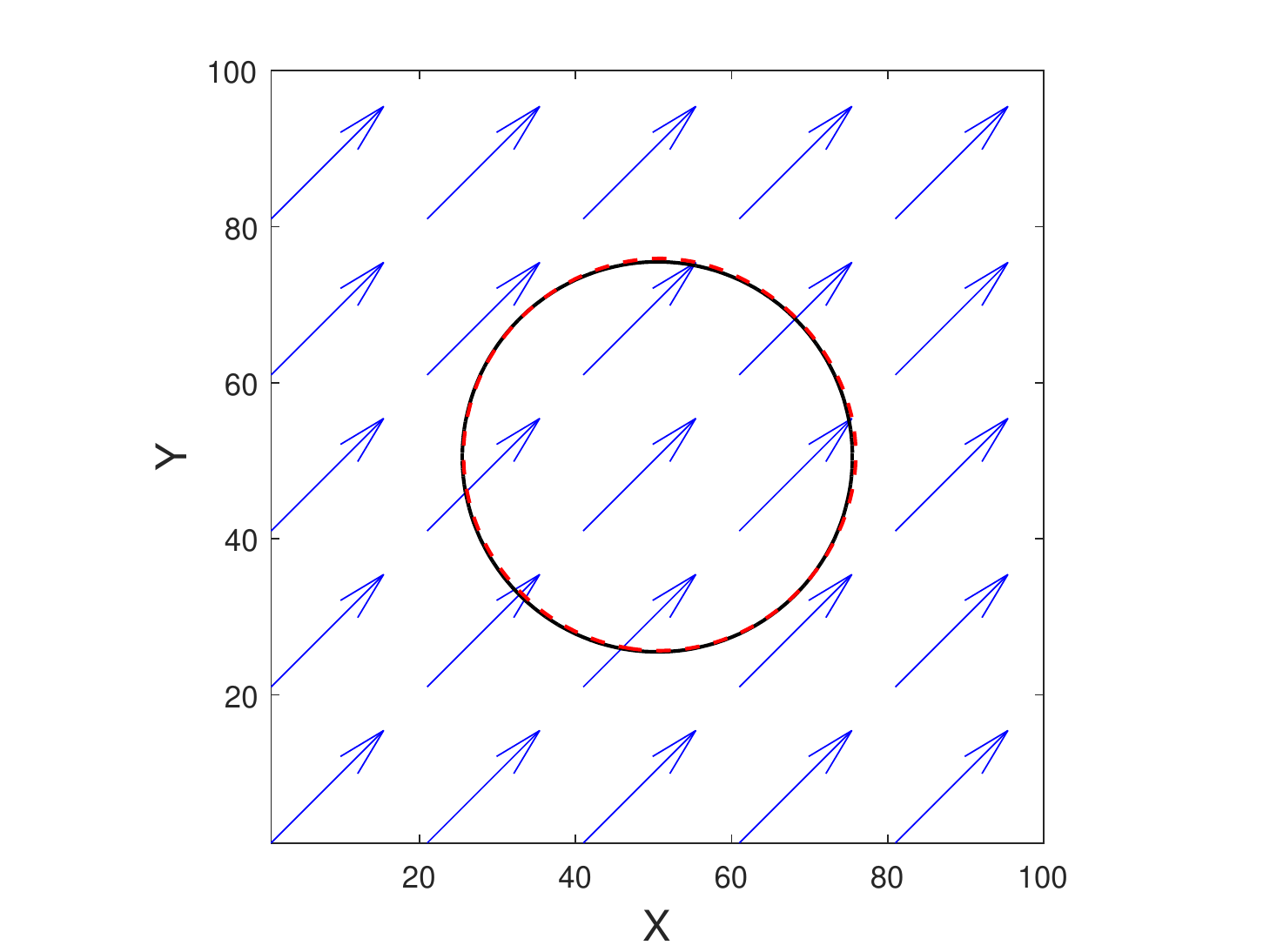}~
\includegraphics[width=0.25\textwidth,trim=20 2 40 10,clip]{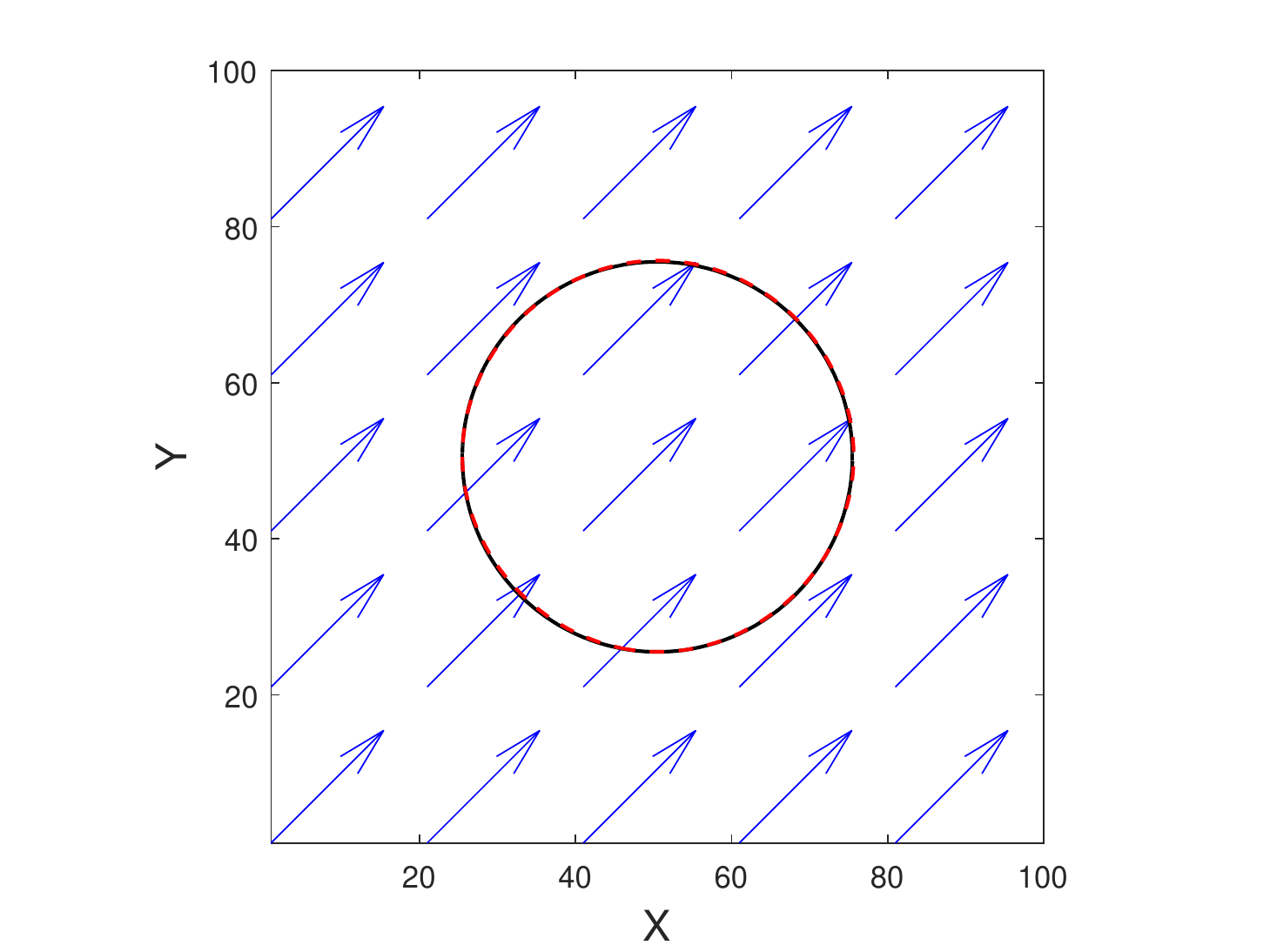}}\\
\subfloat[DUGKS-I]{%
\includegraphics[width=0.25\textwidth,trim=20 2 40 10,clip]{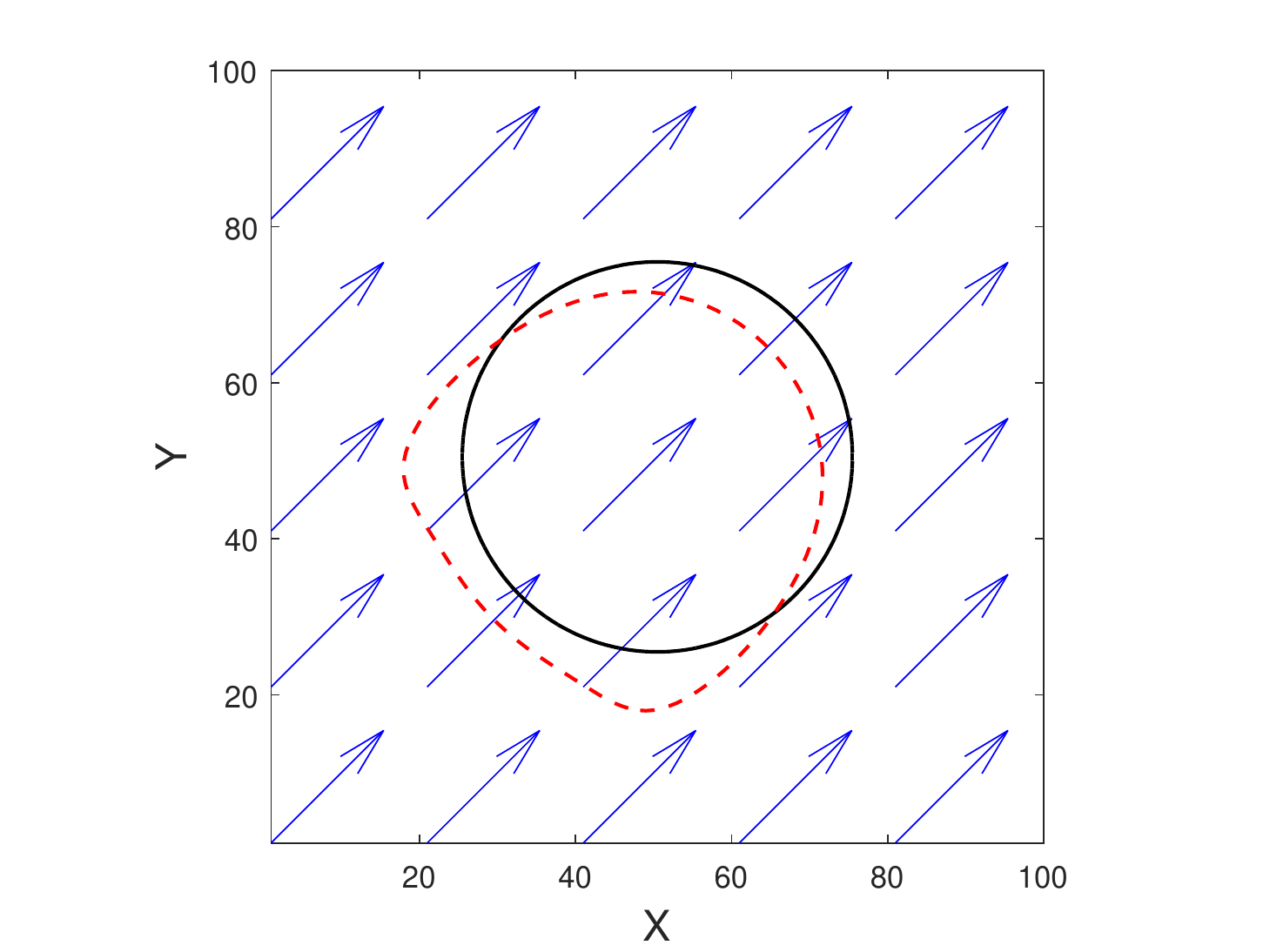}~
\includegraphics[width=0.25\textwidth,trim=20 2 40 10,clip]{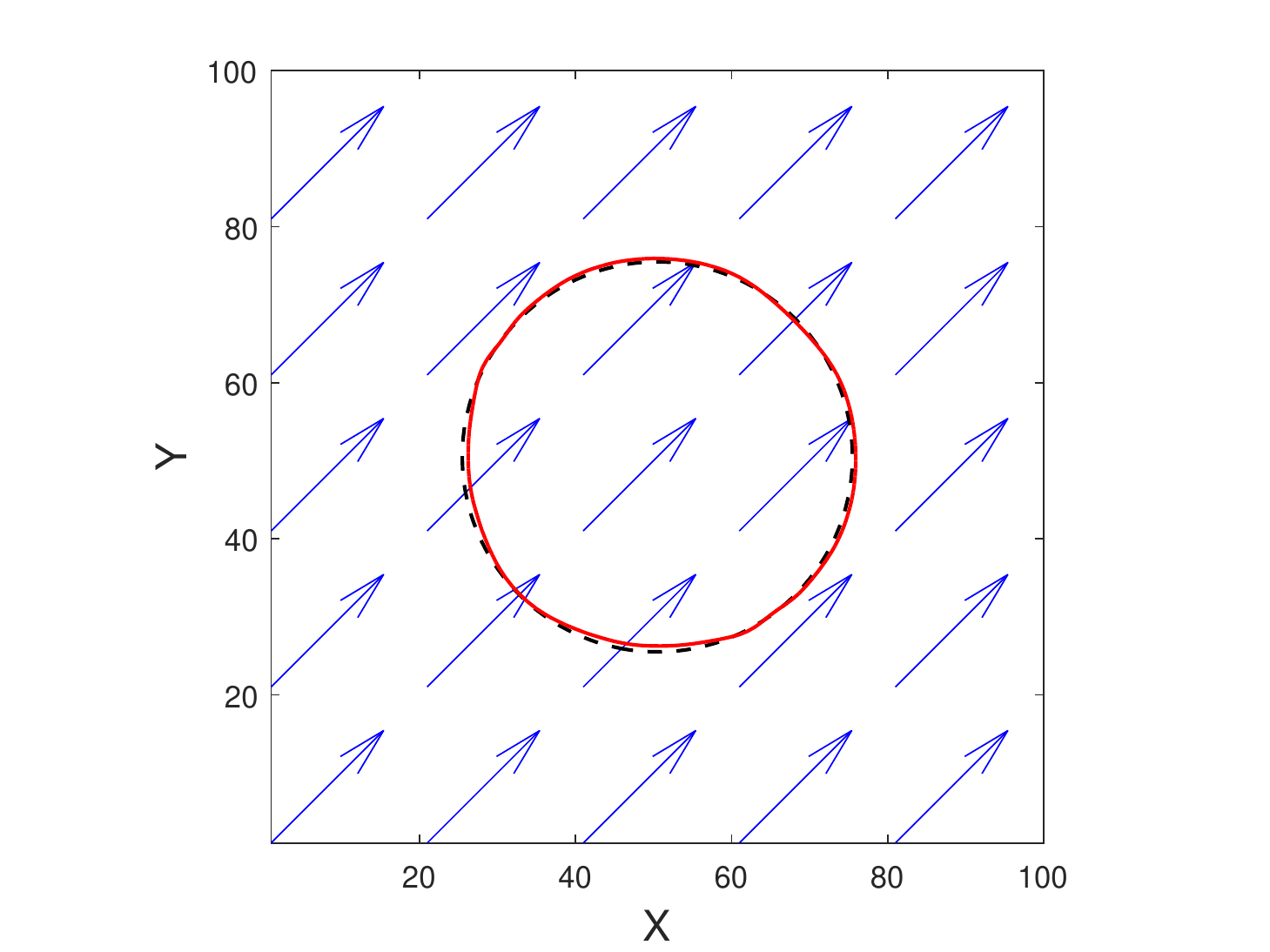}~
\includegraphics[width=0.25\textwidth,trim=20 2 40 10,clip]{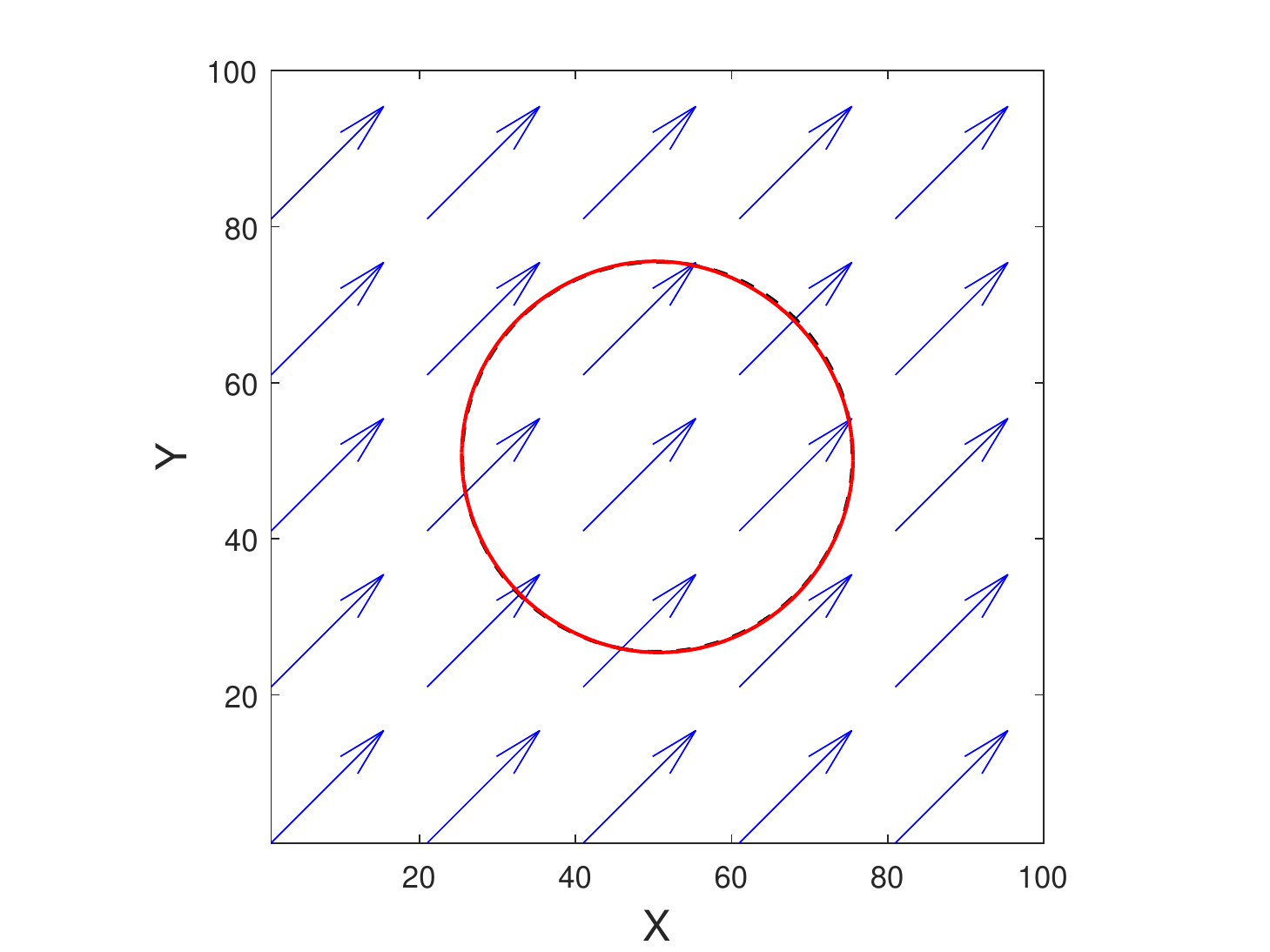}~
\includegraphics[width=0.25\textwidth,trim=20 2 40 10,clip]{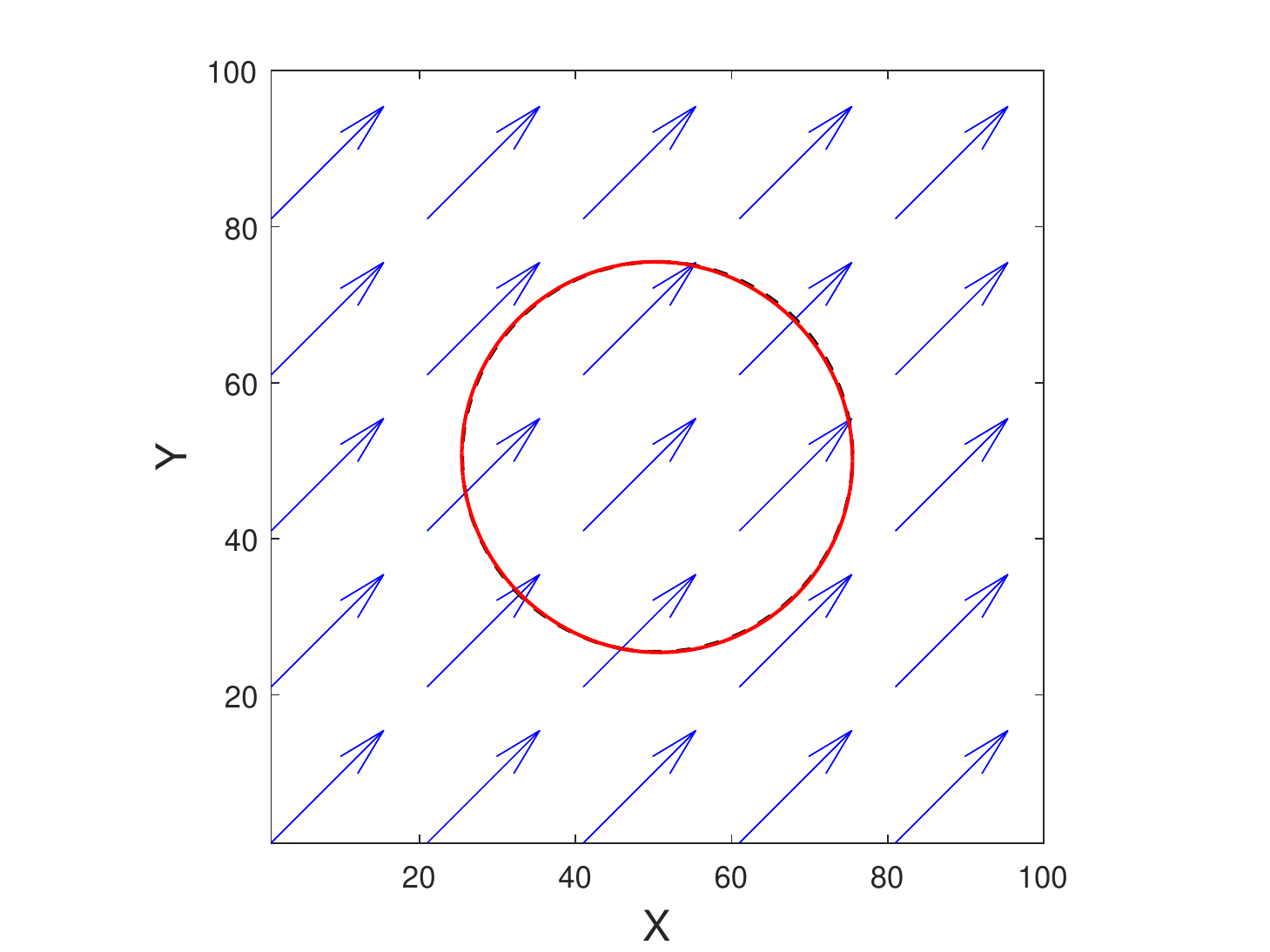}}\\
\subfloat[DUGKS-II]{%
\includegraphics[width=0.25\textwidth,trim=20 2 40 10,clip]{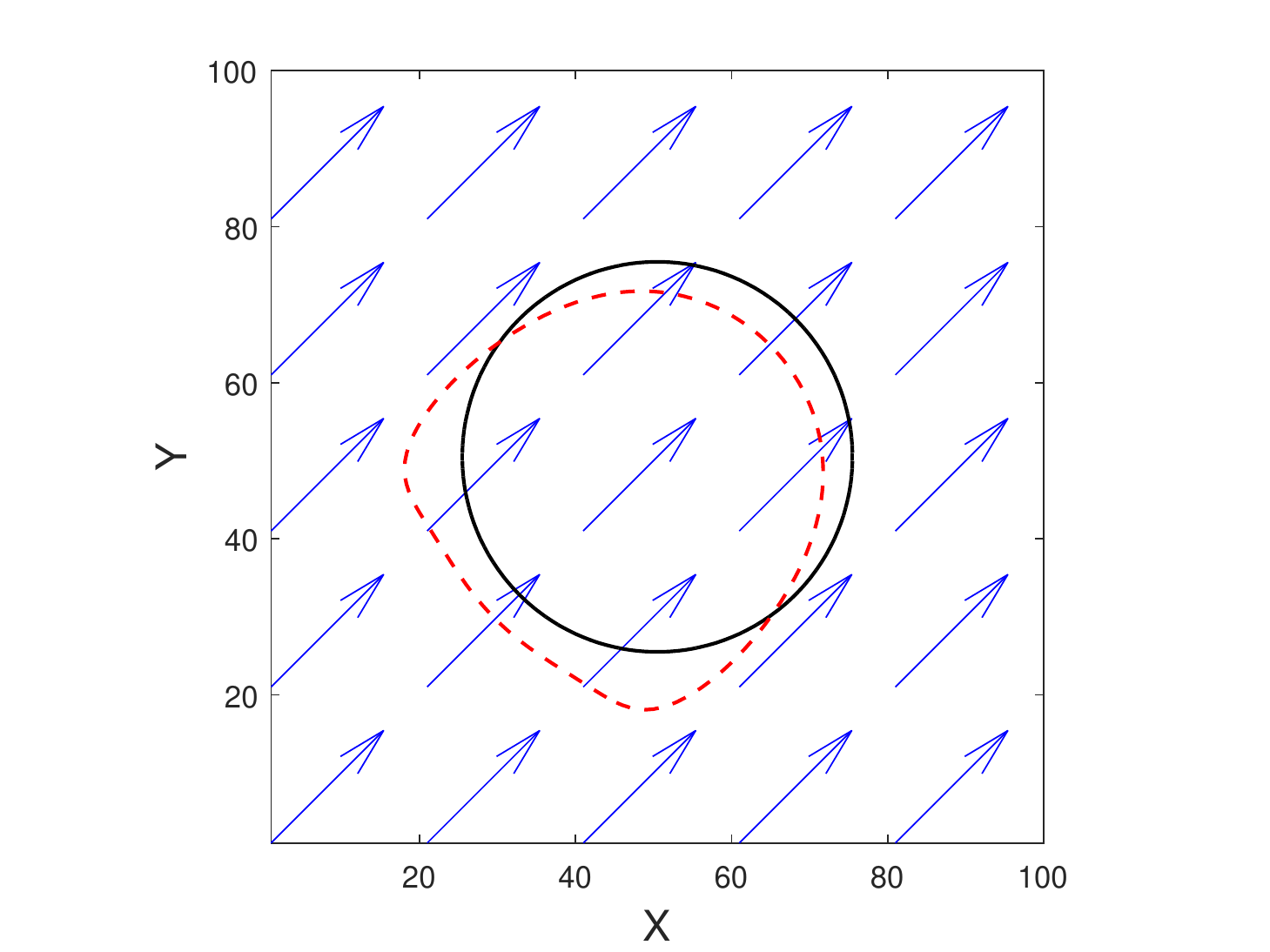}~
\includegraphics[width=0.25\textwidth,trim=20 2 40 10,clip]{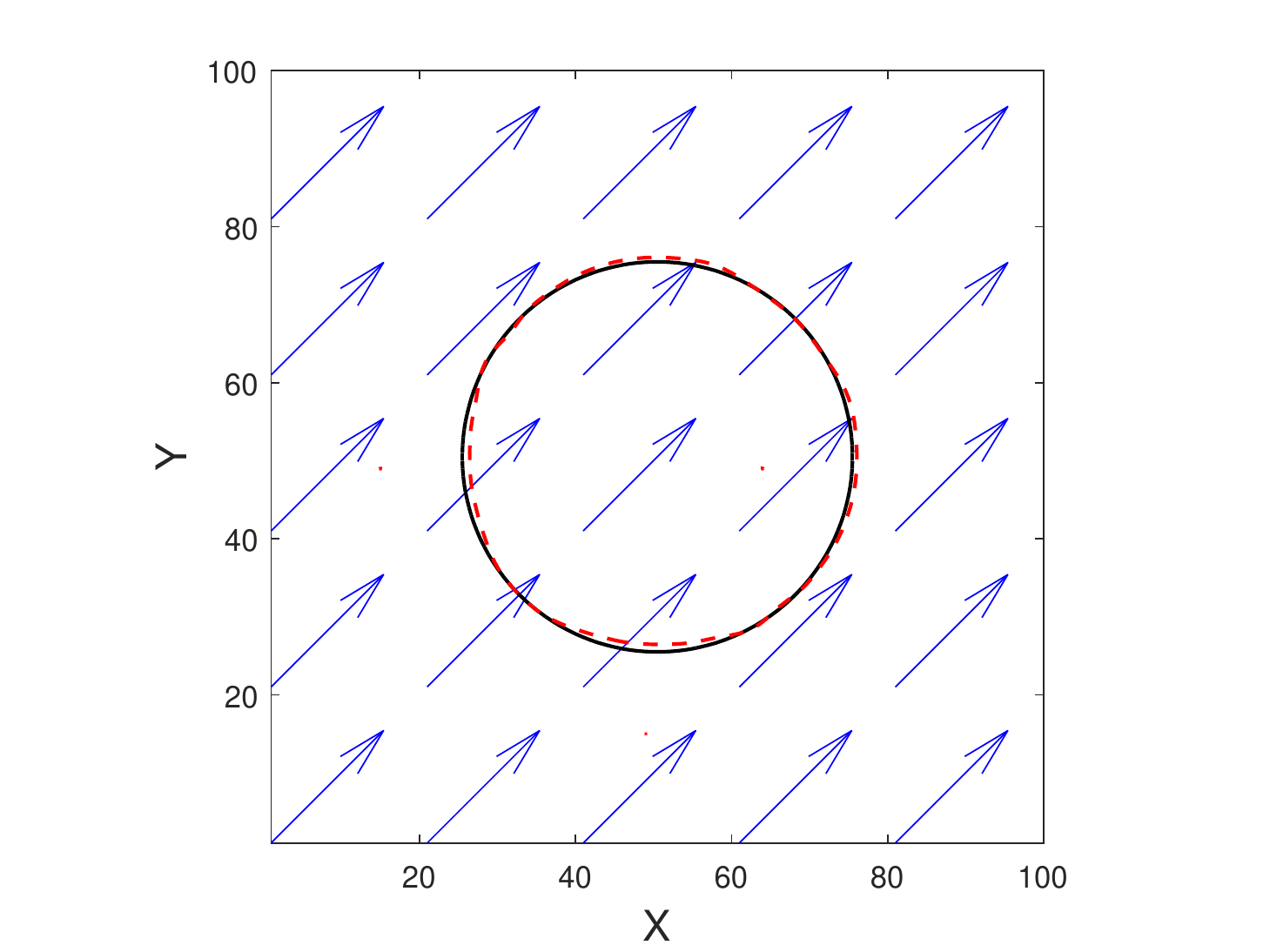}~
\includegraphics[width=0.25\textwidth,trim=20 2 40 10,clip]{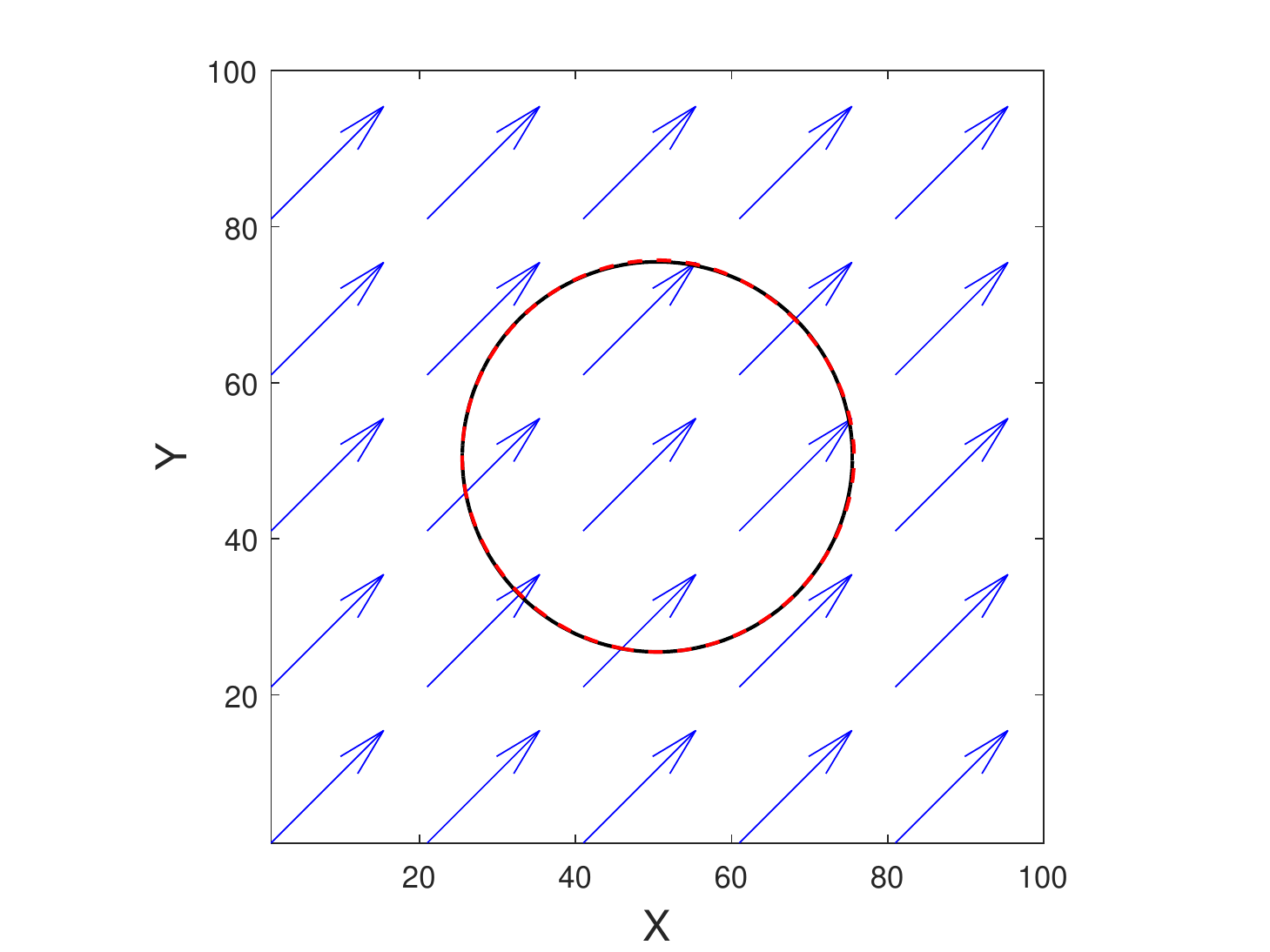}~
\includegraphics[width=0.25\textwidth,trim=20 2 40 10,clip]{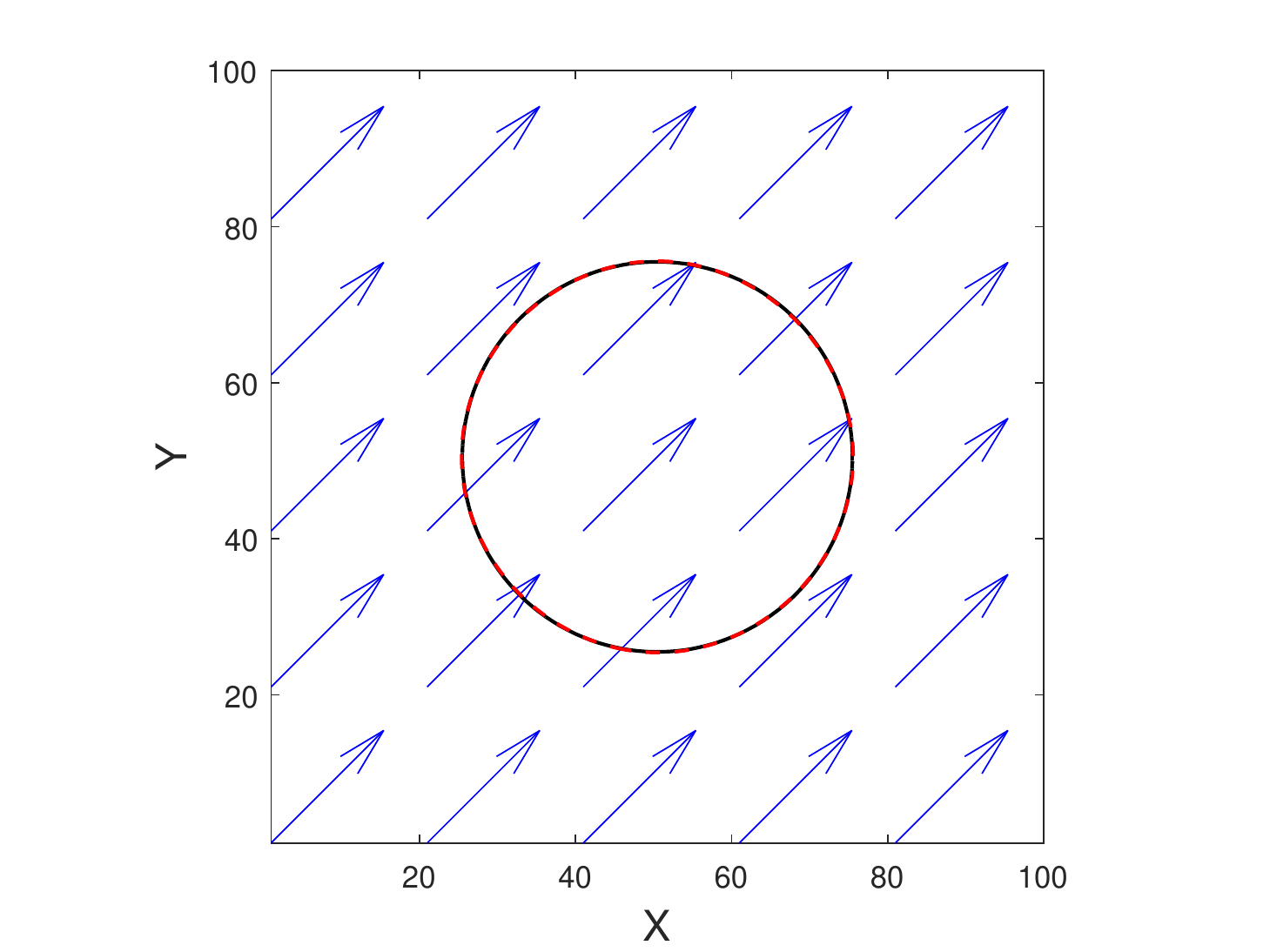}}\\
  \caption{\textcolor{blue}{The phase-field contour ($\phi=0$) of diagonal translation of a circular interface obtained by (a) DUGKS-AC, (b) DUGKS-I and (c) DUGKS-II with different reconstruction schemes at $\text{Pe}=60$}. The reconstruction schemes are 2CDI, 4CDI, WENO-Z3 and WENO-Z5 from left to right. Black solid line denotes the initial profile  and red dashed line represents the reconstructed interface at 10T.}\label{test1:compareInterpolation}
\end{figure}

\begin{figure}[!htb]
\centering
\subfloat[DUGKS-AC]{\includegraphics[width=0.33\textwidth,trim=30 2 30 10,clip]{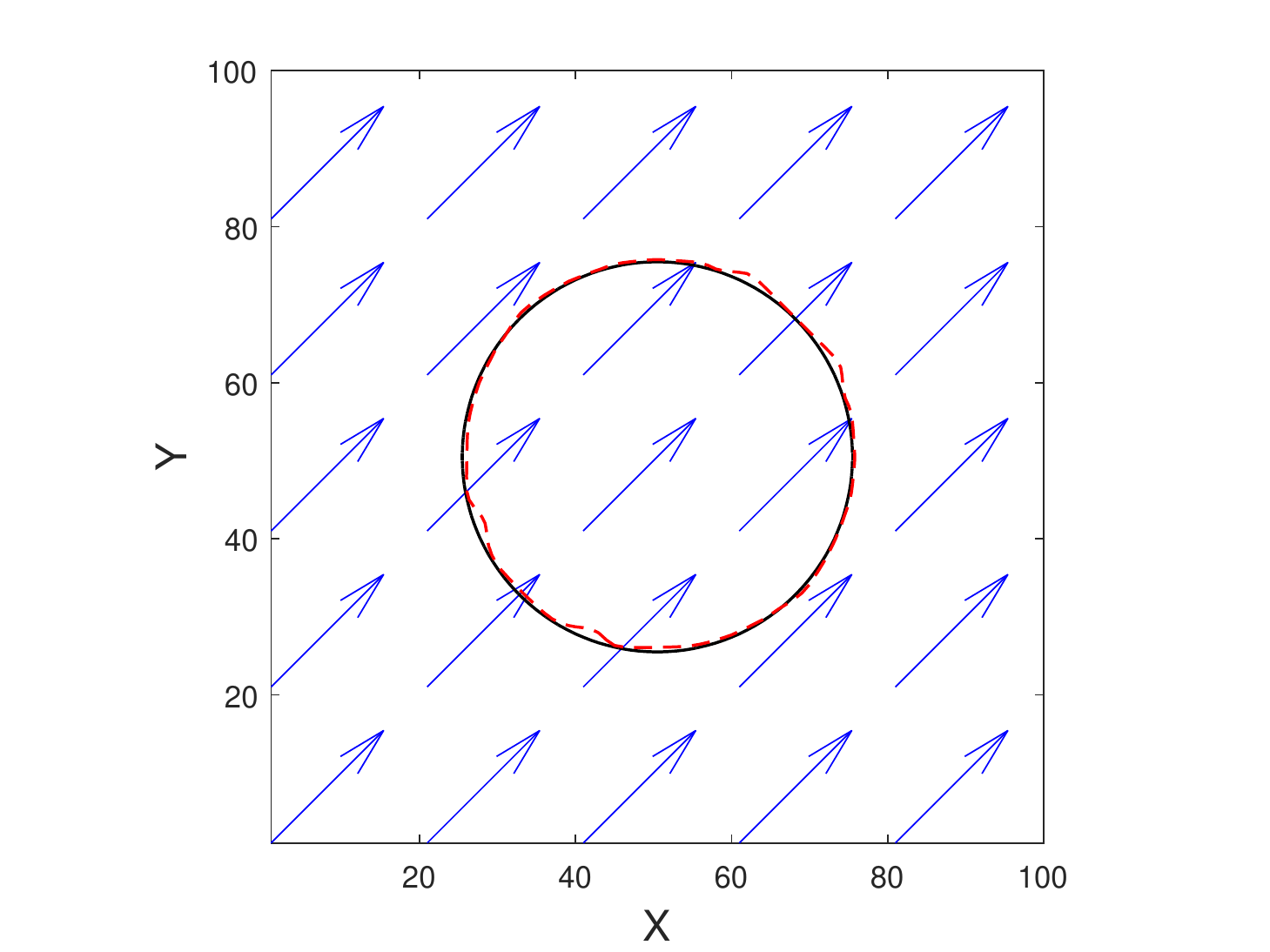}}~
\subfloat[DUGKS-I]{\includegraphics[width=0.33\textwidth,trim=30 2 30 10,clip]{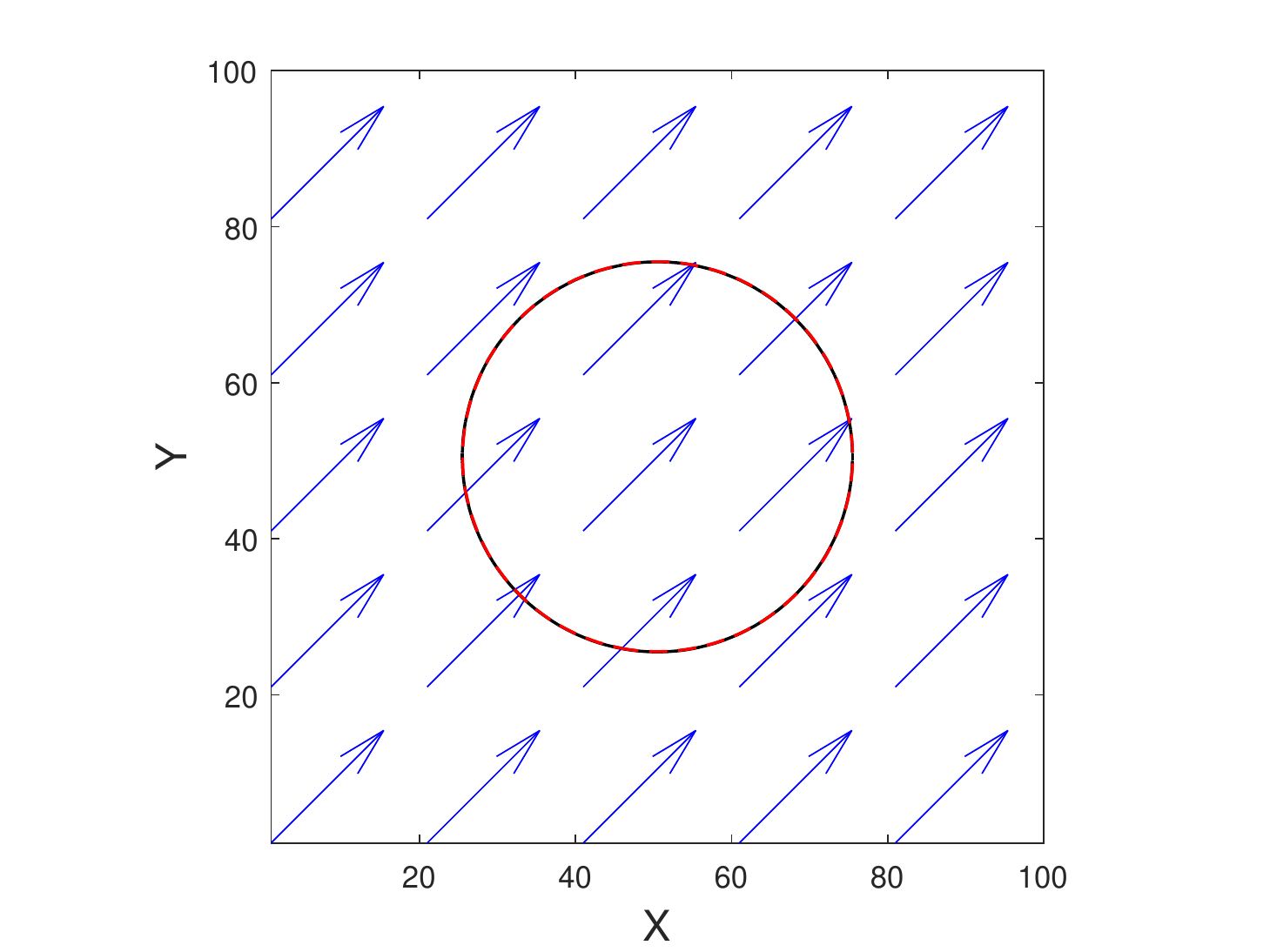}}~
\subfloat[DUGKS-II]{\includegraphics[width=0.33\textwidth,trim=30 2 30 10,clip]{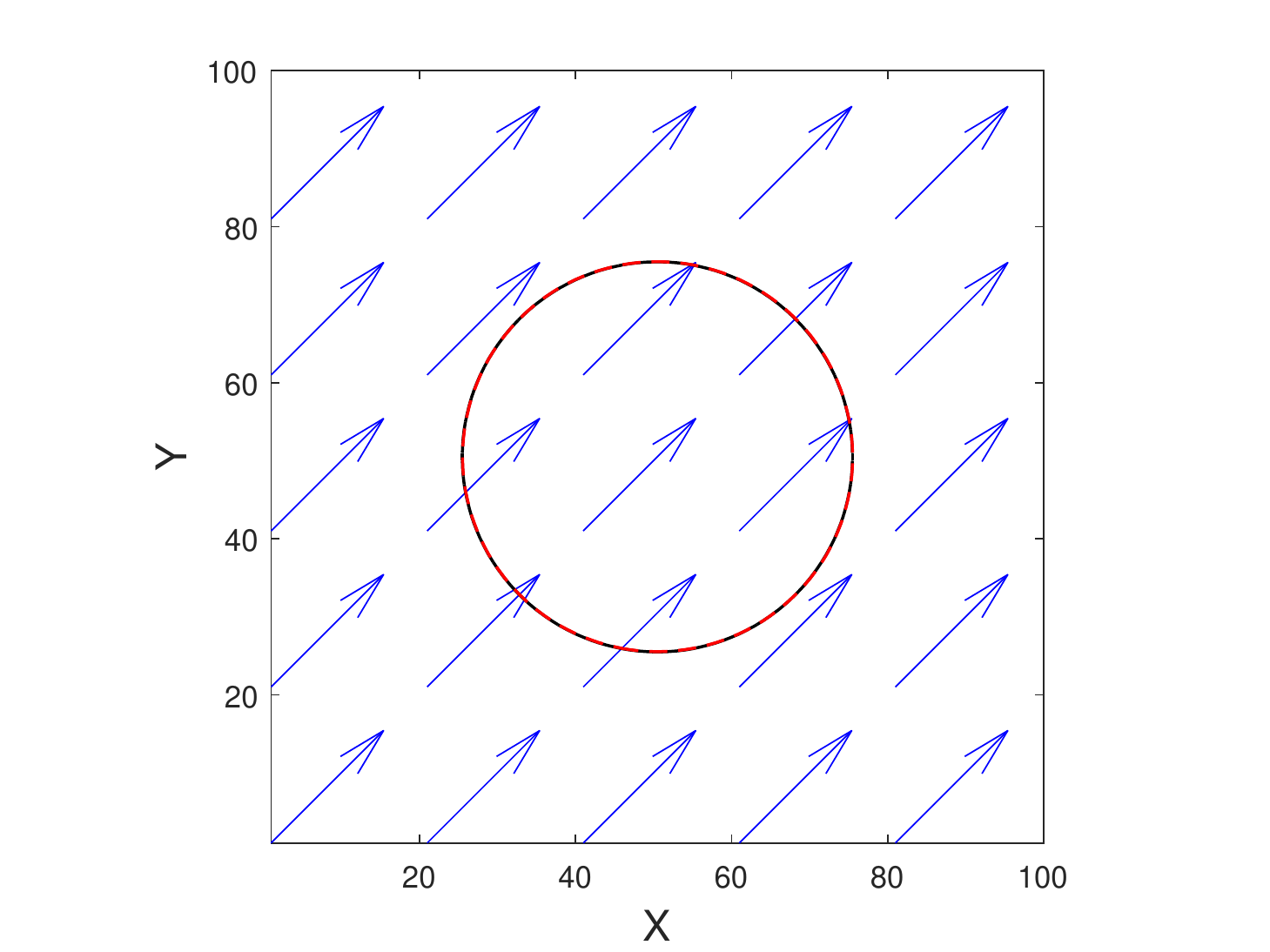}}\\
\caption{\textcolor{magenta}{The phase-field contour ($\phi=0$) of diagonal translation of a circular interface obtained by (a) DUGKS-AC, (b) DUGKS-I and (c) DUGKS-II with WENO-Z5 at Pe=500.  Black solid line denotes the initial profile  and red dashed line represents the reconstructed interface at $t=10T$}.}
\label{test1-Pe500}
\end{figure}

To investigate the effect of the reconstruction scheme for the value of the distribution function at the cell face, we first carried out all DUGKS models with different interpolation schemes, including, second-order central differencing interpolation (2CDI)~\cite{zhang2018discrete}, fourth-order central differencing interpolation (4CDI)~\cite{felker2018fourth},
third-order weighted essentially non-oscillatory (WENO) scheme (referred to as WNEO-Z3)~\cite{jiang1996efficient} and fifth-order WENO (referred to as WENO-Z5)~\cite{shu1998essentially}. The details of these schemes are provided in Appendix~\ref{ap:dugks-reconstruction}.
 We run the code up to $10T$ and the results are shown in Fig.\ref{test1:compareInterpolation}.
It can be found that all DUGKS models with second-order central differencing interpolation fails to capture the interface accurately.
The reconstructed shape of the circle obtained by DUGKS models with both WENO-Z3 and WENO-Z5 agree well with its initial configuration.
The measured $L_2$ error of the order parameter is provided in Table~\ref{tab:test1-interpolation-comparison}.
It can be seen that the value of the $L_2$ error decreases when the higher order reconstruction scheme is employed.
The values of $L_2$ error given by DUGKS-I and DUGKS-II are almost identical and  less than those given by DUGKS-AC.
This implies that the additional terms $M/c_s^2\nabla\cdot(\partial_t(\phi\bm u)+\nabla\cdot(\phi\bm u\bm u))$  have little effect on the results and can be neglected.
\textcolor{magenta}{ To further test the performance of the present DUGKS methods with improved flux evaluation,
we repeated the above case using all three DUGKS methods with WENO-Z5 scheme at different $\text{Pe}$. The calculated circular interface by all methods  at $\text{Pe}=500$ are shown in Fig.~\ref{test1-Pe500}. It can be seen the restored interface  by DGUKS-AC slightly deviate from the exact solution while DUGKS-I and DUGKS-II still agree well with the exact solutions. This implies that the improved flux evaluation is able to improve the accuracy of capturing interface. We calculate the relative errors of three DUGKS methods at various Pe as shown in Table.~\ref{tab:test1-Pe-comparison}. It can be observed that the relative errors given by DUGKS-I and DUGKS-II are nearly identical for all $\text{Pe}$ and much less than those given by DUGKS-AC when $\text{Pe}<1000$. However, the relative error given by DUGKS-AC is smaller that the one given by DUGKS-I and DUGKS-II when $\text{Pe}\geq 1000$.
This may be because the calculated order parameter exhibits severe oscillations due to a small value of the relaxation time (or large Pe number) and the non-dissipative high-order central difference for the spatial derivatives. The accuracy of DUGKS-I and DUGKS-II at large Pe can be improved by using other  discretization schemes (e.g, WENO) for the spatial derivatives in Eq.(\ref{eq:taylor-2order-flux}), which is beyond the scope of this study. In fact, the value of the Peclet number in practical simulations mainly depends on  numerical stability conditions. In the range of Peclet numbers considered, both DUGKS-I and DUGKS-II with the improved flux evaluation are more accurate than DUGKS-AC in capturing interface.}
To compare with the previous LBE method, the above case is also simulated by the LBE-AC in Ref.~\cite{geier2015conservative}. The results are similar to those obtained by DUGKS-I and not shown here. The $L_2$ error of the order parameter obtained by the  LBE is $4.991\times 10^{-3}$, which is slightly less than $6.4\times10^{-3}$ given  by DUGKS-I or DUGKS-II.
Based on the above results, WENO-Z5  is adopted for both DUGKS-I and DUGKS-II to approximate the value of the distribution function at the cell face unless otherwise specified herein.

\begin{table}[!htb]
\centering
\caption{Relative  errors $L_2$ of $\phi$  with different reconstruction schemes for interface diagonal translation at $t=10T$.} \label{tab:test1-interpolation-comparison}
\setlength{\tabcolsep}{1mm}{%
\begin{tabular}{ccccccccc}
\hline
\text{Reconstruction} & 2CDI &  &4CDI &  &WENO-Z3 & &WENO-Z5        \\
\hline
DUGKS-AC    &0.3528   &     &0.1244 &   &0.0278  &    &0.0111    \\
DUGKS-I     &0.3747  &     &0.0999  &   &0.0160 &     &0.0064     \\
DUGKS-II    &0.3747   &     &0.0999  &   &0.0160 &    &0.0064  \\
\hline
\end{tabular}}
\end{table}

\begin{table}[!htb]
\centering
\caption{Relative  errors $L_2$ of $\phi$  with different $\text{Pe}$ for interface diagonal translation at $t=10T$.} \label{tab:test1-Pe-comparison}
\setlength{\tabcolsep}{0.9mm}{%
\begin{tabular}{cccccccccccc}
\hline
Pe          & 50      &  &250     &  &500    &    &1000   & &2000     \\
\hline
DUGKS-AC    & 0.0108  &  & 0.0416 &  & 0.0577 &    &0.0829 & & 0.0901  \\
DUGKS-I     & 0.0077  &  & 0.0032 &  & 0.0059 &    & 0.1147 & & 0.1907  \\
DUGKS-II    & 0.0077  &  & 0.0032  &  &0.0059  &    &0.0948& &0.1906 \\
\hline
\end{tabular}}
\end{table}

The CFL number can be adjusted to improve the accuracy of tracking the interface.
To test this, we repeated the above simulations with different CFL conditions. The
the $L_2$ error calculated by Eq.(\ref{eq:L2-errors}) are presented in Table~\ref{tab:test1-cfl-comparison}.
It is observed that the $L_2$ error decreases with increasing CFL number for both DUGKS-I and DUGKS-II.
For each $\chi$ except for $\chi=0.2$, the value of $L_2$ error given by DUGKS-AC is larger than the one given by DUGKS-I and DUGKS-II. The results of DUGKS-I and DUGKS-II are almost identical again, which indicates that the additional terms in Eq.(\ref{eq:recovered_equation}) have little effect on the numerical results and can be negligible under different CFL conditions.
\begin{table}[!htb]
\centering
\caption{Relative  errors $L_2$ of $\phi$  with different CFL conditions for interface diagonal translation at $t=10T$.} \label{tab:test1-cfl-comparison}
\setlength{\tabcolsep}{0.9mm}{%
\begin{tabular}{cccccccccccc}
\hline
$\chi$ & 0.1&       &0.2 &      &0.4 &        &0.5  &   & 0.8  &&1.0    \\
\hline
DUGKS-AC    &0.0196   &  &0.0117  &  &0.0091  &    &0.0111 & &0.025 && 0.041 \\
DUGKS-I     &0.0196   &  &0.0118  &  &0.0073  &    &0.0064 & &0.0052 && 0.0052  \\
DUGKS-II    &0.0196   &  &0.0118  &  &0.0073  &    &0.0064 & &0.0051 && 0.0051 \\
\hline
\end{tabular}}
\end{table}

Finally, we examine the convergence rate of the proposed DUGKS methods.
 The $L_2$ errors are measured at $t=T$.
The Cahn number $\text{Cn}$ is fixed at $0.015$, which implies that the interface width increases as the mesh is refined.
The rate of convergence is defined as the ratio of successive errors : $\log_2(||\delta \phi_(L_0)||_2/||\delta \phi_{2L_0}||_2$. Since we refined the spatial grids by a factor of 2, the ratio of successive errors increases by a factor of 2.
The $L_2$ errors and ratios of convergence obtained by these definitions are given in Table~\ref{tab:test1-convergence}.
The ratios of convergence obtained by LBE-AC are also presented in Table~\ref{tab:test1-convergence} for comparison.
 It can be seen that the order of accuracy of both DUGKS-I and DUGKS-II is higher than that of DUGKS-AC and the corresponding magnitude of $L_2$ error is also smaller. The $L_2$ errors given by DUGKS-I and DUGKS-II are comparable with those given by LBM-AC.

\begin{table}[!htb]
\centering
\caption{ Error and convergence order for interface diagonal translation ($\chi=0.5$) at $t=T$  and $\text{Cn}=0.015$.} \label{tab:test1-convergence}
\setlength{\tabcolsep}{0.9mm}{%
\begin{tabular}{cccccccc}
\hline
     & $50\times 50 $  &    & $100\times 100 $    &   & $200\times 200$  & & $400\times 400$  \\
\hline
DUGKS-AC  &0.2860  &    &0.1912  &   &0.1090  &   &0.0360   \\
order     & -      &    & 0.56   &   &0.81    &   &1.60     \\
DUGKS-I   &$6.998\times10^{-2}$   &   &$2.793\times10^{-2}$  &  &$4.294\times10^{-3}$  &   &$4.220\times10^{-4} $  \\
order     & -                     &   &1.36                  &  &2.70                 &   & 3.34  \\                                                            DUGKS-II  &$6.997\times10^{-2}$   &   &$2.792\times10^{-2}$  &  &$4.291\times10^{-3}$  &   &$4.210\times10^{-4} $  \\
order     &-                      &   &1.33                  &  &2.70                  &   &3.35    \\
LBE-AC       &$1.036\times10^{-1}$   &   &$3.448\times10^{-2}$  &  &$4.991\times10^{-3}$  &   &$7.66\times10^{-4} $  \\
order     &-                      &   &1.59                  &  &2.79                  &   &2.70    \\

\hline
\end{tabular}}
\end{table}

\subsection{Zalesak's rotation of a slotted disk}%
\begin{figure}[!htb]
\centering
\subfloat[LBE-AC]{\includegraphics[width=0.25\textwidth,trim=30 10 30 10,clip]{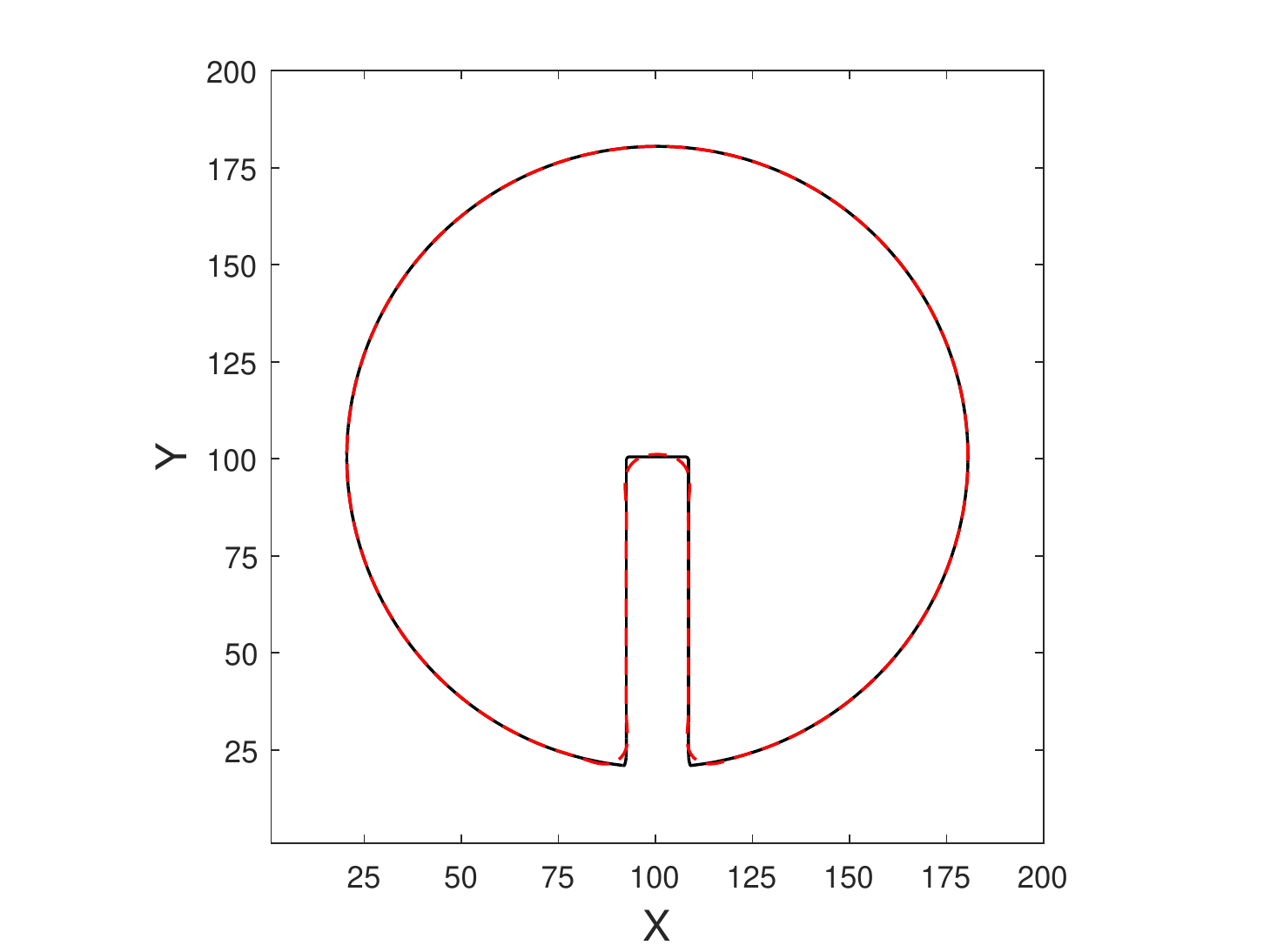}}~
\subfloat[DUGKS-AC]{\includegraphics[width=0.25\textwidth,trim=30 10 30 10,clip]{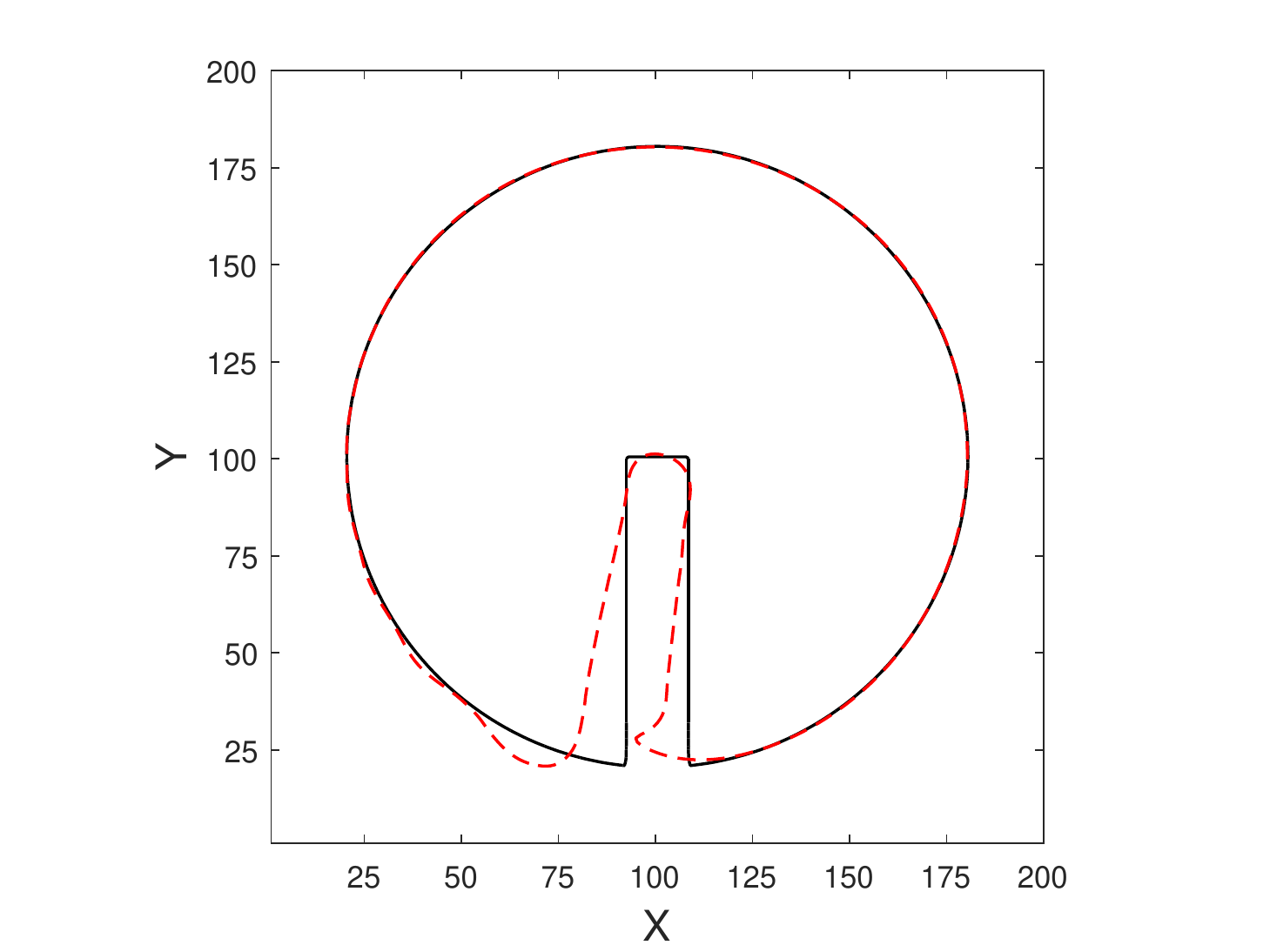}}~
\subfloat[DUGKS-I]{\includegraphics[width=0.25\textwidth,trim=30 10 30 10,clip]{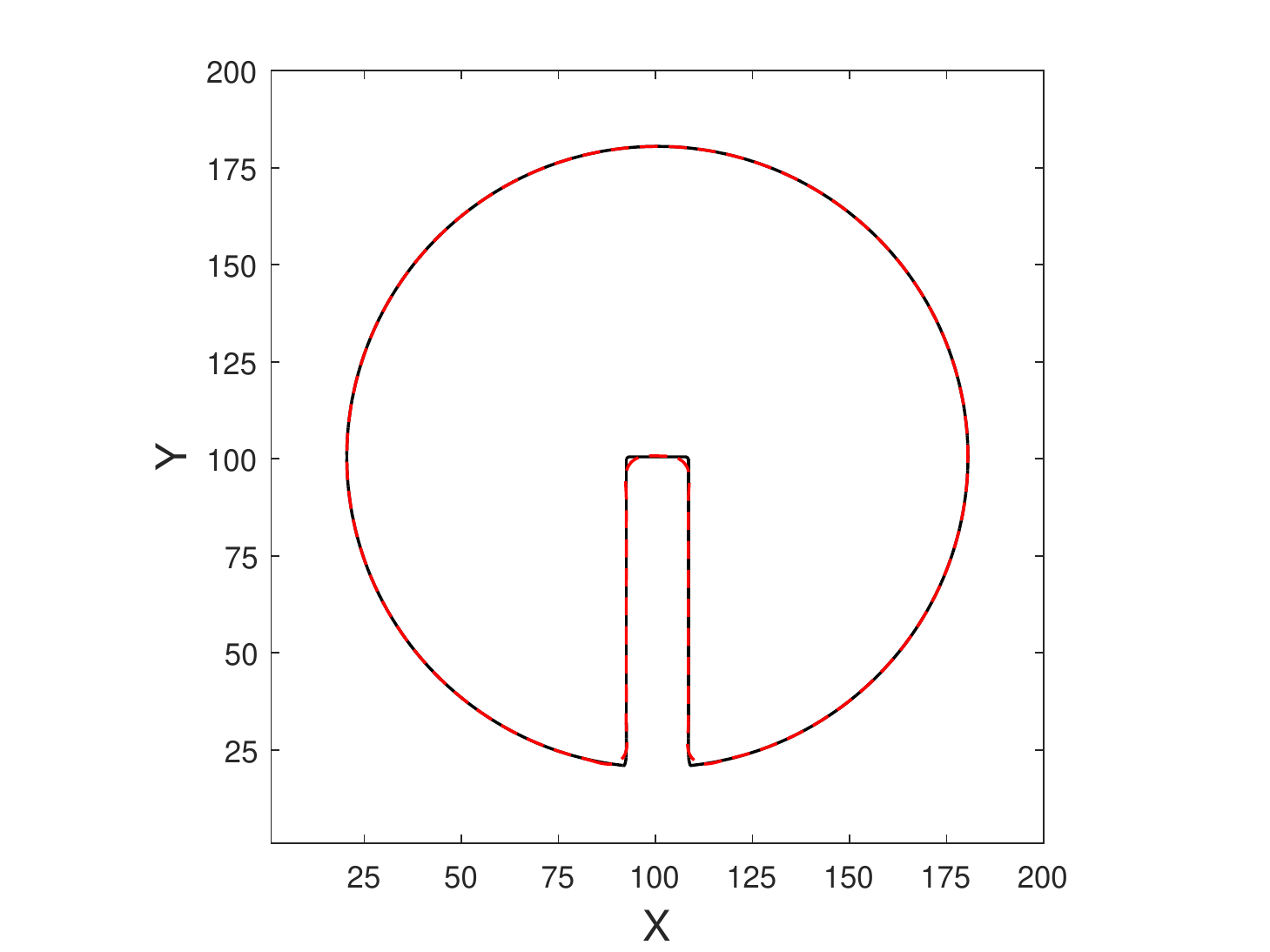}}~
\subfloat[DUGKS-II]{\includegraphics[width=0.25\textwidth,trim=30 10 30 10,clip]{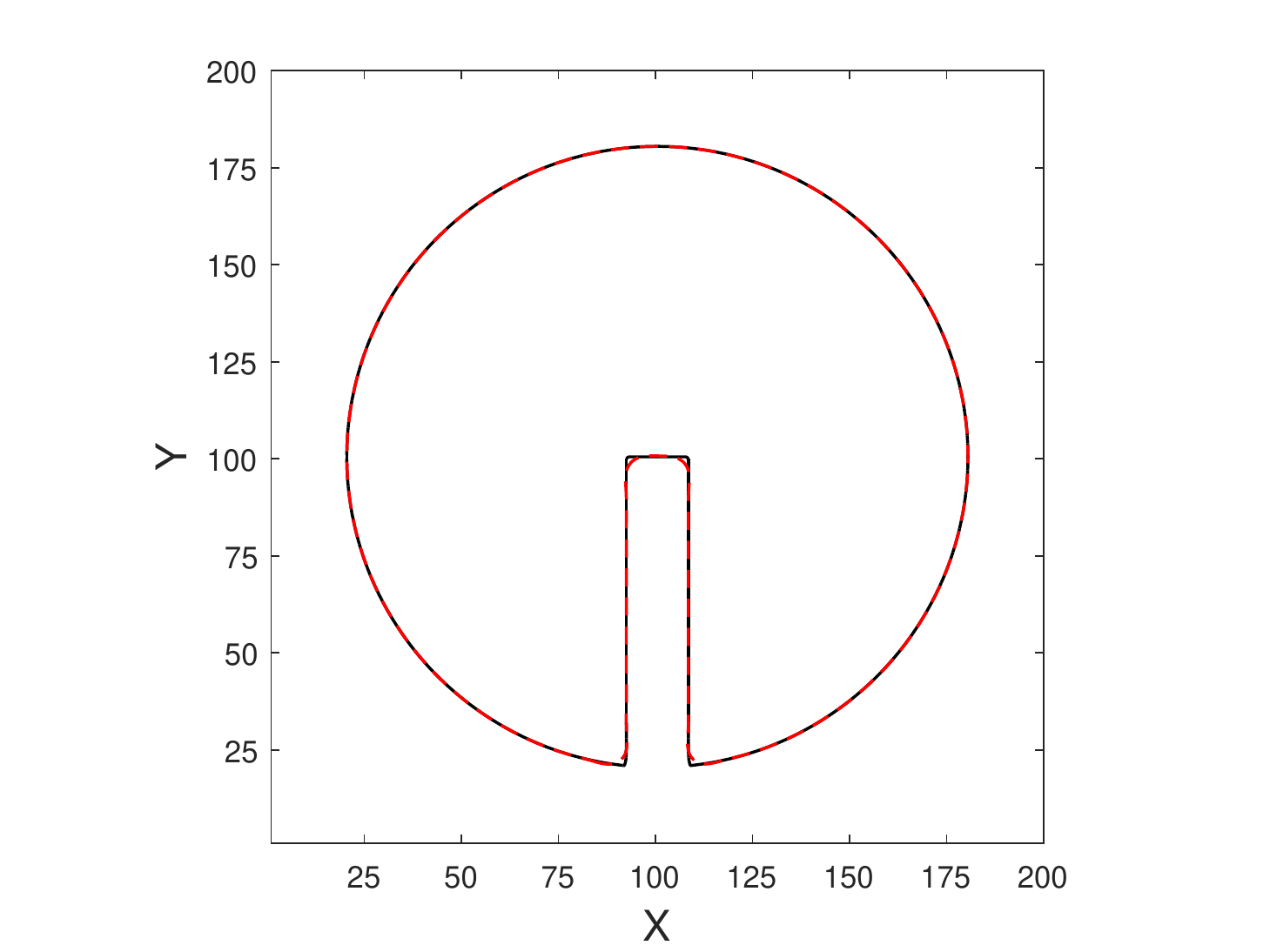}}\\
\caption{\textcolor{blue}{The phase-field contour ($\phi=0$) of Zalesak's disk predicted by (a) LBE-AC, (b) DUGKS-AC, (c) DUKGS-I and (d) DUGKS-II}.  Black solid line denotes the initial profile  and red dashed line represents the reconstructed interface at $t=10T$.}
\label{test2-shape}
\end{figure}

In this subsection, Zalesak's disk rotation is simulated. Initially, a slotted disk with radius $R=0.4L_0$ is placed in the middle of a square domain of size $L_0\times L_0$. The width of the slot is $0.1875R$. The velocity field $\bm u=(u,v)$ is given by
\begin{equation}\label{eq}
u(x,y)=-U_0\pi\left(\frac{y}{L_0}-0.5\right),\qquad v(x,y)=U_0\pi\left(\frac{x}{L_0}-0.5\right).
\end{equation}
After $2T$ with $T=L_0/(U_0\Delta t)$, the disk will return to its initial positions. The parameters are set as
$L_0=200$, $U_0=0.02$ and $\chi=0.5$. Figure~\ref{test2-shape} shows the interface shapes obtained by all four models after  $10T$ and the initial shape of the slotted disk is also presented for comparison. It can be seen that
there are significantly discrepancies between the initial and final shapes of the slotted disk  obtained by DUGKS-AC. By contrast,
the results given by the other three models are similar and agree well with the initial shape and position of the slotted disk.

\subsection{ Vortex deformation of a circle}
\begin{figure}[!htb]
\centering
\subfloat[t=T/2]{
\includegraphics[width=0.25\textwidth,trim=30 5 30 10,clip]{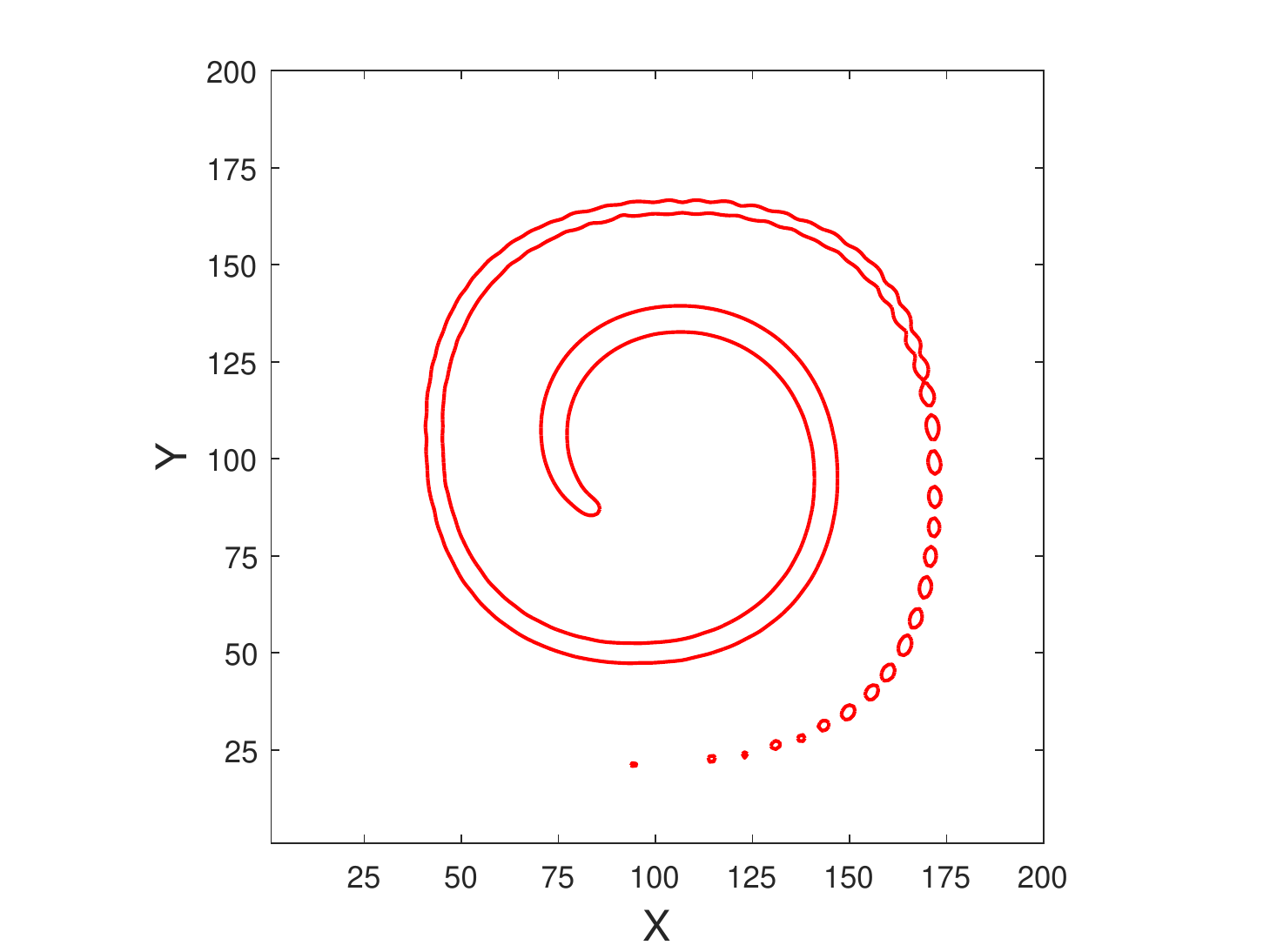}~
\includegraphics[width=0.25\textwidth,trim=30 5 30 10,clip]{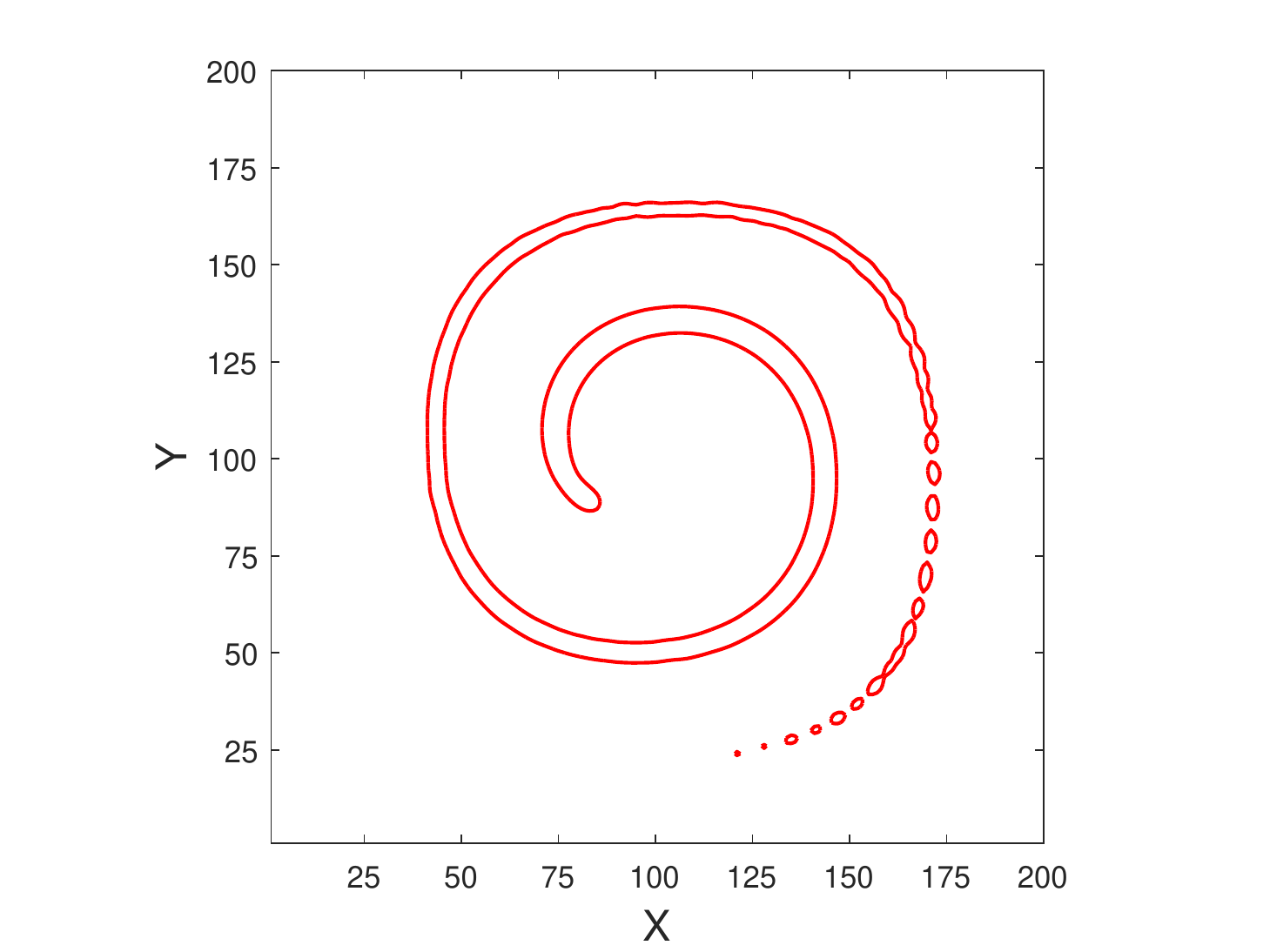}~
\includegraphics[width=0.25\textwidth,trim=30 5 30 10,clip]{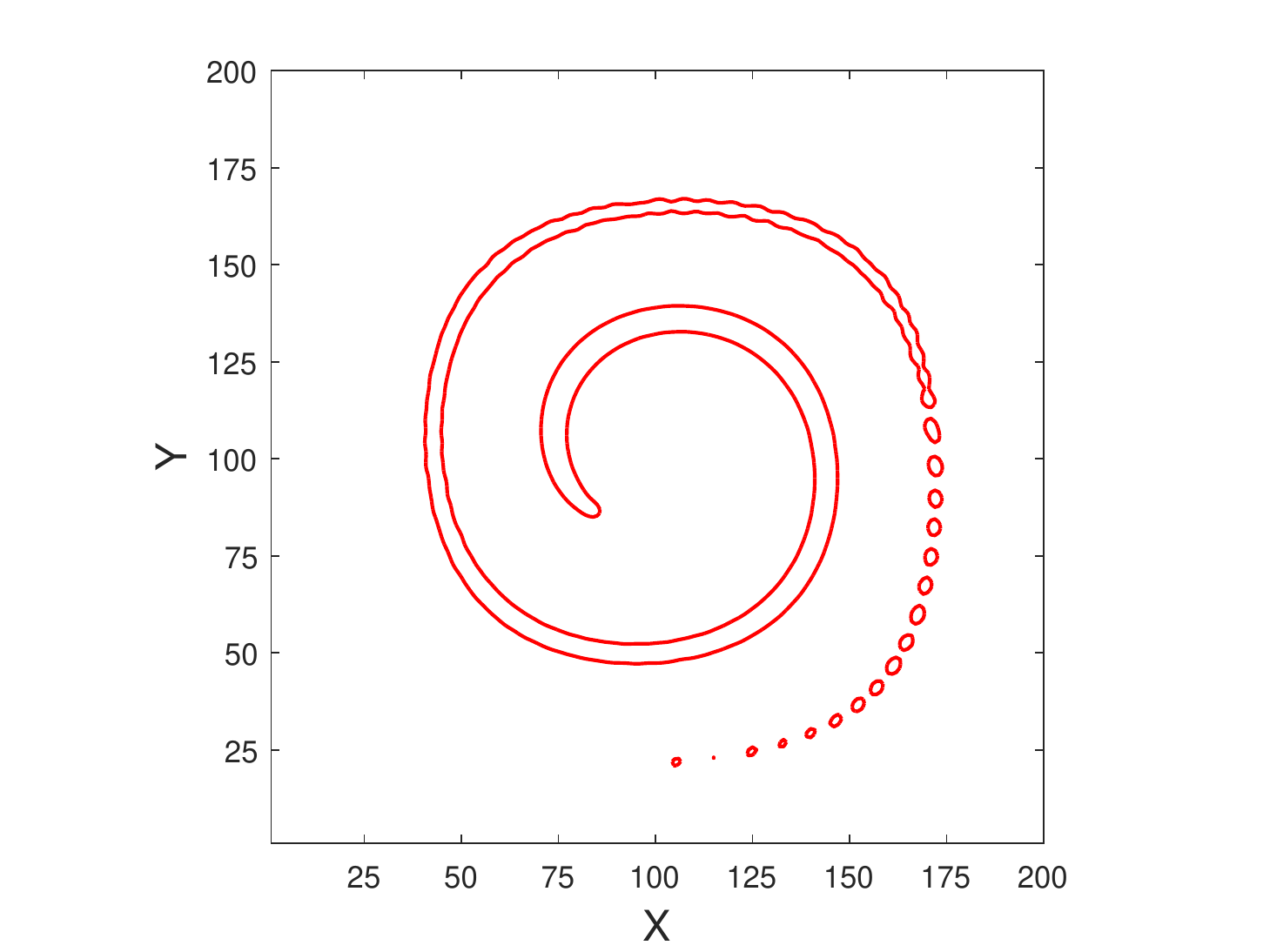}~
\includegraphics[width=0.25\textwidth,trim=30 5 30 10,clip]{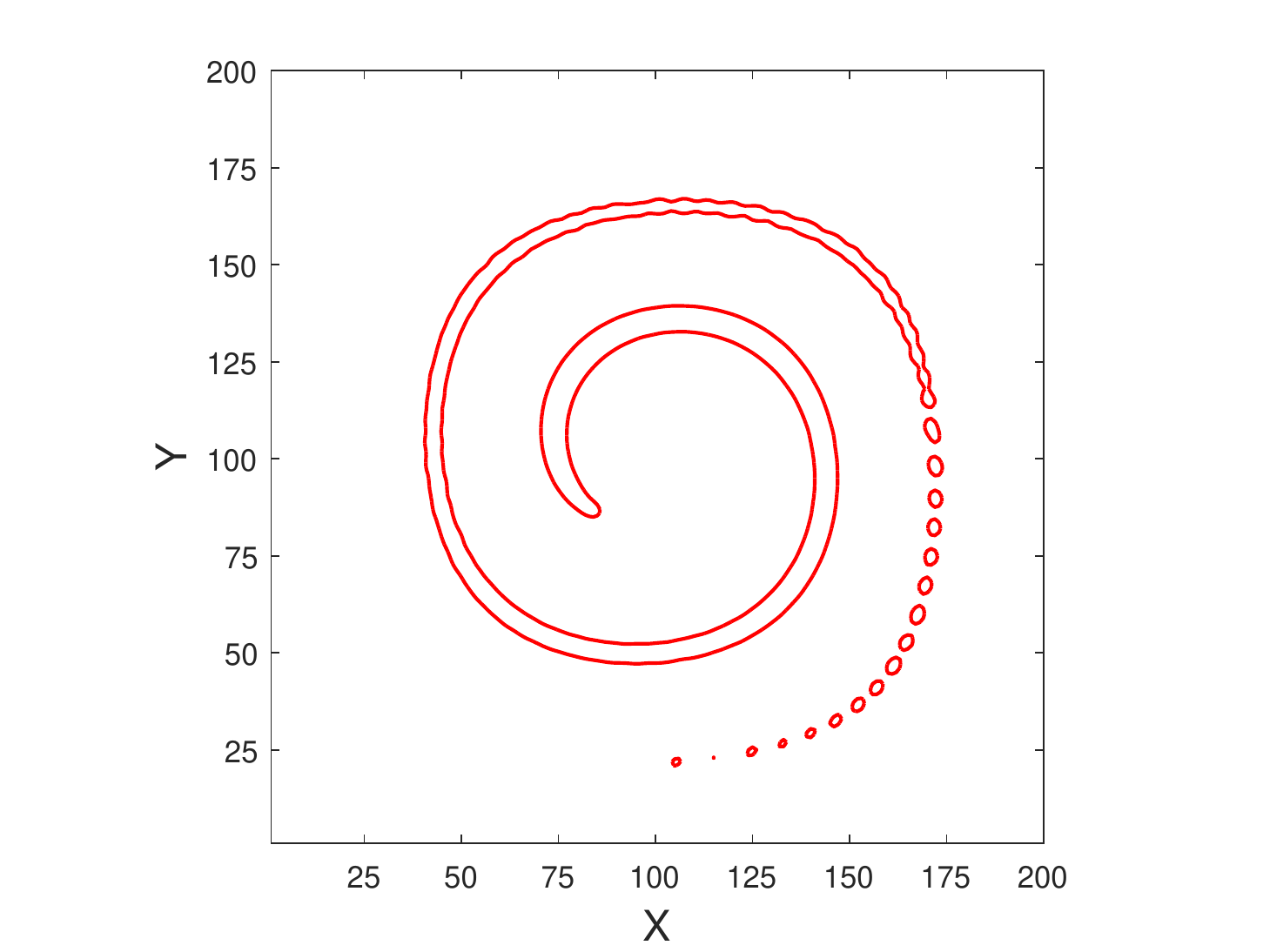}
}\\
\subfloat[t=T]{
\includegraphics[width=0.25\textwidth,trim=30 5 30 10,clip]{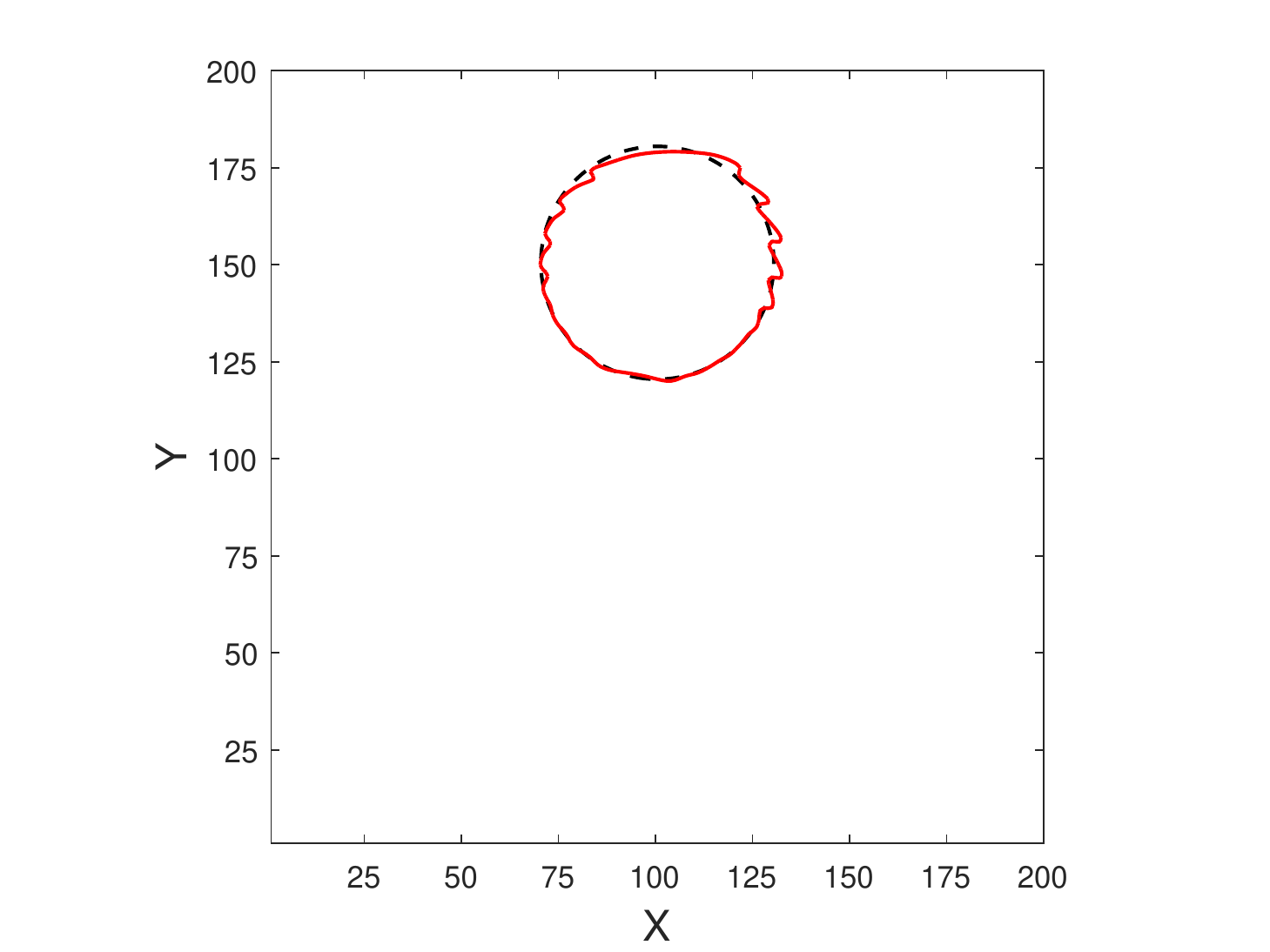}~
\includegraphics[width=0.25\textwidth,trim=30 5 30 10,clip]{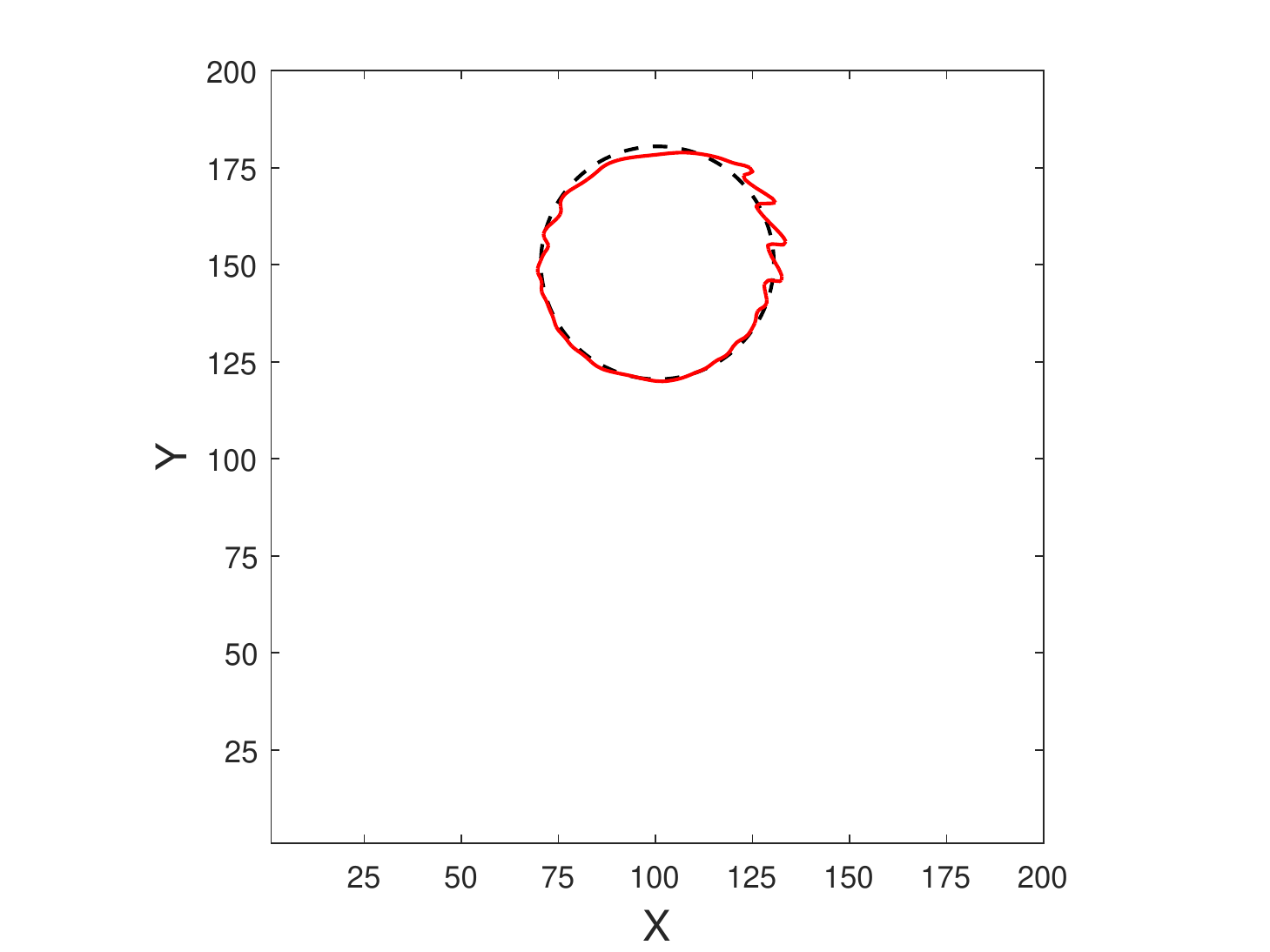}~
\includegraphics[width=0.25\textwidth,trim=30 5 30 10,clip]{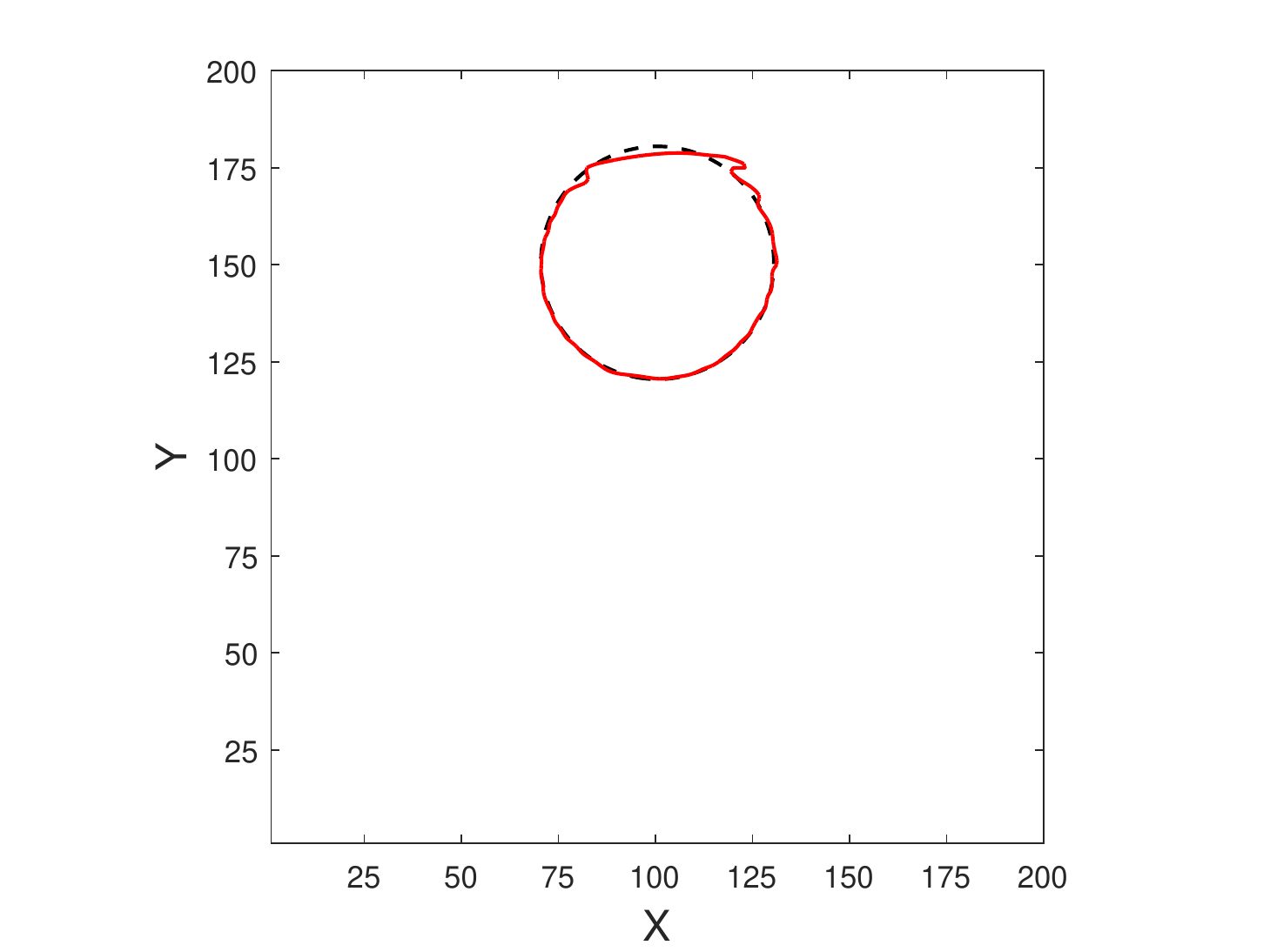}~
\includegraphics[width=0.25\textwidth,trim=30 5 30 10,clip]{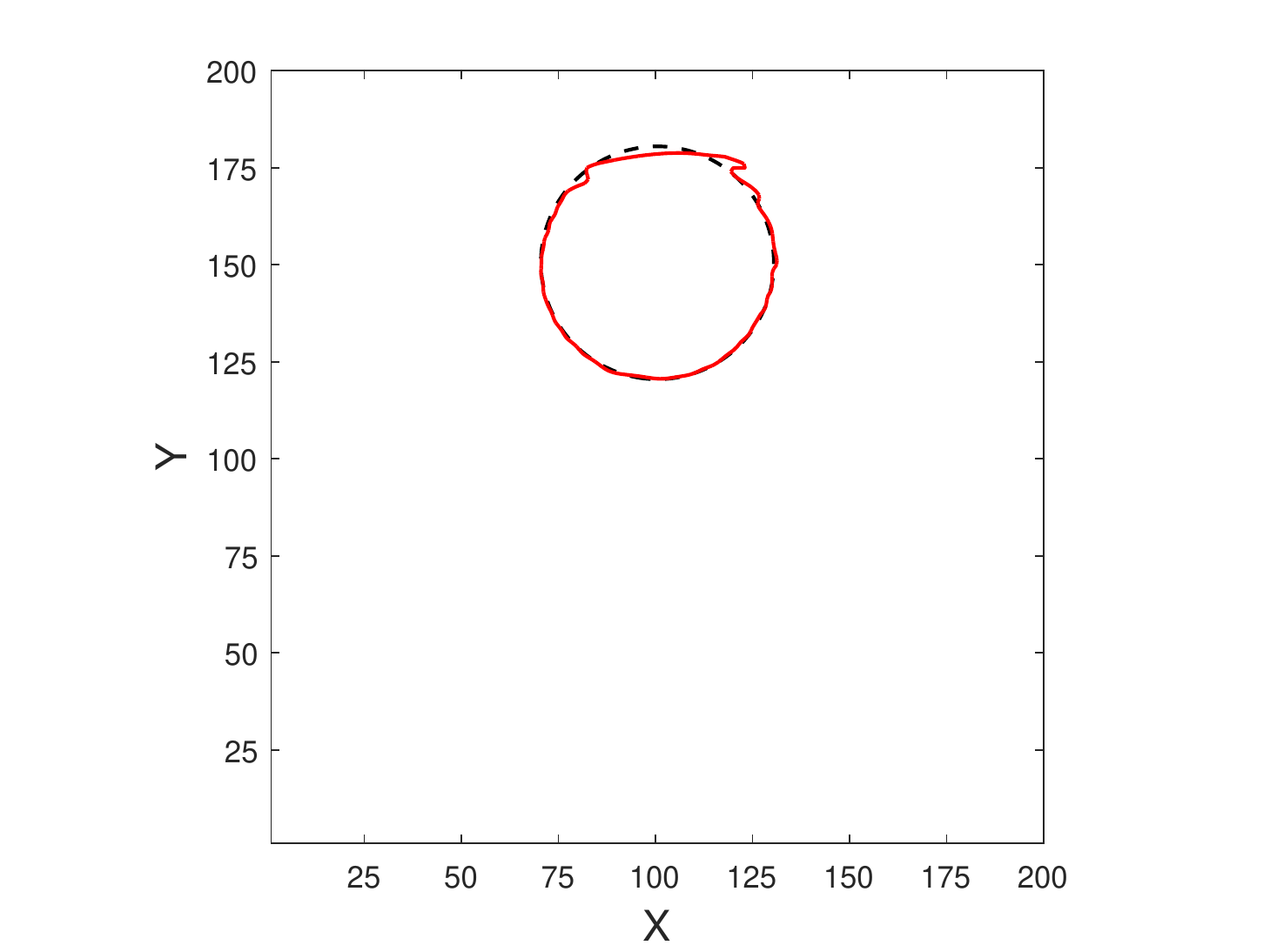}}~\\
\caption{The phase-field contour ($\phi=0$) of vortex deformation of a circle at (a) t=T/2 and (b) t=T. From left to right, the results  are obtained by LBE-AC, DUGKS-AC, DUGKS-I and DUGKS-II.  Black dashed lines denotes the initial profile  and red solid line represents the reconstructed interface.}
\label{fig:test3-Tn2}
\end{figure}

\begin{figure}
  \centering
  \includegraphics[width=0.5\textwidth]{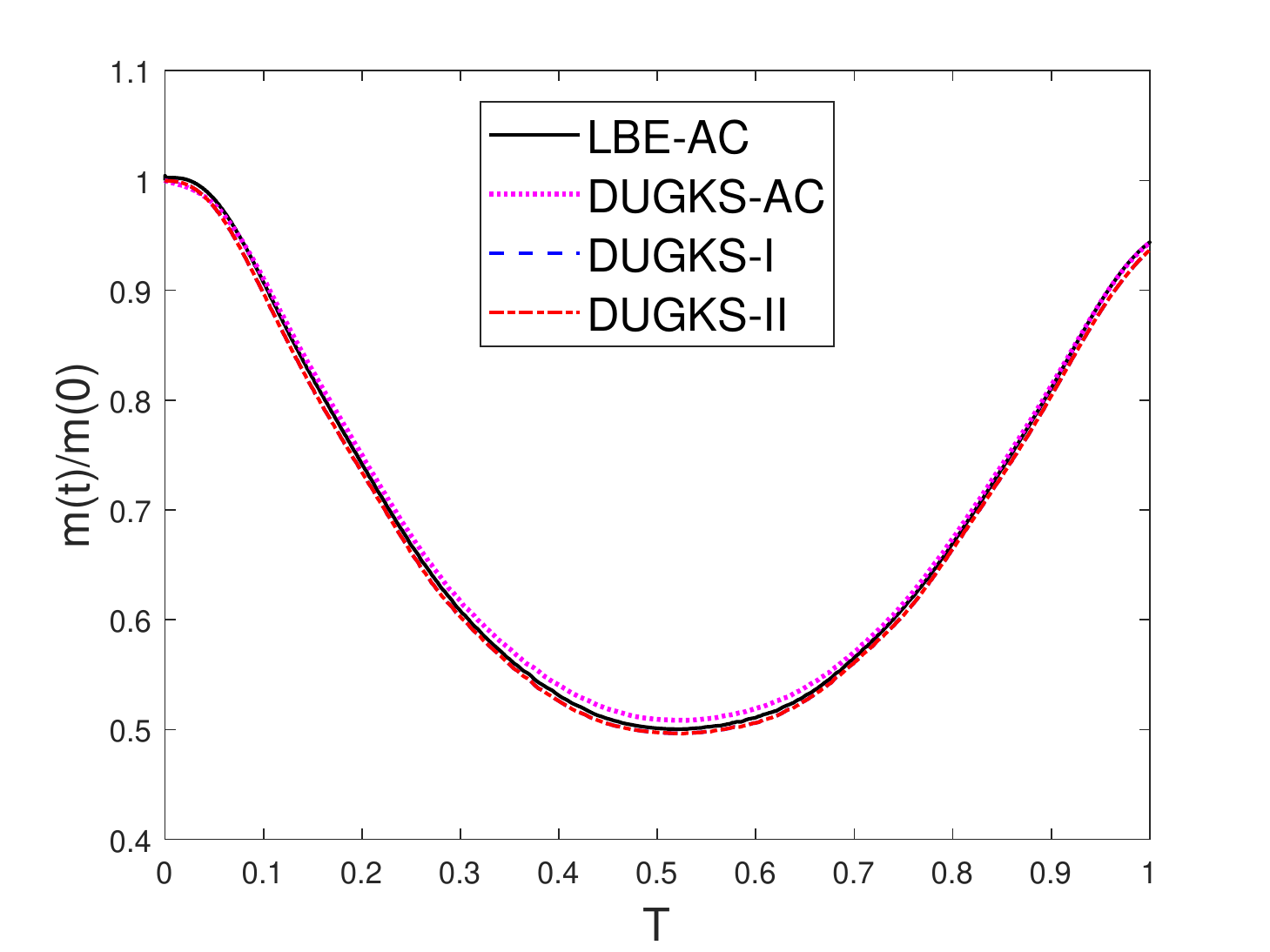}
  \caption{Evolution of the mass of the circle during deformation.}\label{fig:test3-massloss}
\end{figure}

\begin{figure}
  \centering
  \subfloat[]{
  \includegraphics[width=0.5\textwidth]{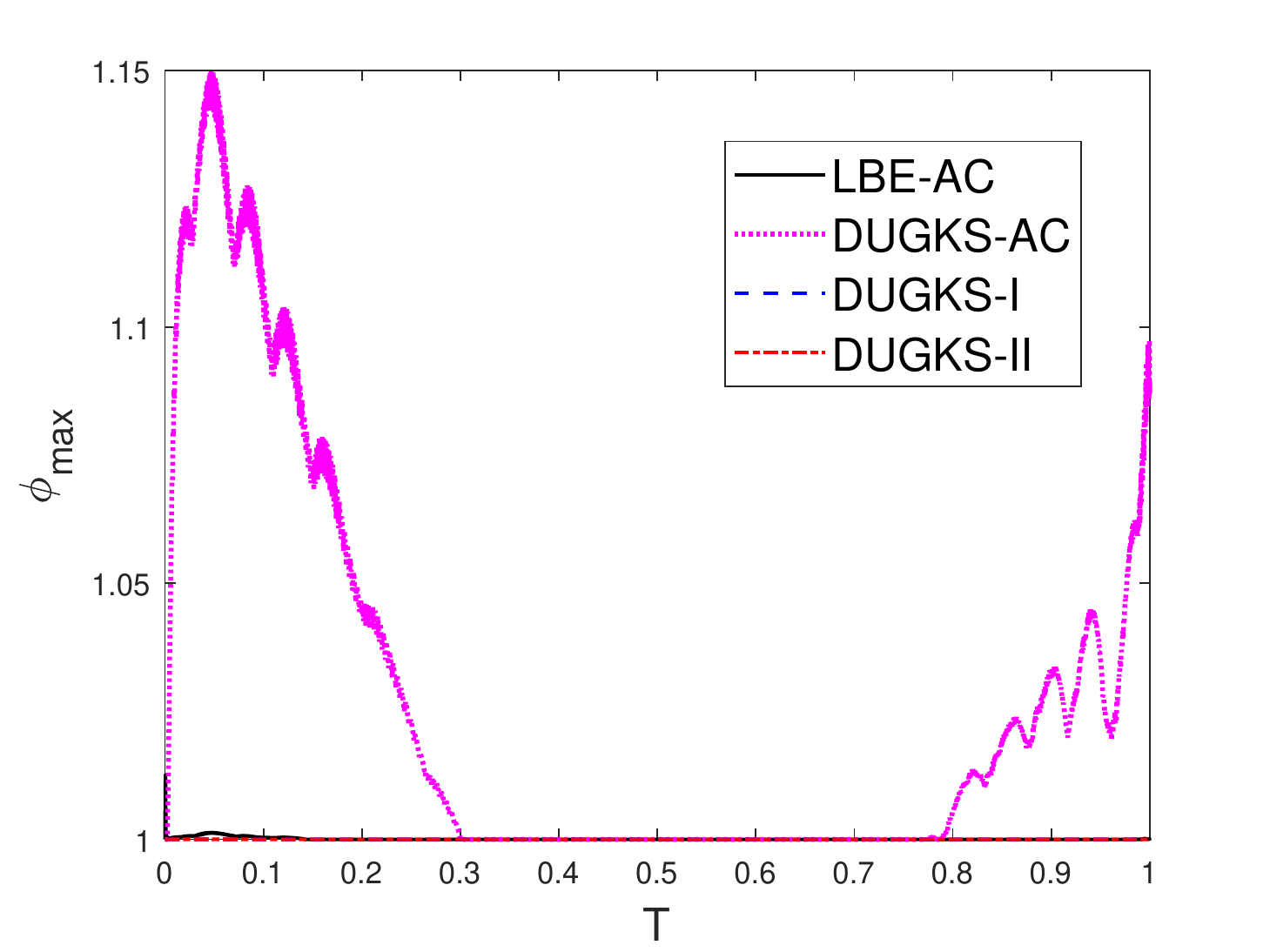}}
  \subfloat[]{
  \includegraphics[width=0.5\textwidth]{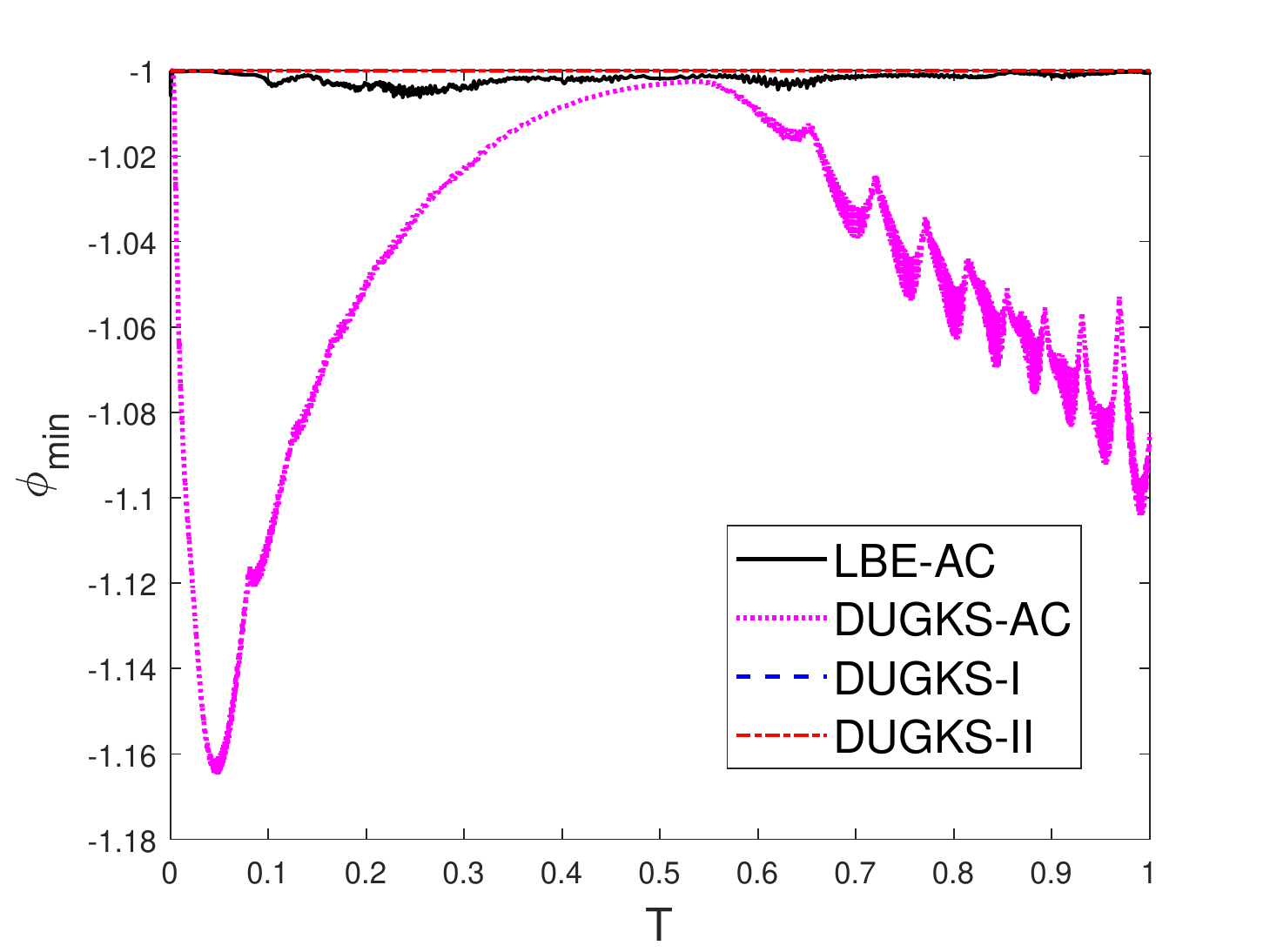}}
  \caption{Evolution of (a) maximum and (b) minimum values of order parameter during deformation.}\label{fig:test3-maxminPhi}
\end{figure}

We further test the capability of the present DUGKS models by simulating a severe deformation of a circular interface.
Initially, a circle with a radius of $R=0.15L$ is centered at $(0.5L,0.75L)$ in a square computational domain $L\times L$. The solenoidal velocity field is given by~\cite{zhang2019high,sun2007sharp}
\begin{equation}\label{eq}
\begin{aligned}
u(x,y,t)=&U_0\sin^2\left(\frac{\pi x}{L_0}\right)\sin\left(\frac{2\pi y}{L_0}\right)\cos\left(\frac{\pi t}{T}\right), \\
v(x,y,t)=&-U_0 \sin^2\left(\frac{\pi y}{L_0}\right)\sin\left(\frac{2\pi x}{L_0}\right)\cos\left(\frac{\pi t}{T}\right),
\end{aligned}
\end{equation}
where $T=nL_0/U_0$ and $n$ is fixed at $8$. The $\cos(\pi t/T)$ term is used to reverse the velocity field smoothly.
 The prescribed velocity field will produce a strong shear flow that can significantly stretch and tear the interface.
 Based on the property of  $\cos(\pi t/T)$,  the circle will undergo the largest deformation at $T/2$ and come back to its initial position at $T$,
at which the errors can be evaluated by Eq.(\ref{eq:L2-errors}).
The parameters are set as $L_0=200$, $U_0=0.02$, $T=2L_0/(U_0 \Delta t)$ and $\chi=0.5$.
Figure~\ref{fig:test3-Tn2} shows the restored interfaces of the circle at $t=T/2$ and $t=T$ for all four models.
At $t=T/2$, the results obtained by all methods are similar.
At $t=T$,  the final circle is distorted and the loss of mass becomes apparent.
In comparison, the final shapes of the interface obtained by both DUKGS-I and DUGKS-II agree better with the initial one than the results given by DUGKS-AC and LBE-AC.
Specifically, the $L_2$ errors at $t=T$  are $0.0666$, $0.0779$, $0.0579$, $0.0579$ for LBE-AC, DUGKS-AC, DUGKS-I and DUGKS-II, respectively.
The history of mass of the circle during deformation is  also measured by $m(t)=\int_{\phi(\bm x,t)>0}\phi(\bm x,t) d\bm x$ and shown in Fig.~\ref{fig:test3-massloss}.
At $t=T$, the mass loss of the circle given by DUGKS-AC, DUGKS-I and DUGKS-II are $5.73\times 10^{-2}$, $5.65\times 10^{-2}$, $6.34\times 10^{-2}$ and $6.34\times 10^{-2}$, respectively.
Compared with DUGKS-AC and LBE-AC, the mass loss for both DUGKS-I and DUGKS-II is slightly large.
Finally, the  order parameter that exceeds its reasonable range  can affect the accuracy and stability of the model, especially for multiphase flows with large density ratios. Hence, we also measured the maximum and minimums values of the order parameter during evolution and the measured results are plotted in Fig.\ref{fig:test3-maxminPhi}. It can be seen that DUGKS-I and DUGKS-II can remain the values of the order parameter within the reasonable range while the values of the order parameter predicted by DUGKS-AC and LBE-AC are beyond the theoretical maximum and minimum values.

\section{CONCLUSIONS}
In this work, two sets of discrete kinetic equation with BGK collision operator are introduced for the conservative Allen-Cahn equation. One without time-dependent terms in the force term is able to correctly recover the CACE up to second-order accuracy and the other with time-dependent terms in the force term can  recover the CACE with some additional terms
$\frac{M_\phi}{c_s^2}\nabla\cdot\left[
 \partial_t (\phi \bm u)+\nabla\cdot(\phi \bm u\bm u) \right]$. As the additional terms  are of the order $\text{Ma}^2/\text{Pe}$, it is expected that they have little effect on the numerical results under small Mach number. Then, the DUGKS scheme as a finite volume method is employed to discretize both kinetic equations for the CACE.
By analysing the discrete velocity  kinetic equation  recovered from the flux evaluation in the previous DUGKS, it is found that  some high order terms that are order of $O(\delta t)$ appear when the force term is involved or the first moment of the collision model has no conservation property, such as, CHE, ACE and the convection diffusion equation.
To correctly recover the target kinetic equation, the improved flux evaluation with  parabolic reconstruction  is proposed in the DUGKS scheme.

To test the performance of the proposed models, three benchmark problems are simulated and the results are also compared with those obtained by the published DUGKS-AC and LBE-AC. Numerical results show that both kinetic models
are capable of capturing the interface with improved accuracy compared with DUGKS-AC.
It is also shown that the calculation of the distribution function at the interface has an important effect on the numerical results. In the considered reconstruction schemes, the WENO scheme is the best.
Meanwhile, the results obtained by DUGKS-I and DUGKS-II are almost identical in all simulations. This implies that the error terms  in DUGKS-I really have little effect on the results and can be neglected . Due to the lack of the calculation of temporal derivative, DUGKS-I is preferred in terms of  computational efficiency.
Numerical results demonstrate that the proposed DUGKS model can greatly improve the accuracy of capturing the interface and the results are comparable with those obtained by LBE-AC.  On the other hand, the current model can effectively control the value of the order parameter within the reasonable range.  The usage of irregular mesh is easily performed due to the finite volume properties in DUGKS. These features could further  improve the numerical stability  and accuracy in multiphase flows with large density ratios, which will be presented in a subsequent paper.

\section*{ACKNOWLEDGEMENTS}
This work was supported by the National Numerical Wind Tunnel program, the National Natural Science Foundation of China (Grant No.51836003,11972142, 51806142, 91852205, 91741101, and 11961131006), NSFC Basic Science Center Program (Award number 11988102),  Guangdong Provincial Key Laboratory of Turbulence Research and Applications (2019B21203001), Guangdong-Hong Kong-Macao Joint Laboratory for Data-Driven Fluid Mechanics and Engineering Applications (2020B1212030001), and Shenzhen Science and Technology Program (Grant No. KQTD20180411143441009). Computing resources are provided by the Center for Computational Science and Engineering of Southern
University of Science and Technology.

\begin{appendix}
\section{Derivation of the ACE from discrete Boltzmann equation with BGK collision model}\label{ap:dukgs-ace}
From the definitions of Eqs.(\ref{eq:macro-gks}) and (\ref{eq:force-gks}), we can have
\begin{subequations}\label{ap:gks-moments}
\begin{align}
\sum_\alpha F_\alpha=0, \quad \sum_\alpha F_\alpha\bm \xi_\alpha=c_s^2\Theta\bm n, \quad
\sum_\alpha F_\alpha\bm \xi_\alpha \bm \xi_\alpha=0, \\
\sum_\alpha f=\sum_\alpha f_\alpha^{eq}=0,\quad \sum_\alpha f_\alpha^{eq}=\phi\bm u, \quad \sum_\alpha f_\alpha \bm \xi_\alpha\bm \xi_\alpha=
c_s^2\phi \bm I + \phi\bm u \bm u,
\end{align}
\end{subequations}

With the help of Eq.(\ref{ap:gks-moments}), the zeroth moment of Eq.(\ref{eq:kineticequaiton}) becomes
\begin{equation}\label{ap:zero-moment}
\partial_t\phi+\nabla\cdot\sum_\alpha(\bm \xi_\alpha f_\alpha)=0.
\end{equation}
The key step is to evaluate the expression of $\sum_\alpha \bm \xi_\alpha f_\alpha$.
From Eq.(\ref{eq:kineticequaiton}), one can have
\begin{equation}\label{ap:relation}
f_\alpha=f_\alpha^{eq}+\tau_f\left[ F_\alpha-\partial_t f_\alpha-\nabla\cdot \bm \xi_\alpha f_\alpha \right].
\end{equation}
It can be found that $f=f_\alpha^{eq}+O(\tau_f)$ and substituting it into Eq.(\ref{ap:relation}) lead to
\begin{equation}\label{aq:relation-feq}
f_\alpha\approx f_\alpha^{eq}+\tau_f\left[ F_\alpha-\partial_t f_\alpha^{eq}-\nabla\cdot \bm \xi_\alpha f^{eq}_\alpha \right]
+O(\tau_f^2\partial).
\end{equation}
Multiplying Eq.(\ref{aq:relation-feq}) by $\bm \xi_\alpha$ and taking summation over the subscript $\alpha$ results in
\begin{equation}\label{aq:relation-feq-sum}
\sum_\alpha\bm \xi_\alpha f_\alpha \approx \phi\bm u-c_s^2\tau_f\left[
\partial_t(\phi \bm u)+\nabla\cdot(\phi \bm u\bm u)+\nabla\phi - \Theta\bm n)\right]+O(\tau_f^2\partial^2).
\end{equation}
Substituting Eq.(\ref{aq:relation-feq-sum}) into Eq.(\ref{ap:zero-moment}) gives
\begin{equation}\label{eq}
\partial_t \phi+\nabla\cdot(\phi\bm u)=\nabla\cdot M(\nabla\phi-\Theta\bm n)+\frac{M}{c_s^2}\nabla\cdot \left(\partial_t(\phi \bm u)+\nabla\cdot(\phi\bm u\bm u)\right)+O(\tau_f^2\partial^2),
\end{equation}
where $M_\phi=c_s^2 \tau_f$ is the mobility.

\section{Truncation error analysis}\label{ap:dugks-reconstruction}
For simplicity, we assume that the grid points $(x_i,y_j)$ are uniformly distributed with the cell size $h=\Delta x=\Delta y$ and cell centers $(x_c,y_c)$.  Then, we consider a Taylor Series expansion of the function $\varphi(x,y)$ about the point $\bm (x_c,y_c)$, i.e,
\begin{equation}\label{ap2:taylor-varphi}
\begin{aligned}
\varphi(x,y) & = \varphi (x_c,y_c)+ \varphi_x (x-x_c)+ \varphi_y (y-y_c)
+\frac{\varphi_{xx}\cdot(x-x_c)^2}{2}
+\varphi_{x,y}(x-x_c)(y-y_c)
+\frac{\varphi_{yy}\cdot(y-y_c)^2}{2}\\
& +\frac{\varphi_{xxx}}{6}(x-x_c)^3
+\frac{\varphi_{xxy}(x-x_c)^2(y-y_c)}{2}
+\frac{\varphi_{xyy}(x-x_c)(y-y_c)^2}{2}
+\frac{\varphi_{yyy}(y-y_c)^3}{6}
 +\frac{\varphi_{xxxx}(x-x_c)^4}{24} \\
&+\frac{\varphi_{xxxy}(x-x_c)^3(y-y_c)}{6}
+\frac{\varphi_{xxyy}(x-x_c)^2(y-y_c)^2}{4}
+\frac{\varphi_{xyyy}(x-x_c)(y-y_c)^3}{6}
+\frac{\varphi_{yyyy}(y-y_c)^4}{24}+\ldots
\end{aligned}
\end{equation}
where $\varphi$ denotes any continuous variable and
$\bar{\varphi}_x=\frac{\partial \varphi(x_c,y_c)}{\partial x}$, $\bar{\varphi}_{xx}=\frac{\partial^2 \varphi(x_c,y_c)}{\partial x^2}$ and so on.
Combining Eqs.(\ref{eq:spatial_average}) and (\ref{eq:cell-face-integral}), we can obtain the following expressions
\begin{equation}\label{ap2:cell-average}
\bar{\varphi}_{i,j}=\varphi_{i,j}+\frac{h^2}{24}(\nabla^2 \varphi)_{i,j}+\frac{h^4}{1920}\left(\nabla^2\varphi +\frac{4}{3}\varphi_{xxyy}  \right)_{i,j}+O(h^6).
\end{equation}
\begin{equation}\label{eq}
\bar{\varphi}_{i+\frac{1}{2},j}=\varphi_{i,j}+\frac{h}{2}(\varphi_x)_{i,j}+\frac{h^2}{12}(\varphi_{xx} )_{i,j}
+\frac{h^2}{24}(\nabla^2 \varphi)_{i,j}+\frac{h^3}{48}(\nabla^2\varphi_x)_{i,j}+O(h^4),
\end{equation}
\begin{equation}\label{eq}
(\bar{\varphi}_x)_{i+\frac{1}{2},j}=(\varphi_x)_{i,j}+\frac{h}{2}(\varphi_{xx})_{i,j}+\frac{h^2}{12}(\varphi_{xxx})_{i,j}
+\frac{h^2}{24}(\nabla^2 \varphi_x)_{i,j}+\frac{h^3}{48}(\nabla^2\varphi_{xx} )_{i,j}+O(h^4),
\end{equation}
\begin{equation}\label{eq}
(\bar{\varphi}_y)_{i+\frac{1}{2},j}=(\varphi_y)_{i,j}+\frac{h}{2}(\varphi_{xy})_{i,j}+\frac{h^2}{12}(\varphi_{xxy})_{i,j}
+\frac{h^2}{24}(\nabla^2 \varphi_y)_{i,j}+\frac{h^3}{48}(\nabla^2\varphi_{xy} )_{i,j}+O(h^4),
\end{equation}
where $\varphi_{i,j}=\varphi(x_i,y_j)$.
The neighboring cell averages can be obtained by the following expression~\cite{shukla2014isotropic}
\begin{equation}\label{aq:cell-variable-relation}
\begin{aligned}
\bar{\varphi}_{i+m,j+n}&=\varphi_{i,j}+h(m \varphi_x+n\varphi_y)_{i,j}
+h^2\left[
\frac{12m^2+1}{24}\varphi_{xx}+mn \varphi_{xy}+\frac{12n^2+1}{24}\varphi_{yy}
\right]_{i,j}\\
&+h^3\left[
m\frac{4m^2+1}{24}\varphi_{xxx}+n\frac{12n^2+1}{24}\varphi_{xxy}+m\frac{12n^2+1}{24}\varphi_{xyy}+n\frac{4n^2+1}{24}\varphi_{yyy}
\right]_{i,j}+O(h^4),
\end{aligned}
\end{equation}
where $m$ and $n$ are integers.
Eq.(\ref{aq:cell-variable-relation}) can be used to establish  the truncation error of the approximation of the values of the variable and its derivative at the cell face. For example,
for the following second-order differentiation formulations,
\begin{equation}\label{ap:second-order-CDM}
\begin{aligned}
(\bar{\varphi}^{C2})_{i+\frac{1}{2},j}=&\frac{\bar{\varphi}_{i,j}+\bar{\varphi}_{i+1,j}}{2}, \\
(\bar{\varphi}^{C2}_x)_{i+\frac{1}{2},j}=&\frac{\bar{\varphi}_{i+1,j}-\bar{\varphi}_{i,j}}{h}, \\
(\bar{\varphi}^{C2}_y)_{i+\frac{1}{2},j}=&\frac{\bar{\varphi}_{i+\frac{1}{2},j+1}-\bar{\varphi}_{i+\frac{1}{2},j-1}}{2h}, \\
\end{aligned}
\end{equation}
the truncation errors are given by
\begin{subequations}\label{eq}
\begin{align}
(\bar{\varphi}^{C2})_{i+\frac{1}{2},j}=\bar{\varphi}_{i+\frac{1}{2},j}+\frac{h^2}{6}(\varphi_{xx})_{i,j}+\frac{h^3}{12}(\varphi_{xxx})_{i,j}+O(h^4), \\
(\bar{\varphi}^{C2}_x)_{i+\frac{1}{2},j}=(\bar{\varphi}_x)_{i+\frac{1}{2},j}+\frac{h^2}{12}(\varphi_{xxx})_{i,j}
-\frac{h^3}{48}(\nabla^2\varphi_{xx})_{i,j}+O(h^4),
 \\
(\bar{\varphi}^{C2}_y)_{i+\frac{1}{2},j}=(\bar{\varphi}_y)_{i+\frac{1}{2},j}+\frac{h^2}{6}(\nabla^2\varphi_{y})_{i,j}
-\frac{h^3}{48}\nabla^2\varphi_{xy}+O(h^4).
\end{align}
\end{subequations}
It can be found that the truncation errors depend on the orientation of the solution with respect to the Cartesian grid, which may contribute to the abnormal interface behavior. To remove the directional derivatives that appear in the lowest order term in the truncation error,  the following reconstruction formulations can be employed
\begin{equation}\label{ap2:isotropical-cd2-face}
\begin{aligned}
(\bar{\varphi}^{C2})_{i+\frac{1}{2},j}^I&=
\frac{(\bar{\varphi}^{C2})_{i+\frac{1}{2},j+1}+4(\bar{\varphi}^{C2})_{i+\frac{1}{2},j}+(\bar{\varphi}^{C2})_{i+\frac{1}{2},j-1}}{6}\\
&=\bar{\varphi}_{i+\frac{1}{2},j}
+\frac{h^2}{6}(\nabla^2\varphi)_{i,j}+\frac{h^3}{12}(\nabla^2\varphi_x)_{i,j}+O(h^4),
\end{aligned}
\end{equation}
\begin{equation}\label{ap2:isotropical-cd2-derivative}
\begin{aligned}
(\bar{\varphi}_x^{C2})_{i+\frac{1}{2},j}^I&=
\frac{(\bar{\varphi}_x^{C2})_{i+\frac{1}{2},j+1}+10(\bar{\varphi}_x^{C2})_{i+\frac{1}{2},j}+(\bar{\varphi}_x^{C2})_{i+\frac{1}{2},j-1}}{12} \\
&=(\bar{\varphi}_x)_{i+\frac{1}{2},j}
+\frac{h^2}{12}(\nabla^2\varphi_x)_{i,j}-\frac{h^3}{48}(\nabla^2\varphi_{xx})_{i,j}+O(h^4),
\end{aligned}
\end{equation}
Similar expressions can be easily obtained in a analogous manner for the values of $\bar{\varphi}_{i,j+\frac{1}{2}}$, $(\bar{\varphi}_x)_{i,j+\frac{1}{2}}$ and $(\bar{\varphi}_y)_{i,j+\frac{1}{2}}$.

If we replace the cell center point $(x_c,y_c)$ by the center of the cell face in Eq.(\ref{ap2:taylor-varphi}), the following
equations can be obtained,
\begin{equation}\label{ap2:face-average}
\begin{aligned}
(\bar{\varphi})_{i+\frac{1}{2},j} &=\varphi_{i+\frac{1}{2},j}+\frac{h^2}{24}(\varphi_{yy})_{i+\frac{1}{2},j}+\frac{h^4}{1920}(\varphi_{yyyy})_{i+\frac{1}{2},j},\\
(\bar{\varphi})_{i,j+\frac{1}{2}} &=\varphi_{i,j+\frac{1}{2}}+\frac{h^2}{24}(\varphi_{xx})_{i,j+\frac{1}{2}}+\frac{h^4}{1920}(\varphi_{xxxx})_{i,j+\frac{1}{2}}. \\
\end{aligned}
\end{equation}
From Eqs.(\ref{ap2:cell-average}) and (\ref{ap2:face-average}), it can be found that the pointwise and face-averaged values are interchangeable when the second-order truncation error is acceptable.

\section{Reconstruction of the distribution function $f_\alpha^+$ at the cell face}\label{ap:dugks-reconstruction}
To calculate the advection flux in Eq.(\ref{eq:evolution-dugks}), the distribution function $\widehat{f}_\alpha^+$ at the cell face must be carefully reconstructed.
In this study, we compared  the second-order linear interpolation, fourth-order line interpolation, third-order WENO and fifth-order WENO for the reconstruction of $\widehat{f}_\alpha^+$ at the cell face.
Taking $(\widehat{f}_\alpha^+)_{i+\frac{1}{2},j}$ in the x-direction as example,
the second-order linear interpolation is given by
\begin{equation}\label{eq}
(\widehat{f}_\alpha^+)_{i+\frac{1}{2},j}=\frac{(\widehat{f}_\alpha^+)_{i,j}+(\widehat{f}_\alpha^+)_{i+1,j}  }{2}
\end{equation}
The fourth-order linear interpolation is~\cite{felker2018fourth,hyman1992high}
\begin{equation}\label{eq}
(\widehat{f}_\alpha^+)_{i+\frac{1}{2},j}=\frac{(7\widehat{f}_\alpha^+)_{i,j}+7(\widehat{f}_\alpha^+)_{i+1,j}
-(\widehat{f}_\alpha^+)_{i+2,j} -(\widehat{f}_\alpha^+)_{i-1,j}
 }{12}.
\end{equation}
\textcolor{magenta}{When $\bm \xi_\alpha>0$, the third order WENO scheme for the value  of $(\widehat{f}_\alpha^+)_{i+\frac{1}{2},j}$ is given by
\begin{equation}\label{eq}
(\widehat{f}_\alpha^+)_{i+\frac{1}{2},j}=w_1\left[
\frac{1}{2}(\widehat{f}_\alpha^+)_{i,j}+\frac{1}{2}(\widehat{f}_\alpha^+)_{i+1,j}
 \right]
+w_2\left[
-\frac{1}{2}(\widehat{f}_\alpha^+)_{i-1,j}+\frac{3}{2}(\widehat{f}_\alpha^+)_{i,j}
\right]
\end{equation}
and the classical weight functions proposed by Jiang and Shu~\cite{shu1998essentially,jiang1996efficient} are as follows
\begin{equation}\label{eq}
w_k=\frac{\bar{\omega}_k}{\bar{\omega}_1+\bar{\omega}_2},\qquad  \bar{\omega}_k=\frac{\gamma_k}{(\epsilon+\zeta_k)^p},\qquad k=1,2.
\end{equation}
where $\gamma_1=2/3$ and $\gamma_2=1/3$ are the optimal  weights. The power parameter $p\geq 1$ is used to enhance the relative ratio between the smoothness indicator $\zeta_k$. The sensitivity parameter $\epsilon>0$ is used to avoid divisions by zero.
  The smoothness indicators are defined as $\zeta_1=(
(\widehat{f}_\alpha^+)_{i+1,j}-(\widehat{f}_\alpha^+)_{i,j}
)$, $\zeta_2=(
(\widehat{f}_\alpha^+)_{i,j}-(\widehat{f}_\alpha^+)_{i-1,j}
)$.
In the third-order WENO-Z scheme~\cite{don2013accuracy}, the non-linear weights are defined as
\begin{equation}\label{eq}
\bar{\omega}_k=\gamma_k\left[1+\left(\frac{\tau_z}{\epsilon+\zeta_k}\right)^p \right], k=1,2, \qquad \tau_z=|\zeta_1-\zeta_2|.
\end{equation}
We take $\epsilon=10^{-6}$ and $p=1$ in our numerical simulations. The formulations for negative wind case are symmetric with respect to the point $x_{i+\frac{1}{2}}$}.  

\textcolor{magenta}{If the fifth order finite difference WENO scheme is employed, $ (\widehat{f}_\alpha^+)_{i+\frac{1}{2},j}$ is obtained by
\begin{equation}\label{eq}
(\widehat{f}_\alpha^+)_{i+\frac{1}{2},j}=w_1(\widehat{f}_\alpha^+)_{i+\frac{1}{2},j}^{(1)} +w_2(\widehat{f}_\alpha^+)_{i+\frac{1}{2},j}^{(2)}+w_3 (\widehat{f}_\alpha^+)_{i+\frac{1}{2},j}^{(3)}
\end{equation}
where $(\widehat{f}_\alpha^+)_{i+\frac{1}{2},j}^{(k)}$ are three third order fluxes on three different stencils given by
\begin{equation}\label{eq}
\begin{aligned}
(\widehat{f}_\alpha^+)_{i+\frac{1}{2},j}^{(1)}=& \frac{1}{3} (\widehat{f}_\alpha^+)_{i-2,j}
-\frac{7}{6}(\widehat{f}_\alpha^+)_{i-1,j} +\frac{11}{6} (\widehat{f}_\alpha^+)_{i,j}, \\
(\widehat{f}_\alpha^+)_{i+\frac{1}{2},j}^{(2)}=& -\frac{1}{6} (\widehat{f}_\alpha^+)_{i-1,j}
+\frac{5}{6}(\widehat{f}_\alpha^+)_{i,j} +\frac{1}{3} (\widehat{f}_\alpha^+)_{i+1,j}, \\
(\widehat{f}_\alpha^+)_{i+\frac{1}{2},j}^{(3)}=& \frac{1}{3} (\widehat{f}_\alpha^+)_{i,j}
+\frac{5}{6}(\widehat{f}_\alpha^+)_{i+1,j} -\frac{1}{6} (\widehat{f}_\alpha^+)_{i+2,j},
\end{aligned}
\end{equation}
and the classical nonlinear weights $w_k$ are given by~
\begin{equation}\label{eq}
  w_k=\frac{\bar{\omega}_k}{\sum_{k=1}^{3}\bar{\omega}_k},\quad  \bar{\omega}_k=\frac{\gamma_k}{(\epsilon+\zeta_k)^2},
\end{equation}
where the optimal  weights are given by $\gamma_1=0.1, \gamma_2=0.6, \gamma_3=0.3$. The smoothness indicators $\zeta_k$ are given by
\begin{equation}\label{eq}
\begin{aligned}
\zeta_1 &=\frac{13}{12}(
(\widehat{f}_\alpha^+)_{i-2,j}-2(\widehat{f}_\alpha^+)_{i-1,j}+(\widehat{f}_\alpha^+)_{i,j}
)^2
+\frac{1}{4}(
(\widehat{f}_\alpha^+)_{i-2,j}-4(\widehat{f}_\alpha^+)_{i-1,j}
+3(\widehat{f}_\alpha^+)_{i,j}
)^2,
\\
\zeta_2 &=\frac{13}{12}(
(\widehat{f}_\alpha^+)_{i-1,j}-2(\widehat{f}_\alpha^+)_{i,j}+(\widehat{f}_\alpha^+)_{i+1,j}
)^2+\frac{1}{4}(
(\widehat{f}_\alpha^+)_{i-1,j}-(\widehat{f}_\alpha^+)_{i+1,j}
)^2,
\\
\zeta_3 &=\frac{13}{12}(
(\widehat{f}_\alpha^+)_{i,j}-2(\widehat{f}_\alpha^+)_{i+1,j}+(\widehat{f}_\alpha^+)_{i+2,j}
)^2+\frac{1}{4}(
3(\widehat{f}_\alpha^+)_{i,j}-4(\widehat{f}_\alpha^+)_{i+1,j}
+(\widehat{f}_\alpha^+)_{i+2,j}
)^2.
\end{aligned}
\end{equation}
In the WENO-Z scheme, the nonlinear weights are defined as
\begin{equation}\label{eq}
  \bar{\omega}_k=\gamma_k\left( 1+\left(\frac{\tau_z}{\epsilon+\zeta_k}\right)^p\right),\qquad \tau_z=|\zeta_1-\zeta_3|.
\end{equation}
The interested reader is referred to Refs.\cite{shu1998essentially,borges2008improved,jiang1996efficient,don2013accuracy} for details.
Similar expressions can be easily obtained in a analogous manner for the value of $(\widehat{f}_\alpha^+)_{i,j+\frac{1}{2}}$.}
\end{appendix}
\section*{References}
\bibliography{mybib}
\end{document}